\documentclass{arxiv_cls}
\usepackage{pg2022}

\usepackage[T1]{fontenc}
\usepackage{dfadobe}  

\usepackage{cite}  %
\BibtexOrBiblatex
\electronicVersion
\PrintedOrElectronic
\ifpdf \usepackage[pdftex]{graphicx} \pdfcompresslevel=9
\else \usepackage[dvips]{graphicx} \fi

\usepackage{egweblnk} 

\usepackage{egweblnk}
\usepackage[dvipsnames]{xcolor}
\usepackage{lipsum}
\usepackage{todonotes}

\definecolor{myblue}{rgb}{0.0156, 0.2431, 0.5843}
\definecolor{gray}{rgb}{0.7,0.7,0.7}
\definecolor{orange}{rgb}{1.0,0.5,0.0}
\newcommand*{\rv}{\textcolor{black}}

\newif\ifcomments
\commentstrue
\commentsfalse

\newcommand{\tobias}[1]{\ifcomments \todo[color=red!50]{T: #1} \else \fi}

\newcommand{\yun}[1]{\ifcomments \todo[color=yellow!50]{Y: #1} \else \fi}
\newcommand{\yuninline}[1]{\ifcomments \todo[inline,color=yellow!50]{Y: #1} \else \fi}
\newcommand{\martin}[1]{\ifcomments \todo[color=blue!30]{MN: #1} \else \fi}

\newcommand{\soeren}[1]{\ifcomments \todo[color=teal!30]{S: #1} \else \fi}
\newcommand{\soereninline}[1]{\ifcomments \todo[inline,color=teal!30]{S: #1} \else \fi}

\title[MixedMetroMap]%
      {\rv{Shape-Guided} Mixed Metro Map Layout}%

\author[T. Batik \& S. Terziadis \& Y.-S. Wang \& M. N{\"o}ellenburg \& H.-Y.~Wu]
{\parbox{\textwidth}{\centering 
T. Batik
$^{1}$\orcid{0000-0002-9764-9409},
S. Terziadis
$^{1}$\orcid{0000-0001-5161-3841},
Y.-S. Wang
$^{2}$\orcid{0000-0003-2550-2990},
M. N{\"o}llenburg
$^{1}$\orcid{0000-0003-0454-3937},
and H.-Y.~Wu
$^{3,1}$\orcid{0000-0003-1028-0010} 
}
\author[T. Batik \& S. Terziadis \& Y.-S. Wang \& M. N{\"o}llenburg \& H.-Y.~Wu]
{\parbox{\textwidth}{\centering 
T. Batik$^{1}$\orcid{0000-0002-9764-9409},
S. Terziadis$^{1}$\orcid{0000-0001-5161-3841},
Y.-S. Wang$^{2}$\orcid{0000-0003-2550-2990},
M. N{\"o}llenburg$^{1}$\orcid{0000-0003-0454-3937},
and H.-Y.~Wu$^{3,1}$\orcid{0000-0003-1028-0010} 
}
\\
{\parbox{\textwidth}{\centering 
$^1$TU Wien, Austria 
$^2$National Yang Ming Chiao Tung University, Taiwan 
$^3$St. P{\"o}lten University of Applied Sciences, Austria
      }
}
}
}

\usepackage{amssymb}
\usepackage{enumerate}       
\usepackage{amsmath}
\usepackage{tabularx}
\usepackage{appendix}
\usepackage{svg}
\usepackage{hyperref}
\usepackage[shortlabels]{enumitem}

\usepackage{lineno}
\EGpagenumber

\newcommand{\connectionsSmooth}{C_{\text{shape}}'}
\newcommand{\stationsSmooth}{S_{\text{shape}}'}
\newcommand{\connectionsOctiMixed}{\Tilde{C}_{\text{octo}}}
\newcommand{\connectionsSmoothMixed}{\Tilde{C}_{\text{shape}}}

\begin{document}

\teaser{
\centering{
 \setlength{\tabcolsep}{4pt}
 \begin{tabular}{ccc}
    \includegraphics[page=2,width=0.29\linewidth]{figures/teaser/Taipei/Taipei_label.pdf} &
    \includegraphics[page=3,width=0.29\linewidth]{figures/teaser/Taipei/Taipei_label.pdf} &
    \includegraphics[width=0.29\linewidth]{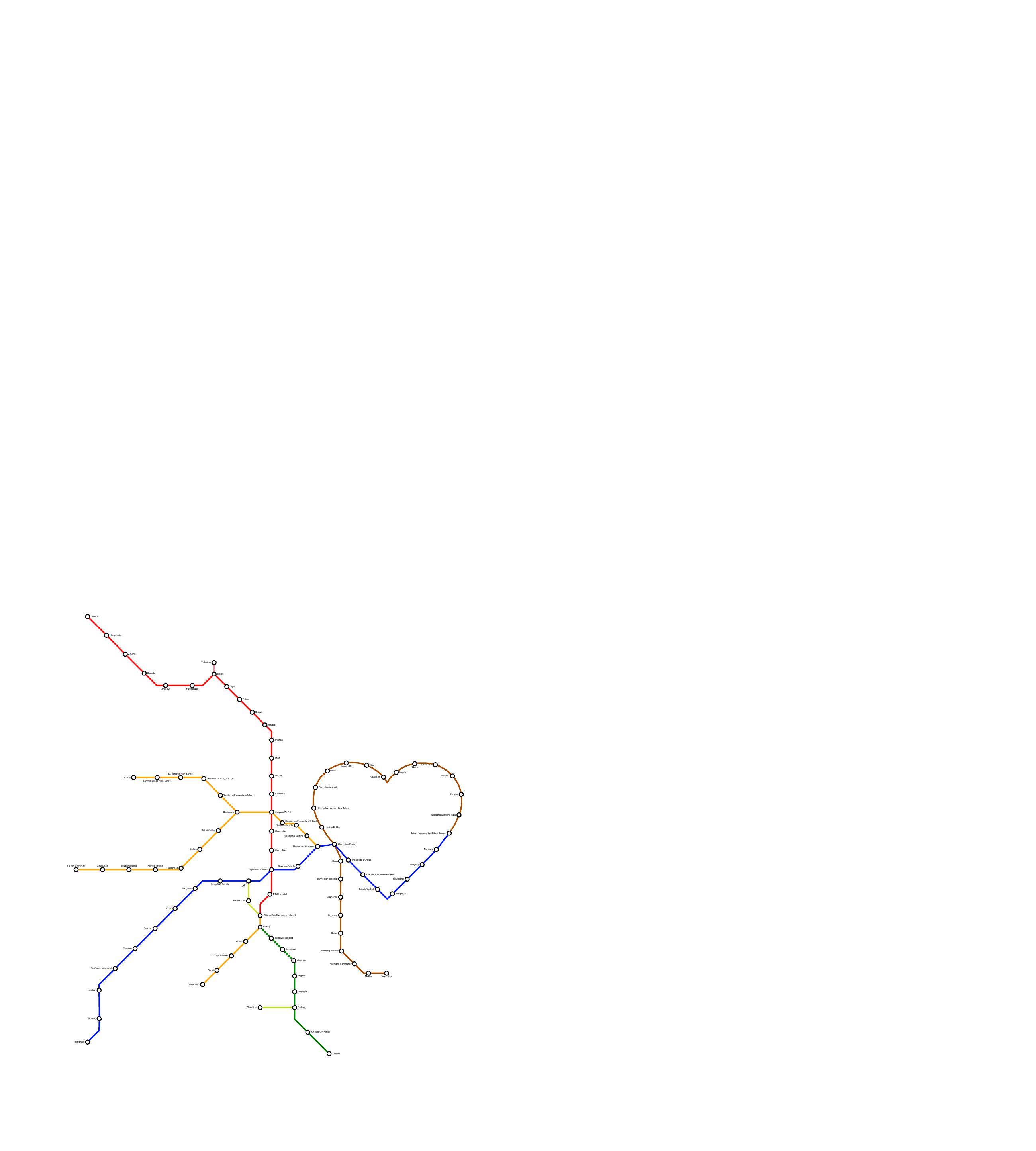} \\
   (a) & \rv{(b)} & (c) \\
 \end{tabular}
}
\caption{An example of Taipei metro maps, including (a) the input transit network and the guide shape \rv{(gray)}, and \rv{the output (b) without and (c) with the shape.}}
\label{fig:teaser}
}

\maketitle
\begin{abstract}
Metro or transit maps, are \rv{schematic representations} of transit networks to facilitate effective route-finding.
These maps are often advertised on a web page or \rv{pamphlet highlighting routes from source to destination stations}.
To visually support such route-finding, designers often distort the layout by embedding symbolic \rv{shapes} (e.g., circular routes) in order to guide readers' attention \rv{(e.g., Moscow map and Japan railway map)}. However, manually producing such maps is labor-intensive and the effect of \rv{shapes} remains unclear.
In this paper, we propose an approach to generalize such mixed metro maps that take user-defined shapes as an input. 
In this mixed design, lines that are used to approximate the shapes are arranged symbolically, while the remaining lines follow classical layout convention. 
A three-step algorithm, including (1) detecting and selecting routes for shape approximation, (2) shape and layout deformation, and (3) aligning lines on a grid, is integrated to guarantee good visual quality.
Our contribution lies in the definition of the mixed metro map problem and the formulation of design criteria so that the problem can be resolved systematically using the optimization paradigm.
Finally, we evaluate the performance of our approach and perform \rv{a user study to test if the embedded shapes are recognizable or reduce the map quality.}


\end{abstract}  

\section{Introduction}
\label{sec:intro}

A metro or transit map used to show transportation line services is an intuitive, schematic representation of a transit network.
Here, a schematic representation is a simplified network geometry (e.g., straightened lines, uniform spacing of stations, etc.), to facilitate effective way-finding activities~\cite{r-wytesmd-14}.
\rv{Such nicely arranged representations make transit maps popular visual metaphors for network visualization in physics, biology, engineering, etc.~\cite{wu-ascsnm-19}.
However, manually creating schematic maps} is not straightforward and requires intensive iterative processes. 
Automatic approaches have been thus investigated to solve this high-complexity problem~\cite{wu-stml-2020}, \rv{however, most existing approaches aim for one} single style. 
For example, an \rv{\emph{octolinear} (also \emph{octilinear}) layout} limits all edge orientations to horizontal, vertical, or diagonal at $45^\circ$~\cite{nw-dlhqm-11, srmw-amlumo-11, wc-fmm-11}.
A \emph{curvilinear layout} constrains metro lines as continuous and smooth curves~\cite{fhnrsw-dmmubc-12}.
Other styles such as \emph{concentric circles}~\cite{flw-cmm-2014} that \rv{align metro lines along concentric circles,} or \emph{multilinear}~\cite{nickel-tdmmm-2020} designs \rv{that relax octolinear layout by allowing multiple edge directions were also proposed and their algorithmic complexity and perceptual effectiveness were investigated.}

\begin{figure}[t]
    \centering{
    \setlength{\tabcolsep}{1pt}
    \begin{tabular}{cc}
        \includegraphics[width=0.61\linewidth]{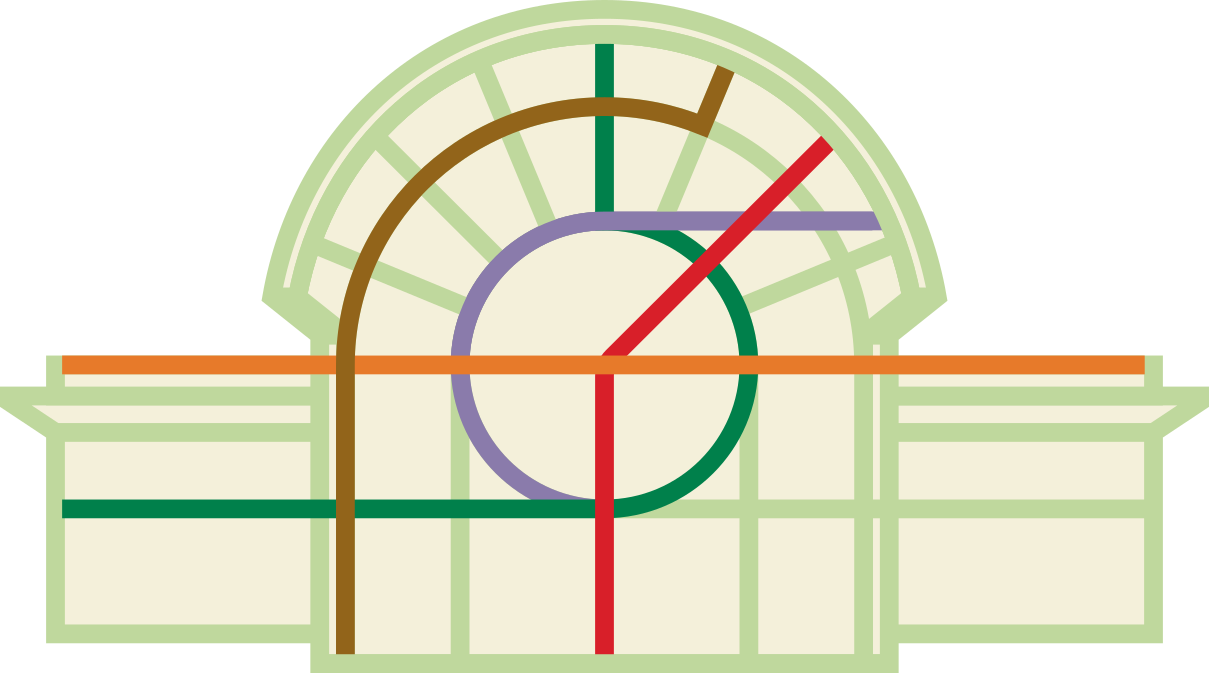} &
        \includegraphics[width=0.34\linewidth, trim=0 20 0 40,clip]{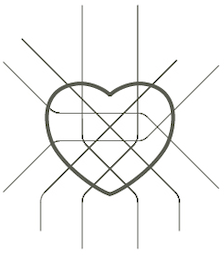} \\
        (a) Schematic Mapping Workshop logo & 
        (b) \rv{Embedded heart} \\
    \end{tabular}
    \caption{Examples of shape-embedded layouts, including (a) the logo of Schematic Mapping Workshop 2019 \rv{(courtesy of Dr. Maxwell J. Roberts)}, and (b) an example illustration from One Metro World~\cite{c-omw-16} \rv{(courtesy of Mr. Jug Cerovi{\'c})}.}
    \label{fig:example}
    }
\end{figure}

However, in reality, if we take a close look at the handcrafted metro maps by professional designers~\cite{c-omw-16,zerozero, transitmaps},
multiple styles are often incorporated in a single diagram.
Designers often distinguish primary and secondary information to emphasize the difference between specific lines in the system.
To achieve this, some lines in the map are arranged using a simple geometry (e.g., circle, etc.) or an iconic shape (e.g., heart, etc.), while the remaining lines follow the layout convention. 
The official Moscow map with a circular route highlighted is a typical example~\cite{moscow}, 
while maps with more complicated shapes~\cite{zerozero} are more often used for advertising purposes or special events \rv{\cite{kwan2016pyramid, wang2019shapewordle}}.
Figure~\ref{fig:example} shows more examples, where a designer embedded the Vienna metro map to create a workshop logo~\cite{schematic19}, and Cerovi{\'c} explains the influence of shapes in mental map development~\cite{c-omw-16}. 
Nevertheless, as summarized by Wu et al.\cite{wu-stml-2020}, creating transit maps is an iterative process, and an automatic approach provides opportunities for scientists to effectively investigate advantages and disadvantages of various metro maps.

There exist automatic approaches for synthesizing metro maps with more than one style; however, supported geometry (i.e., shape) is limited to either polylines~\cite{wtly-tmla-12} or circles~\cite{wptaly-dammll-15} to the best of our knowledge.
\rv{This paper presents a more general approach to the \emph{mixed layout problem}, which allows the integration of specific shapes and classical metro map layout. This is achieved by introducing} a user-defined \emph{guide shape}, represented as single or multiple polylines, as input. 
Figure~\ref{fig:teaser} shows an example, where we embed a heart shape into the Taipei metro system.
This is done with a three-step approach by firstly matching routes for shape approximation, then deforming the network layout, and finally, aligning lines on a grid.
Note that each individual step here has been categorized as a challenging problem~\cite{funke2015compass,wu-stml-2020,lvw-sgdgo-16}.
To guarantee the visual quality of the generated maps, the guide shape should be recognizable in the final layout, while the remaining part of the layout should still fulfill the classical octolinear design criteria~\cite{nw-dlhqm-11}.
Our contributions in the paper are summarized:
\begin{itemize}
    \item A general \rv{definition and solution for the mixed metro map problem beyond the state-to-the-art}.
    \item An algorithmic approach to perform shape matching, shape approximation, and metro line alignment \rv{as a whole}.
    \item Quantitative and qualitative evaluation \rv{to test shape recognizability and the corresponding influence on route planning tasks}.
\end{itemize}

The paper is structured as follows:
In Section~\ref{sec:related}, we summarize the state-of-the-art literature relevant to our approach.
We then begin with a definition of the proposed problem, followed by an explanation of the design criteria and a high-level description of our algorithm (Section~\ref{sec:overview}).
The method used to solve the proposed problem is detailed in Sections~\ref{sec:route}-\ref{sec:align}.
\rv{We collect several results (Section~\ref{sec:result}), examine the approach performance and quality (Section~\ref{sec:discuss}), and} conclude this paper and propose open topics in Section~\ref{sec:conclude}.

\section{Related Work}
\label{sec:related}

Metro maps are designed to help passengers navigate transit lines when taking trains in a rail system. Since train routes are fixed and passengers mainly need to know at which station to get on or off a train, some sort of distortions are allowed in metro maps for improving readability. Since exact geographic information is no longer needed, cartographers usually enlarge downtown areas to label station names and prevent visual clutters \cite{ovenden2008paris,schwetman2014harry}. The stations on metro lines are re-positioned to fulfill several criteria, such as even spacing of stations \cite{degani2013tale,lloyd2017modernism}, minimum direction changes in routes \cite{roberts2014schematic}, and the relative positions of stations \cite{roberts2017preference}. Generally, the layout of a metro map can be curvilinear, multilinear \cite{roberts2017preference}, k-linear (including octolinear, hexalinear, and tetralinear) \cite{nickel-tdmmm-2020}, and even based on concentric circles \cite{roberts2016radi,niedermann2019efficient}. In addition to the network map, text or image labels are placed around stations for passengers to connect the map and the real world \cite{ovenden2015transit}. The labels are expected to be overlap-free, close to their corresponding stations, and have consistent orientations if they represent neighboring stations.

\begin{figure*}[thb]
    \centering{
    \setlength{\tabcolsep}{0pt}
    \begin{tabular}{ccccc}
        \includegraphics[width=0.20\linewidth]{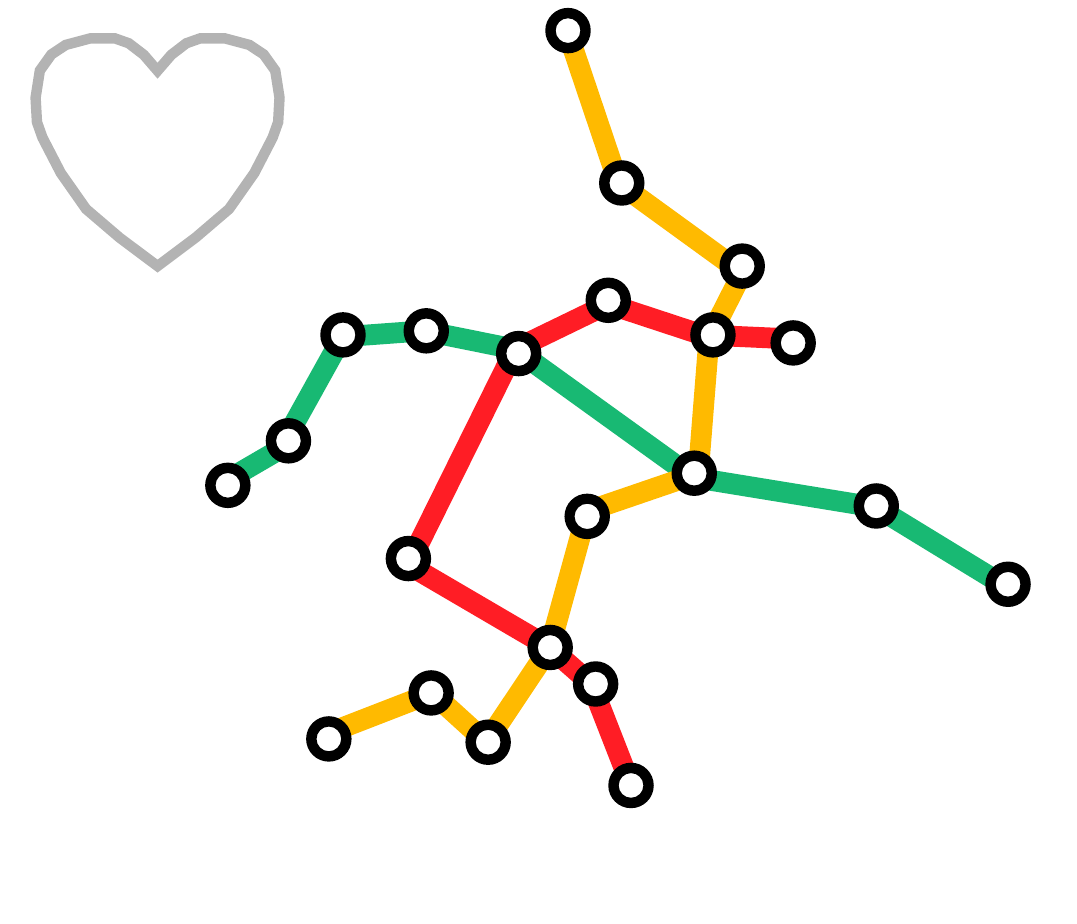} &
        \includegraphics[width=0.20\linewidth]{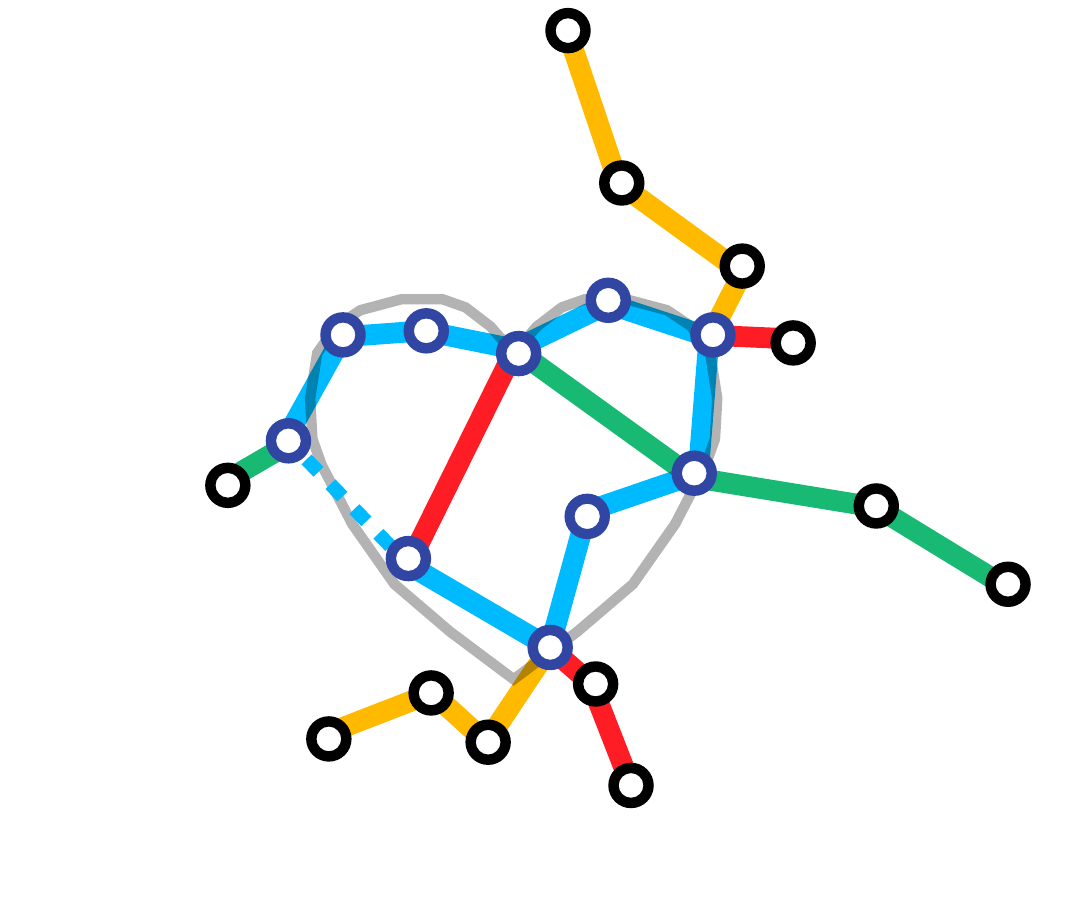} &
        \includegraphics[width=0.20\linewidth]{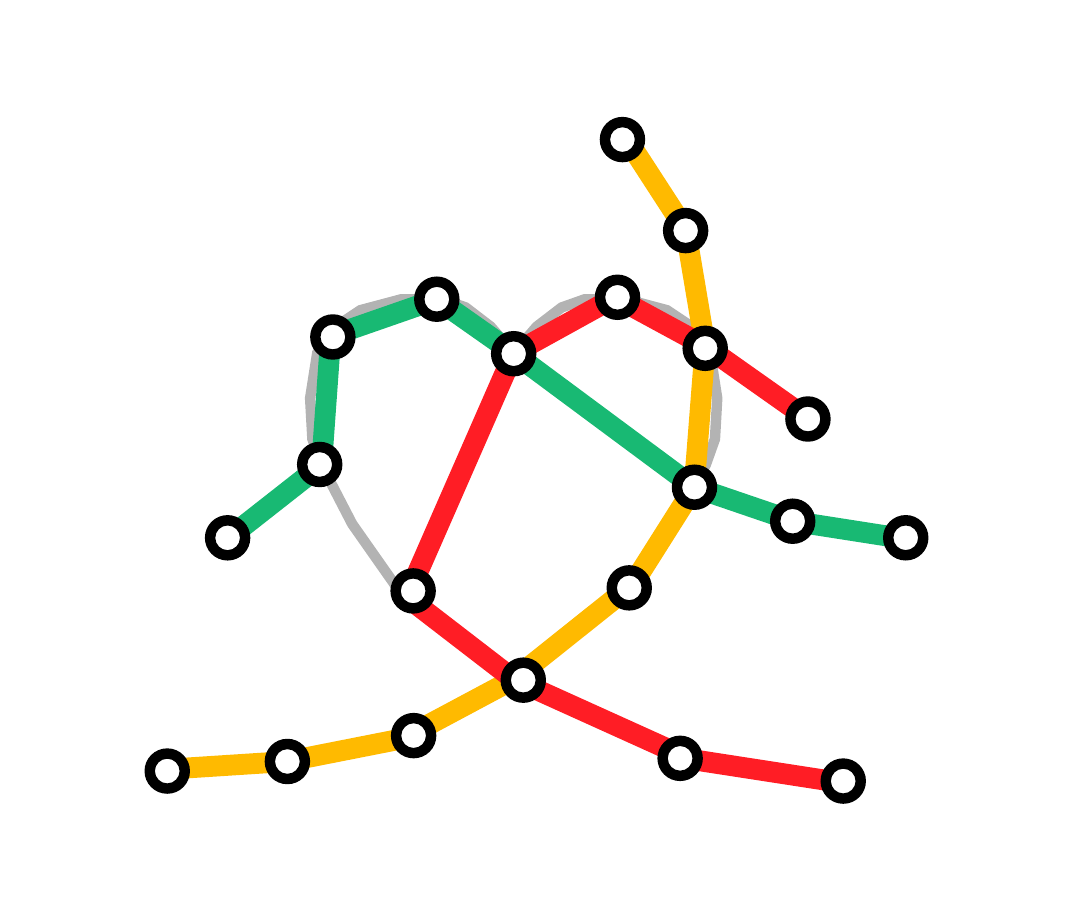} &
        \includegraphics[width=0.20\linewidth]{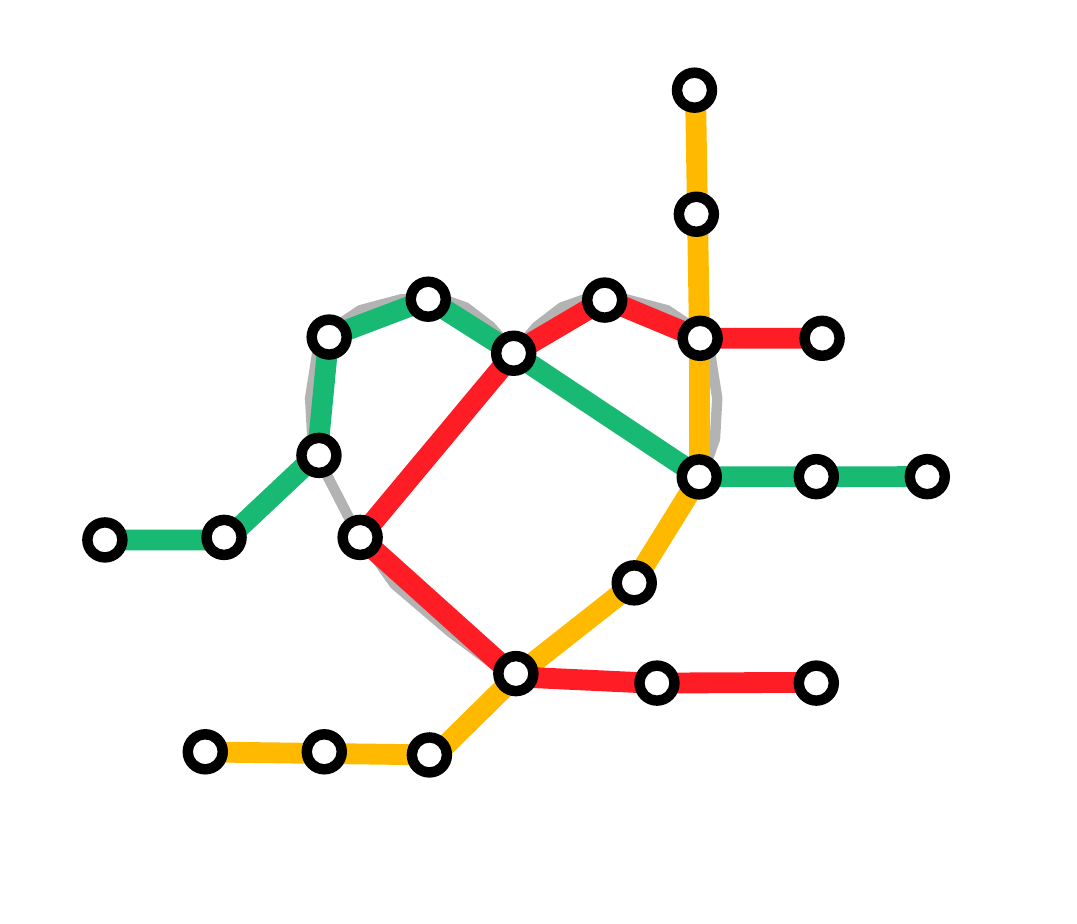} &
        \includegraphics[width=0.20\linewidth]{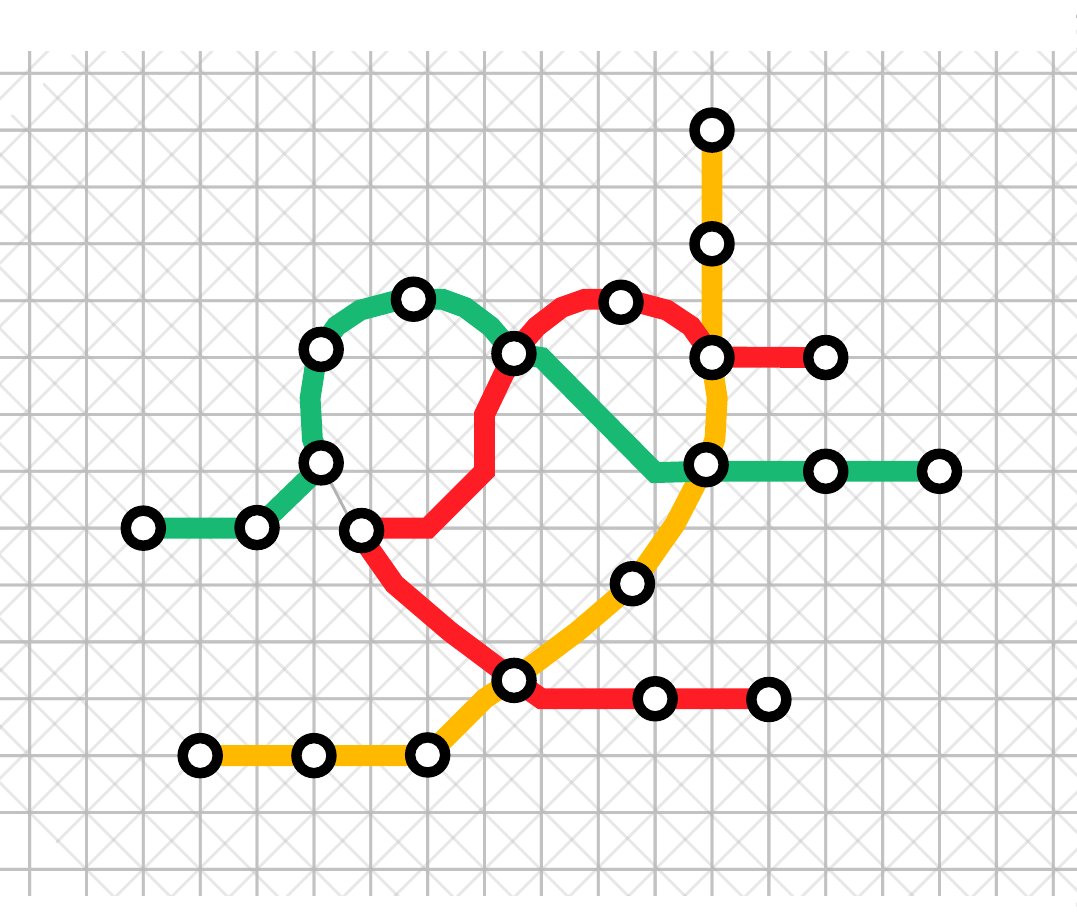} \\
        (a) Input network and shape & 
        (b) Route matching & 
        (c) Smooth deformation & 
        (d) Octolinear deformation & 
        (e) Grid alignment \\
    \end{tabular}
    \caption{Algorithmic pipeline of our layout approach. (a) The input metro network and a user-specified shape. (b) Shape and route matching, shown in blue. (c) Shape fitting using a smooth layout. (d) Shape fitting using an octolinear layout. (e) Final layout after aligning stations and lines on a modified grid, which includes the shape.
    }
    \label{fig:overview}
    }
\end{figure*}

\textbf{Schematic Layout Approaches.}
So far the most commonly used metro map layout is octolinear, which was introduced by Henry Beck for the London Underground in 1933. It is known that computing an optimal octolinear layout is NP-hard~\cite{nollenburg2005automated}. 
Several categories of methods were then introduced to pursue high-quality metro map layouts within a reasonable computation time. The first category is line simplification, which attempts to find a set of linked line segments to represent each path on the map \cite{neyer1999line,merrick2006path,dwyer2008fast,cabello2005schematization,li2010stroke}. During the simplification process, the deviation between each station and the closest line segment is bounded, and the number of line segments is minimized. The second is local optimization. Methods in this category iteratively move stations to nearby positions to align edges in octolinear directions \cite{avelar2000generating,avelar2007convergence}. Several strategies, such as simulated annealing \cite{mark2006automated}, ant colony optimization \cite{ware2013ant}, magnetic force field \cite{chivers2014octilinear}, and least-squares optimization \cite{van2018realtime} were applied to prevent the algorithm from falling into local minimums. The third category is to formulate a metro map layout problem into a mixed-integer linear program \rv{\cite{nw-dlhqm-11,wptaly-dammll-15}}, which distinguishes hard constraints that must be fulfilled and soft constraints that are globally optimized. Typically, the octolinear layout is mandatory \rv{since edge orientation is formulated using a set of binary variables in the model}, whereas other aesthetic requirements can be relaxed if they conflict with others.
Finally a recent result computes shortest paths between candidate positions of stations on a predefined grid~\cite{BastBS20,BastBS21}.
This was formulated as an ILP, but heuristical solutions through repeated applications of shortest path algorithms can be obtained very quickly.
In addition to methods belonging to the above-mentioned categories, the works of \cite{wc-fmm-11} and \cite{wp-imme-16} applied a two-step deformation technique to achieve the goal. Their methods first compute curvilinear layouts of metro maps and then rotate each edge to the closest octolinear direction. Each step solves a least-squares optimization problem. \rv{Compared to previous works, our approach allows embedding a guide shape with arbitrary edge directions into a schematic metro map, which is challenging to integrate with their frameworks.}

\textbf{Station Names Labeling.}
Station names are essential in metro maps, yet the labeling problem itself is also NP-hard \cite{garrido2001labeling,niedermann2018algorithmic}. Several previous methods solve the layout and the labeling problems in two steps to reduce the computation cost \cite{wc-fmm-11,wu2011zone}.
Niedermann and Haunert in particular present a general framework for labeling network maps~\cite{niedermann2018algorithmic}.
Since occlusions could be inevitable when the layout does not provide enough space for labeling, the methods of \cite{niedermann2018algorithmic,yoshida2018progressive,takahashi2019mental} systematically scale certain edges on the map to solve the problem. There are also methods considering both layout and labeling problems simultaneously when computing metro maps \cite{stott2005automatic,srmw-amlumo-11,nw-dlhqm-11}. Please refer to \cite{wu-stml-2020} for details as our work does not focus on station names labeling problem.

\newcommand{\network}{N}
\newcommand{\stations}{S}
\newcommand{\connections}{C}
\newcommand{\orientations}{\mathcal{C}}
\newcommand{\layout}{\mathcal{L}}
\newcommand{\map}{\mathcal M}
\newcommand{\transits}{T}
\newcommand{\wiring}{\mathcal T}
\newcommand{\shape}{P}

\section{Overview}
\label{sec:overview}

Our approach was developed by firstly investigating several real-world examples~\cite{zerozero,moscow,c-omw-16} and existing design criteria~\cite{r-wytesmd-14,wu-stml-2020}, and then summarizing common design principles for our mixed metro map layout. 
Secondly, we transform these design principles to aesthetic constraints. 
Finally, these aesthetic constraints are formulated into equations for a  mathematical model that can be solved systematically.
We will demonstrate these design principles, followed by an introduction of our approach pipeline achieving these goals.

\subsection{The Mixed Metro Map Problem} \label{ssec:problem}

We define the mixed metro map problem in a more general fashion by relaxing the user-specified shape constraints introduced by Wu et al.~\cite{wtly-tmla-12,wptaly-dammll-15}.
Note that to retain consistency, in this paper, we follow the terminology and symbols used in the recent transit map survey~\cite{wu-stml-2020}.
A transit network is defined as $\network=(\stations,\connections)$, 
where $\stations=\{v_1, v_2, ..., v_n\}$ represents a vertex set of stations and $\connections=\{e_1, e_2, \ldots, e_m\}$ is an edge set describing station connectivity.
In addition to classical networks, a \emph{path cover $\transits$} of $N$ is used to describe a set of paths, in $\network$, indicating the different service lines in the transportation system.
Note that each connection %
$e\in\connections$ belongs to one or more paths in $\transits$.
Moreover, a \emph{guide shape} $\shape$, which can consist of one or more open or closed polylines $\shape=\{p_1, p_2, ..., p_k\}$, is introduced to express the user's design goal. 
\subsection{Design Principles} \label{ssec:design}

Based on criteria collected for layout approaches~\cite{wu-stml-2020} and open criteria raised by the design studio~\cite{zerozero,c-omw-16}, we summarize the design principles for our mixed metro maps as follows:

\martin{what about preserving topology? that's usually also a mandatory constraint...}
\soeren{It is not elegant, but I added this to criteria D1 now, so we do not need to change all references in the text. Is this an acceptable solution for now? @Yun, should be fine for now.}

\begin{enumerate}[\bfseries D1] 
\setlength{\itemindent}{6pt}
    \item \textbf{Constrained Layouts:}
    To increase legibility~\cite{wu-stml-2020,r-wytesmd-14} line configurations are often simplified to an octolinear design~\cite{nw-dlhqm-11}.
    In such a layout, the preservation of the input topology is often assumed a natural constraint~\cite{wu-stml-2020}.
    \item \textbf{Topographic Accuracy and Relative Positions of Stations:} The topographic accuracy has been discussed as an important factor in a schematic representation~\cite{r-wytesmd-14} to preserve users' mental map of a city. Relative positions between pairs of stations should be maintained. 
    \item \textbf{Spatially-separated Stations:} To avoid overlaps between stations, a minimum distance between stations is commonly incorporated. This distance is preferably uniform~\cite{g-mbum-94,nw-dlhqm-11}.  
    \item \textbf{Simplification of Trajectories:} Line orientations are expected as continuous as possible~\cite{g-mbum-94,nw-dlhqm-11}. 
\end{enumerate}

The aforementioned design rules have been defined previously~\cite{wu-stml-2020} and are applied to real-world metro maps in general. 
In this paper, we further propose design criteria, which allow us to achieve the desired affect when embedding the guide shape in the final layout.

\begin{enumerate}[\bfseries D1] 
\setcounter{enumi}{4}
\setlength{\itemindent}{6pt}
    \item \textbf{Translation and Scale Invariant Shape Embedding:} The shape can be translated and scaled when embedding it, while we exclude rotation to avoid adding another layer of recognition complexity~\cite{tarr-meoisr-1989}.
    \item \textbf{Shape Representation:} The edges in the transit network should be appropriately selected to approximate the shapes. Perceptually, the embedded shape in the layout should be recognizable. 
    \item \textbf{Hybrid Edge Orientation:} Edges that are used to approximate the guide shape should follow the shape structure, while the remaining edges follow the aforementioned classical design.
\end{enumerate}

\subsection{Mixed Metro Map Pipeline} \label{ssec:pipeline}

Figure~\ref{fig:overview} gives a conceptual overview of the proposed algorithmic pipeline.
Initially, the user needs to select a transit network and provide a target shape as input to our approach. 
Figure~\ref{fig:overview}(a) shows an example, where the transit network is drawn with colored lines and a user-specified shape (heart) is marked in gray. 
In the second step, as shown in Figure~\ref{fig:overview}(b), we scale and translate the guide shape in order to find an appropriate sub-network to approximate this shape (Section~\ref{sec:route}).
Once the guide shape is uniformly scaled and translated, we adapt a two-step deformation procedure inspired by Wang et al.~\cite{wc-fmm-11,wp-imme-16} in order to create a relatively well-representative network embedding. 
Here, we first generate a smooth layout (Figure~\ref{fig:overview}(c)) to balance mutual distance between stations and straighten lines with degree $2$ stations, followed by synthesizing a nearly octolinear layout (Figure~\ref{fig:overview}(d)). 
This optimization process is detailed in Section~\ref{sec:deform}.
In the last step, to guarantee that the metro lines are fully aligned on a grid (Figure~\ref{fig:overview}(e)), we introduce a grid alignment approach while retaining the mixed layout structure (see Section~\ref{sec:align}).

\newcommand{\frechetdistance}{\delta}
\newcommand{\frechetdistancePart}{\delta_{partial}}
\newcommand{\connectionPath}{W}

\section{Route Matching and Shape}
\label{sec:route}

We compute a metro map layout that exhibits the given guide shape. Limited by the metro network's route structure, determining the optimal size and position of the guide shape is essential. In addition to size and position, however, we do not rotate either the guide shape or the metro map when computing the map layout because their orientations are restricted. For example, the upright direction of a \emph{``heart''} shape is meaningful to humans; and the relative positions of stations help passengers navigate themselves in an urban area. An up-side-down metro map would conflict with users' mental map and decrease the map's usability.
To determine the position of a guide shape, we first determine a path $\connectionPath$ consisting of metro connections $e \in \connections$. Then, we scale and translate the guide shape $\shape$ to align with the path $\connectionPath$. Note that the guide shape's size and position will not be updated during the deformation process.

We focus on a\rv{n} automatic approach for aligning guide shapes and metro maps\rv{, which determines a} path inside the transit network visually similar to the guide shape $\shape$. %
However, as outlined in Section \ref{sec:intro}, there are use cases (artistic or advertising purposes), where specific lines must lie on iconic shapes. 
We optionally let designers define $\connectionPath$ manually.

\subsection{Automatic Route Matching}
\label{ssec:matching-automatic-scenario}
\begin{figure}[t]
    \centering{}
    \includegraphics[width=0.72\linewidth]{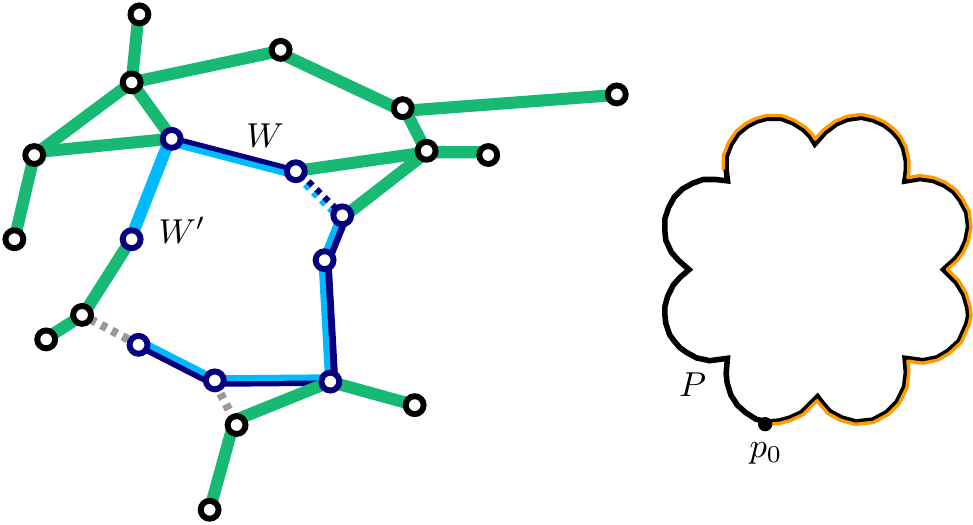}
    \caption{Dummy edges are shown with dotted lines. $\connectionPath'$ is shown with a light blue stroke and $\connectionPath$ with a dark blue. The complete guide shape $\shape$ is shown in black and a subsection of the guide shape is drawn with an orange stroke.
    }
    \label{fig:routing}
\end{figure}

We search for a path $\connectionPath = (e_1, e_2, ..., e_k)$ on the transport network that is visually similar to the user-defined guide shape.
Since the guide shape $\shape$ can be arbitrary and the path $\connectionPath$ similar to $\shape$ may not exist, we insert dummy edges to the network for enhancing visual quality.
Specifically, if the geographic distance between two stations $v_i$ and $v_j \in \stations$ is smaller than a predefined threshold and no edge connecting $v_i$ and $v_j$ exists, we insert a dummy edge to connect the two stations (Figure~\ref{fig:routing}).

We adapt the shape-preserving Dijkstra algorithm~\cite{funke2011path} to find a path $\connectionPath$ that can approximate $\shape$. 
The cost of a path in the shape-preserving Dijkstra algorithm, in contrast to the normal shortest path algorithm, is
given by 
similarity between the path and $\shape$. %
Specifically, we quantify the difference between the path $\connectionPath$ and the $\shape$ using the direction-based integral Fréchet distance~\cite{de2011go} between $\connectionPath$ and $\shape$, which we write as $\frechetdistance(\connectionPath, \shape)$.
We also use the concept of the \emph{partial similarity}~\cite{de2011go} $\frechetdistancePart(\connectionPath, \shape)$, which is defined by a subpath $\shape'$ of $\shape$, which minimizes $\frechetdistance(\connectionPath, \shape')$ (see Figure~\ref{fig:routing}).

The cost of a connection is given by the difference between the slope of the metro connections and the slope of the corresponding section of the guide shape. So in case the difference between those slopes is small, the additional cost of the connection for the path is small, too.
Rather than considering the Euclidean distance between points in the original shape-preserving Dijkstra algorithm \cite{funke2011path}, the direction-based integral Fréchet distance is beneficial for our use case because it is scale and translation invariant (i.e., $\shape$ is given with an arbitrary size and position). 
It provides a robust mapping between $\shape$ and $\connectionPath$, and small variations between two lines are not disproportionately penalized.
\rv{Experiments can be found in Appendix~\ref{sup:path}.}

In our implementation, we grow the path $\connectionPath$ from a station on the metro map until none of the inserted stations can reduce the Fréchet distance. Let $W_0$ be the initial path and $W_k$ be the path composed of $k+1$ stations. Namely, the growing process can be expressed as $W_0 \rightarrow W_1 \rightarrow W_2 \rightarrow ... \rightarrow W$. At each step $k$, we examine all neighboring stations of $W_k$ and insert the station $\mathbf{v}$ into a queue if it can reduce the direction-based integral Fréchet distance. In other words, $\frechetdistance(\connectionPath_k + \mathbf{v},\shape) < \frechetdistance(\connectionPath_k, \shape)$. Then, for each station $\mathbf{u}$ in the queue, we compute the partial matching $\frechetdistancePart(\connectionPath_k + \mathbf{u}, \shape)$ to obtain a subsection of the guide shape $\shape$ that is the most similar to $\connectionPath_k + \mathbf{u}$. The station $\mathbf{u}$ that has the shortest distance $\frechetdistancePart$ will be inserted to extend the path from $W_k$ to $W_{k+1}$.
Note that we only consider subsections of the guide shape $\shape$ 
with the intent, that $\connectionPath_{k+1}$ approximates $\shape$ better than $\connectionPath_k$ in each iteration.
The algorithm repeats until the queue is empty.

Using paths with a few dummy edges to represent the guide shape is preferable.
To achieve this goal, we give dummy edges a high cost and real metro edges a small one. 
Since the starting vertex of path $\connectionPath$ is unknown, we run the Dijkstra method from each potential starting station and select the path $\connectionPath$ that has the least cost $\frechetdistance(\connectionPath,\shape)$.
If the guide shape consists of multiple polylines, for example the eye-formed shape shown in Figure \ref{fig:result-paris-eye}, we perform the route matching only for the the polyline containing the top left vertex, and all other polylines are placed at the correct scaling and offset to the matched one.

\begin{figure}[t]
    \centering{
    \setlength{\tabcolsep}{1pt}
    \begin{tabular}{cc}
        \includegraphics[width=0.47\linewidth]{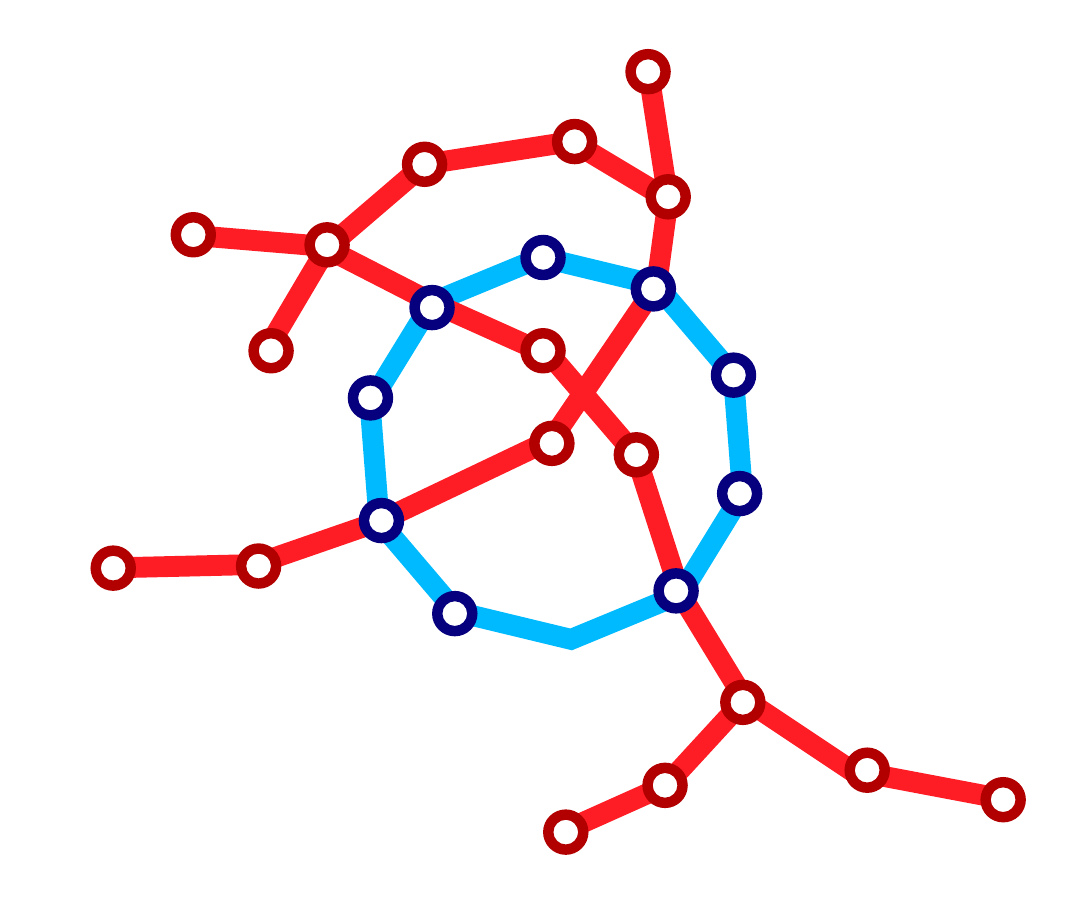} &
        \includegraphics[width=0.47\linewidth]{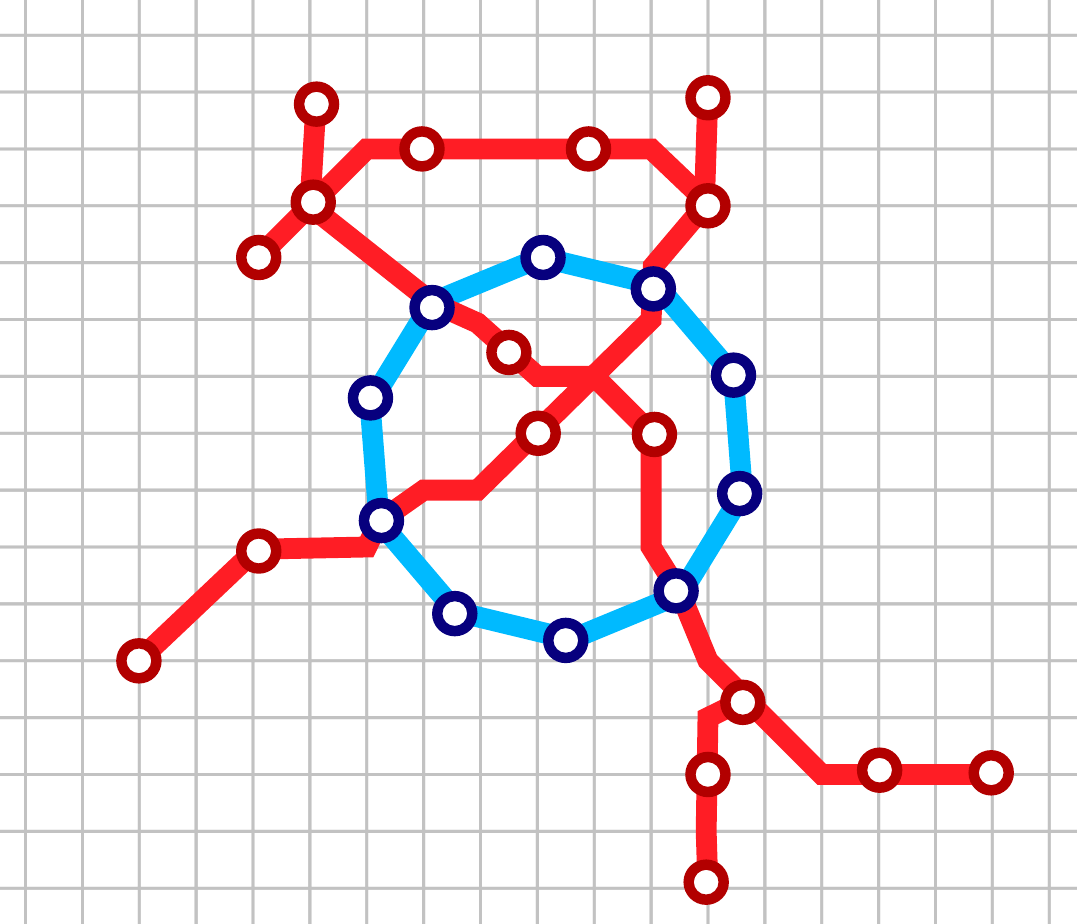} \\
        (a) Smooth deformation & 
        (b) Grid alignment \\
    \end{tabular}
    \caption{Shape stations $\rv{v'_i} \in \stationsSmooth$ and the corresponding edges that approximate \rv{a} circular guide shape are visualized in blue. Stations and connections that do not approximate the guide shape are shown in red.
    }
    \label{fig:motiv_vs_octi}
    }
\end{figure}
\soeren{is there a blue station missing in figure 5 a?}

Once the path $\connectionPath$ is obtained, we transform the metro network to align $\connectionPath$ with the guide shape $\shape$. As mentioned previously, we assume that the guide shape and the metro map are in the right orientation, and thus the transformation contains only translation and scaling. For simplicity, the alignment is achieved based on the bounding boxes of $\shape$ and $\connectionPath$.

\section{Deformation}
\label{sec:deform}
\begin{figure}[t]
    \centering
    \includegraphics[width=0.75\linewidth]{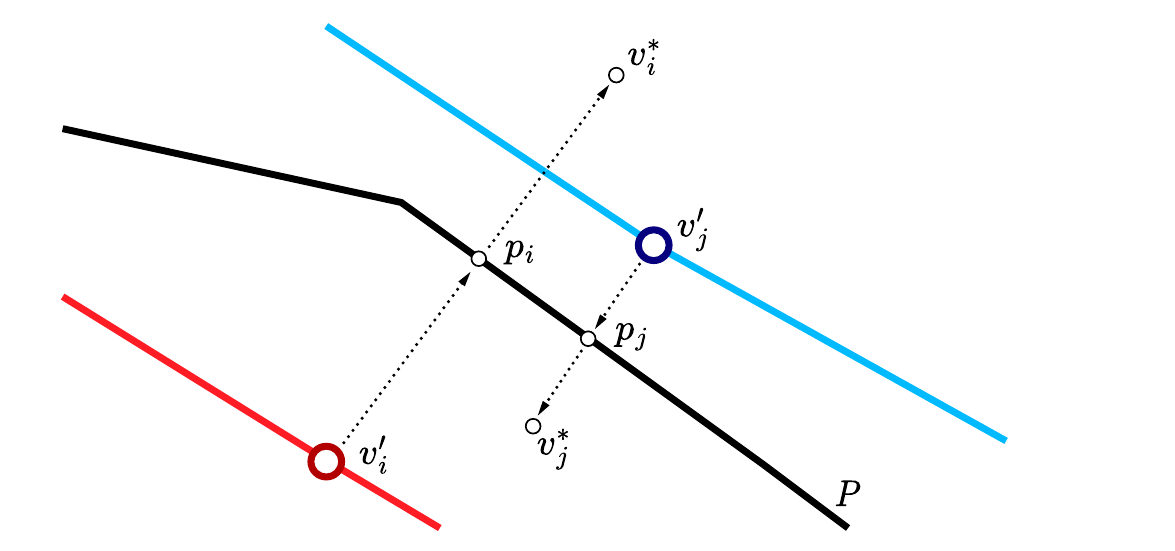}
    \caption{The station $v_j'$, rather than $v_i'$, is used to approximate the guide shape. This guarantees that no other metro line is closer to the guide shape. Therefore, $v_j'$ is added to $\stationsSmooth$, but not $v_i'$. %
    }
    \label{fig:motiv_vs_octi_auto}
\end{figure}

After aligning the guide shape $\shape$ with the metro network, we deform the metro network to fulfill the design principles outlined in Section \ref{ssec:design}. 
This process is inspired by the two-step approach of Wang et al.~\cite{wc-fmm-11}.
First, we create a smooth layout (Figure \ref{fig:overview}.c) that aims to space stations evenly, avoid sharp bends, and maximize angular resolution. 
Additionally, the guide shape $\shape$ is approximated by aligning metro stations with segments of $\shape$ (Figure~\ref{fig:motiv_vs_octi_auto}).
The smooth, as well as the mixed layout, are created using the least squares optimization by minimizing constraints iteratively.

In the following sections, we denote the geographic position of a station by $v_i \in \stations$, the transformed position in the smooth optimization stage by $v'_i \in \stations'$, and in the mixed stage by $\Tilde{v}_i \in \Tilde{\stations}$. We assume that the input metro network is planar. Otherwise, we planarize the network by inserting dummy stations if two connections intersect.

To approximate the guide shape, a subset of stations $v'_i \in \stationsSmooth$ are pushed toward the guide shape segments that are located closest to them. As a result, the edges $e'_i \in \connectionsSmooth$ connecting those stations can represent the guide shape $\shape$.
Figure \ref{fig:motiv_vs_octi_auto} illustrates the connections and stations that should and should not approximate the guide shape.%
It deserves noting that we use $\connectionPath$ only to align the guide shape with the network, rather than to define $\connectionsSmooth$.
The reason is that the automatically computed path $\connectionPath$ represents only the overall shape of the guide shape and may miss details.
Therefore, we update $\stationsSmooth$ at each step of the smooth deformation. 
This is done as follows: let $v_i^*$ be a reflection of $v_i'$ over the closest point $p_i$ on the guide shape.  
We assign $v_i'$ to $\stationsSmooth$ if the line segment between $v_i'$ and $v_i^*$ does not intersect any of the edges in $\connections'$, as illustrated in Figure \ref{fig:motiv_vs_octi_auto}. Otherwise, such an intersection implies that another station and metro line must be closer to the guide shape.

\subsection{Smooth Layout}
We optimize four constraints to compute the smooth layout,
\begin{equation}\label{equ:smooth}
    \Omega_{smooth} = w_c\Omega_c + w_l \Omega_l + w_a \Omega_a + w_p\Omega_p \text{.}
\end{equation}
Specifically, $\Omega_c$ forces the layout to approximate the guide shape. $\Omega_{l}$ causes uniform edge lengths, $\Omega_{a}$ maximizes angular resolutions, and $\Omega_{p}$ minimizes the distance of the position of a metro station $v_i \in \stations$ to its geographical location. The corresponding weight to balance the potentially contrary constraints is denoted by $w$.  
\tobias{Think N(v) is not a good choice. We denote the Network with N; Line 153}
\rv{Let $D(v_i')$ be the degree of $v_i'$}. The constraint $\Omega_c$ penalizes the distance of smooth stations and the polyline and is given as 
\begin{equation}
    \Omega_{c} = \sum_{\rv{v'_i} \in \stationsSmooth} \rv{D(v'_i)} {\vert p_i - v'_i}\vert^2 \text{.}
\end{equation}

To implement the Design principles D1 to D4 (outline in Section \ref{ssec:design}), three energy terms are applied to all stations $v_i \in \connections$. 
The constraint $\Omega_l$ forces stations to be evenly spaced. 
This eliminates the information about the geographic distance between stations, but creates a more uniform and balanced layout.
$\Omega_l$ is given by 
\begin{gather}
    \Omega_l = \sum_{\{i,j\} \in \connections} \vert(v'_i - v'_j) - s_{ij} R_{ij}(v_i - v_j)\vert^2 \text{ ,} \nonumber\\
    \text{where} \quad   
     s_{ij} = \frac{L}{\vert v_i - v_j\vert}
    \quad \text{and} \quad
    R_{ij} = 
    \begin{bmatrix}
    \cos \theta_{ij} && -\sin \theta_{ij} \\
    \sin \theta_{ij} && \cos \theta_{ij}
    \end{bmatrix}
    \text{,}
\end{gather}
$L$ denotes the target length of the edges. This length is equal for all edges where stations are not dummy nodes, and equals the average length of the metro connections in the initial layout. 
In case $v_i$ or $v_j$ of $\{i,j\} \in \connections$ is a dummy node and not a regular station, the target length for $\{i,j\} \in \connections$ is $L/2$.
Besides, $\theta$ describes the unknown angle of a connection $\{i,j\} \in \connections'$ and $R$ is a rotation matrix, ensuring that the rotation of the connection is not penalized.

The constraint $\Omega_a$ aims to maximize the angle between connections sharing a station. 
This separates the different metro lines in case of station with a degree $> 2$ (e.g. an interchange), making it easier for the user to distinguish two metro lines. 
In case of a station with degree $ = 2$, the metro lines are straightened naturally. 
$\Omega_a$ is formulated as
\begin{gather}
    \Omega_a = \sum_{v_i' \in \stations'}
                \sum_{\{i,j\}, \{i,k\} \in \connectionsSmooth}
                \vert 
                    v'_i - (v'_j + c'_{ij} + \tan(\frac{\pi - \rv{\theta_i}}{2})c'_{jk})
                \rvert ^2 \text{,}
\end{gather}
\rv{where $\enspace c'_{ij} = \frac{1}{2}(v'_k - v'_j)$.}
$\rv{\theta_i}$ depends on the number of outgoing connections of $v\rv{'}_i$, namely, $\rv{\theta_i} = 2 \pi /{\rv{D(v'_i)}}$. 

To preserve the overall geographic structure of the metro system, we add the energy term $\Omega_p$. This term can avoid large deformations of the network and avoid a conflict with the mental map of users. 
For example, to avoid a layout where stations located in the north of a city are moved to the south. 
$\Omega_p$ is given as: 
\begin{equation}
 \Omega_p = \sum_{v_i \in \stations} \vert v_i' - v_i\vert^2 \text{.}
\end{equation}

\subsection{Mixed Layout}\label{sec:mixed_layout}
Following the smooth optimization stage, we create a mixed layout. 
Similar to the previous step, we differentiate between connections $\Tilde{e}_j \in \connectionsSmoothMixed$ that aim to approximate the guide shape and octolinear connections $\Tilde{e}_i \in \connectionsOctiMixed$ that do not contribute to the recognisability of the shape.
Octolinear edges should have an octolinear slope (e.g. a multiple of $\frac{\pi}{4}$), whereas shape connections approximate $\shape$ and should align to a section of the guide shape $\shape$.
The differentiation between octolinear and shape connections is based on the assignment used for the smooth stage. 
Therefore, if a connection was in $\connectionsSmooth$ in the smooth layout, then this edge is treated as a shape edge $\Tilde{e_i} \in \connectionsSmoothMixed$.
Otherwise, the edge is an octolinear one. %

\begin{figure*}[t]
    \centering{
    \setlength{\tabcolsep}{0pt}
    \begin{tabular}{cccc}
        \includegraphics[width=0.25\linewidth]{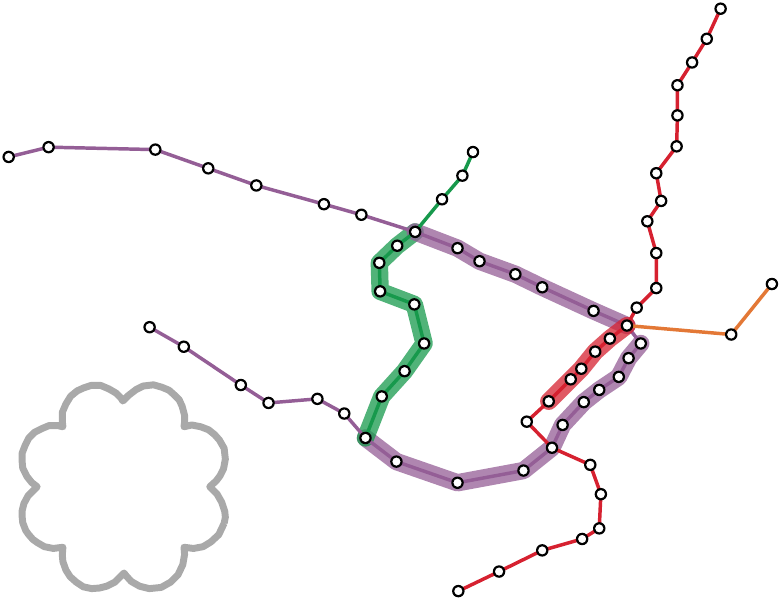} &
        \includegraphics[width=0.25\linewidth]{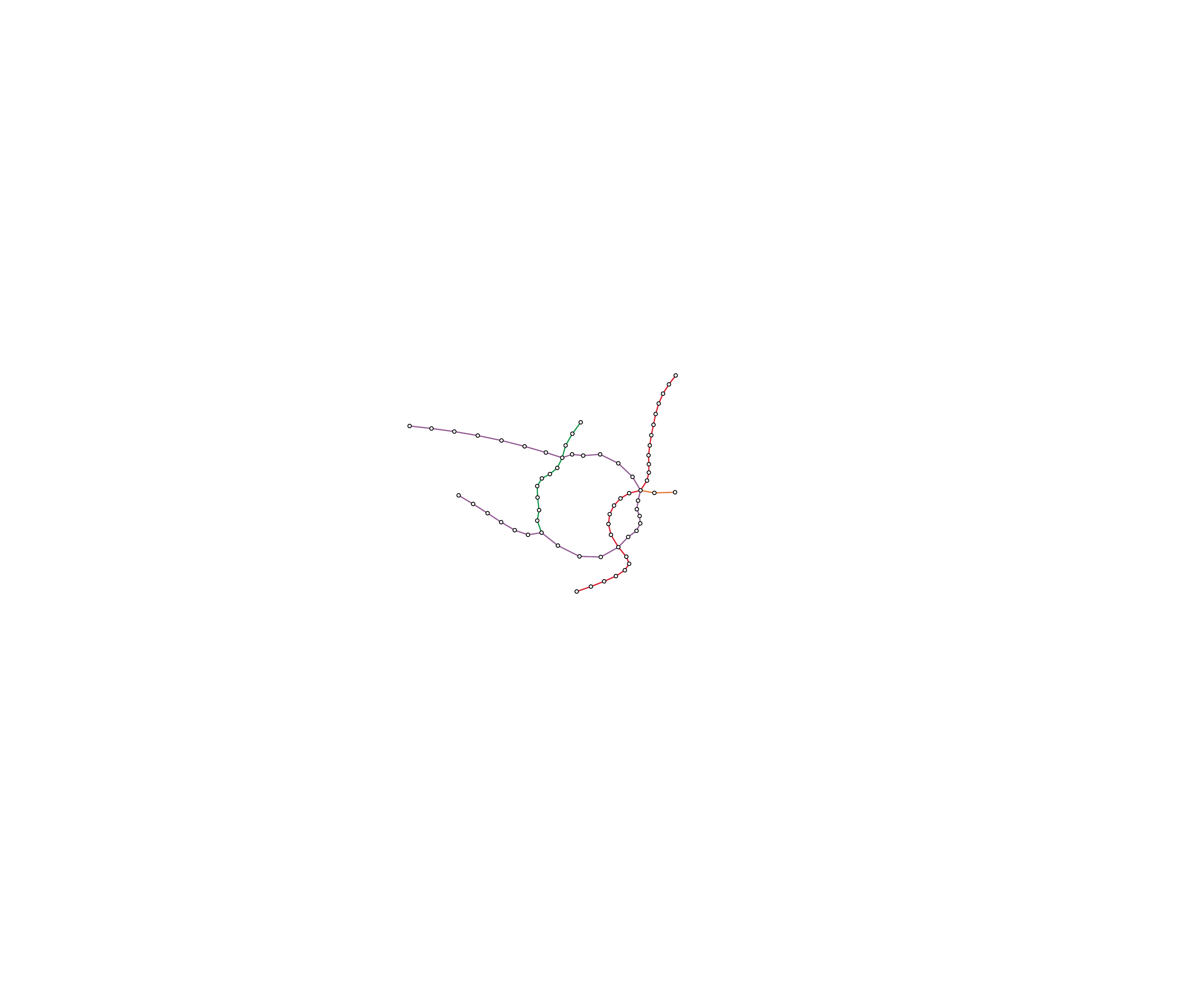} &
        \includegraphics[width=0.25\linewidth]{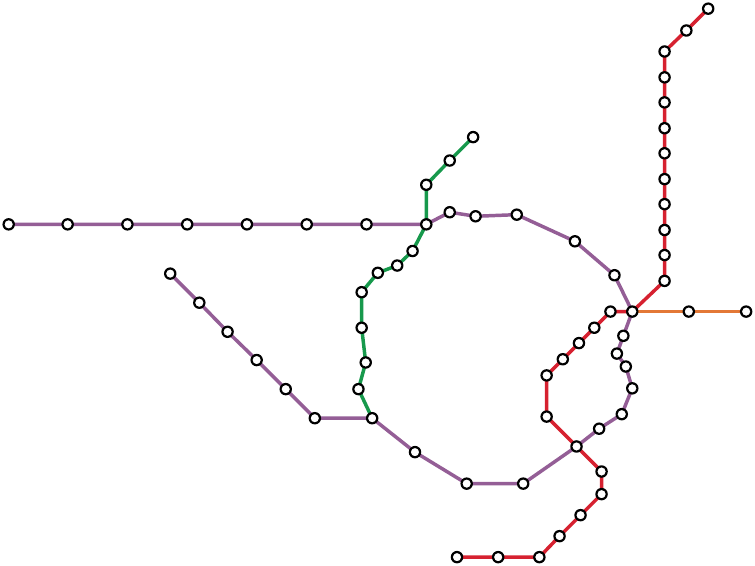} &
        \includegraphics[width=0.25\linewidth]{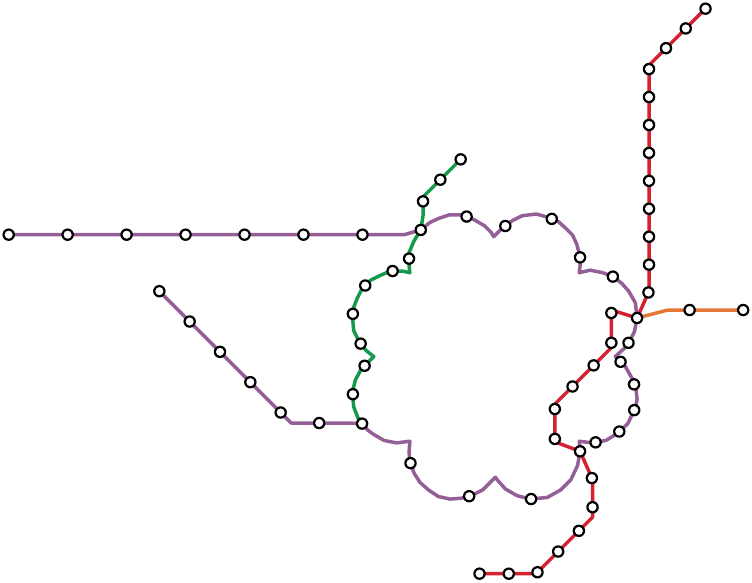} \\
        (a) & (b) & (c) & (d)
    \end{tabular}
    \caption{Montreal metro map created with a cloverleaf-shaped guide inspired by the city's logo. (a) Geographically accurate layout with the automatically computed path $P$ highlighted. The input guide shape is shown on the bottom left. (b) The smooth layout, (c) the mixed layout and (d) the final grid-aligned layout.}
    \label{fig:result-montreal}
    }
\end{figure*}

To determine the mixed layout, each connection $\Tilde{e_i} \in \connectionsOctiMixed$ is assigned to its closest octolinear slope.
In case of a conflict, the Hungarian algorithm is applied to reassign the octolinear slope of each edge by minimizing the sum of rotations. 
\yuninline{@Yu-shuen, check please.}
\rv{We minimize the term}
\begin{gather}
    \Omega_{o} = \sum_{\{i,j\} \in \connectionsOctiMixed} 
        \vert 
            (\theta'_{ij} + \Delta\theta_{ij}) - \Tilde{\theta}_{ij}
        \vert^2 \text{,}
\end{gather}
to rotate octolinear edges $\Tilde{e_i} \in \connectionsOctiMixed$, where
$\theta'_{ij}$ denotes the angle of the connection after the smooth optimization, $\Delta\theta_{ij}$ is the difference between $\theta'_{ij}$ and the target angle, and $\Tilde{\theta}_{ij}$ is the current slope of the connection $\{i,j\} \in \connectionsOctiMixed$.
Further we apply the constraints $\Omega_{p}$ and  $\Omega_{c}$ equally as for the smooth layout. 
The mixed layout is than computed by minimizing
\begin{equation}\label{equ:mixed}
    \Omega_{mixed} = w_{o}\Omega_{o} + w_{p}\Omega_{p} + w_{c}\Omega_{c} \text{.}  
\end{equation}

During the smooth and the mixed deformation stages, the planarity is checked after each optimization step.
In case of an intersection the corresponding stations are moved back to the position at the previous iteration. An energy term is applied to avoid the same intersection in the following optimization step. 
If the distance between a station and a connection is below a pre-defined threshold, we apply an energy term to move them apart.
\rv{A parameter testing is shown in Appendix~\ref{sup:parameter}.}

\martin{general question: the description of course gives a rough idea of what's going on, but there are many details and parameter settings that are not described. Do we publish or plan to publish the code so our results become reproducible? Or should we have more details in the supplemental material? Like how did we set weights, how was it tuned?
@Yun, I don't have the info now. Can only add it in the next round.}

\section{Grid Alignment}
\label{sec:align}

\newcommand{\grid}{G}

While the previous mixed layout step tries to align the shape edges with the shape and the remaining edges with an octolinear direction, the resulting layout may still contain some edges that are not fully octolinear yet.
Our goal is to compute a final layout, which is as similar to the mixed layout as possible, while all edges of $C_{octo}$ are exactly aligned with one of the octolinear directions and all edges of $C_{shape}$ are tracing the guide shape.

This issue can not easily be resolved by simply snapping every edge to the closest octolinear direction.
However we can adapt an existing layout algorithm developed by Bast et al.~\cite{BastBS20}, which computes octolinear layouts of metro maps given as geographical inputs on predefined grids, and renders edges as piece-wise octolinear curves.

We will proceed by giving a description of the original approach first and then we will describe all adaptions we made to accommodate for our setting.

\subsection{The Octi framework}
In their Octi framework for metro map layout, Bast et al.~\cite{BastBS20,BastBS21} consider a geographic network $N$ as input.
Let $x_{min}$, $y_{min}$, $x_{max}$ and $y_{max}$ be the minimal maximal $x$ and $y$ coordinate over all stations in $N$ respectively.
A grid graph $\grid = (V_\grid, E_\grid)$ (with cell diagonals) is created, with a cell size $d$ equal to the average distance between two stations in $N$ multiplied by a factor $f_{d}$ with $\lceil(x_{max} - x_{min})/d\rceil$ columns and $\lceil(y_{max} - y_{min})/d\rceil$ rows.
The dimensions are chosen, s.t., $N$ can be placed on top of $\grid$ and the axis-aligned bounding box of $N$ is completely contained in that of $\grid$.
Now every edge $\{a, b\}$ of the grid is replaced with a path $a, a_b, b_a, b$ of length three. The two nodes $a_b$ and $b_a$ are called the port of $a$ in the direction of $b$ and the port of $b$ in the direction of $a$, respectively, while $a$ and $b$ are called the sink nodes of their respective ports.
Note that every port node has as many ports as its original degree.
Finally for every sink node, edges are introduced to create the complete graph on its port nodes.
Every edge between two ports of different sink nodes has weight $c_{cop}$, every connection of a port to its own sink node has a (sufficiently large) weight $c_{sink}$ and every connection $\{a_b, a_c\}$ in the complete graph between port nodes of a node $a$ have one of four possible values, inversely proportional to the angle $\measuredangle bac$.

Let $x(s)$ and $y(s)$ be the $x$ and $y$ coordinates of a station $s$ in $N$.
Now the candidate set $V^c(s)$ of a station $s$ is defined as all sink nodes of $\grid$, which have a distance of $r$ or smaller to $(x(s), y(s))$.

By iterating over all edges of $N$ in a computed order $\Sigma$, the edges are routed as paths through the grid.
For the first edge $\{s, t\}$, this is done -- conceptually -- by temporarily adding \emph{virtual} vertices $a_s$ and $a_t$ to $\grid$ and adding the edges $(a_x, k)$ for $k \in V^c(x)$ and $x \in \{s, t\}$.
Every such edge is relative to the distance of $x$ and $k$. 
Now a shortest path $\Pi_{s,t}$ between $a_s$ and $a_t$ is computed, $a_s$ and $a_t$ are removed from $\grid$, the positions for $s$ and $t$ are fixed to the first and last sink node of $\Pi_{s,t}$ and the piece-wise octolinear path between $s$ and $t$ is obtained by concatenating all sink nodes of $\Pi_{s,t}$ or the sink nodes of port nodes in $\Pi_{s,t}$.
This is now done for all edges of $\Sigma$.
Should an endpoint of an edge already be fixed to a grid node, no virtual vertex is added and the shortest path is instead computed to or from the already fixed position.

Two main factors require adaption of this process, namely the additional design principle D7, i.e., the presence of the guide shape and
that we do not use an input map with geographical station positions, but
the mixed layout, which lets us assume that our input is already reasonably close to the final layout we want to compute.

\subsection{Adaptions}
We reduce the size of the map by removing stations and reinserting them afterwards, similar to Section~\ref{sec:mixed_layout}.
While the original algorithm~\cite{BastBS20} uses a similar approach and retains any vertex with degree three or higher, we also retain every vertex of $S_{octo}$ of degree two, if its two adjacent edges are not assigned to the same (or opposing) octolinear angles, since the mixed layout already aims -- in accordance with design principle D4 -- to simplify line trajectories.
Moreover, the mixed layout also, by design principle D6, approximates $\shape$ and we therefore also keep all stations in $S_{shape}$.

Next, all computed paths are paths in $\grid$.
Recall that $\shape$ is given as a polygonal shape.
To ensure that the distances along the shape are comparable to the average distances in the grid, $\shape$ is subsampled, however, no vertices are removed towards this goal to ensure that no essential features of $\shape$ are lost.
We need a suitable representation of $\shape$ in $\grid$.
This is done by overlaying $\shape$ onto $\grid$, removing all edges of $\grid$, which cross an edge of $\shape$ and all vertices, which are closer than a small threshold value $d_{min}$ to a vertex of $\shape$.
Then we reconnect
a node $p_i \in \shape$ with a node $a\in \grid$ if their distance $d'$ is $d_{low} \leq d' \leq d_{up}$, where $d_{low} < d < d_{up}$.
We choose $d_{low}$ and $d_{up}$, s.t.
the connecting edges \rv{are} roughly the same size as the cell size of the grid.
Since the nodes on the guide shape are not perfectly aligned with nodes in $\grid$ the connecting edges are not \rv{necessarily} octolinear.
\rv{We} could approximate the shape using nodes of $\grid$, however we emphasize design principle D6 over D7 and allow some edges to deviate from the octolinear or shape aligned directions.

\begin{figure}[t]
    \centering{
    \setlength{\tabcolsep}{1pt}
    \begin{tabular}{cccc}
        \includegraphics[page=1]{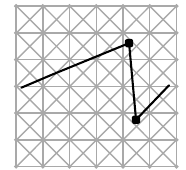} &
        \includegraphics[page=2]{figures/grid_explanation/grid_overlay.pdf} &
        \includegraphics[page=3]{figures/grid_explanation/grid_overlay.pdf} &
        \includegraphics[page=4]{figures/grid_explanation/grid_overlay.pdf} \\
        (a) $\shape$ on $G$ & 
        (b) Subsampling & 
        (c) Removal & 
        (d) Connections \\
    \end{tabular}
    \caption{
    (a) $\shape$ is overlayed on $G$.
    (b) $\shape$ is subsampled to ensure similar distances on $\shape$ as on $G$.
    (c) Edges of $G$, which intersect $\shape$ and vertices of $G$, which are too close to a vertex of $\shape$ are removed.
    (d) Vertices of $\shape$ are reconnected with vertices of $G$, which are within a certain distance range.}
    \label{fig:motiv_vs_octi}
    }
\end{figure}

\begin{figure*}[t]
    \centering{
    \setlength{\tabcolsep}{0pt}
    \begin{tabular}{cc}
        \includegraphics[width=0.5\linewidth]{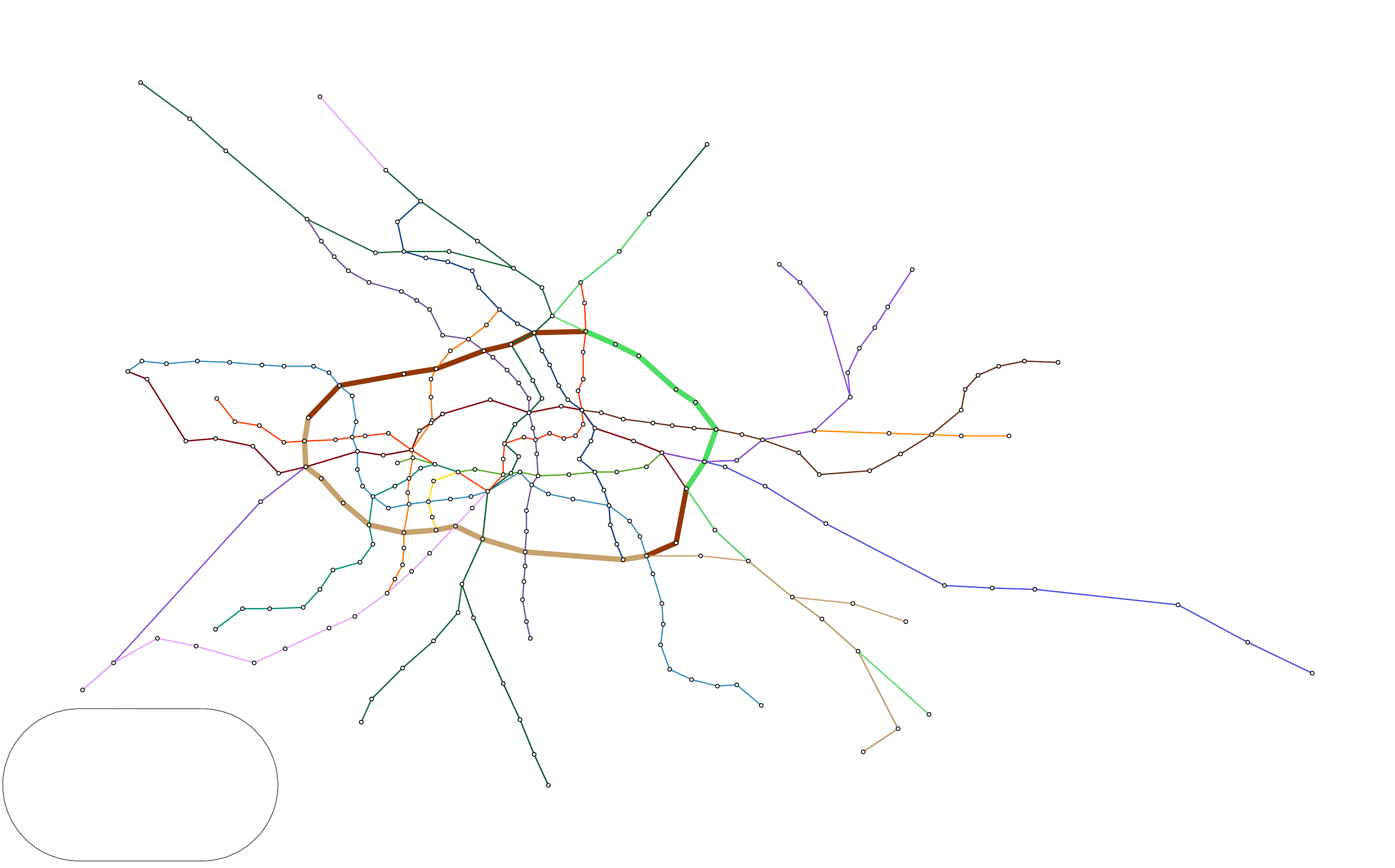} &
        \includegraphics[width=0.5\linewidth]{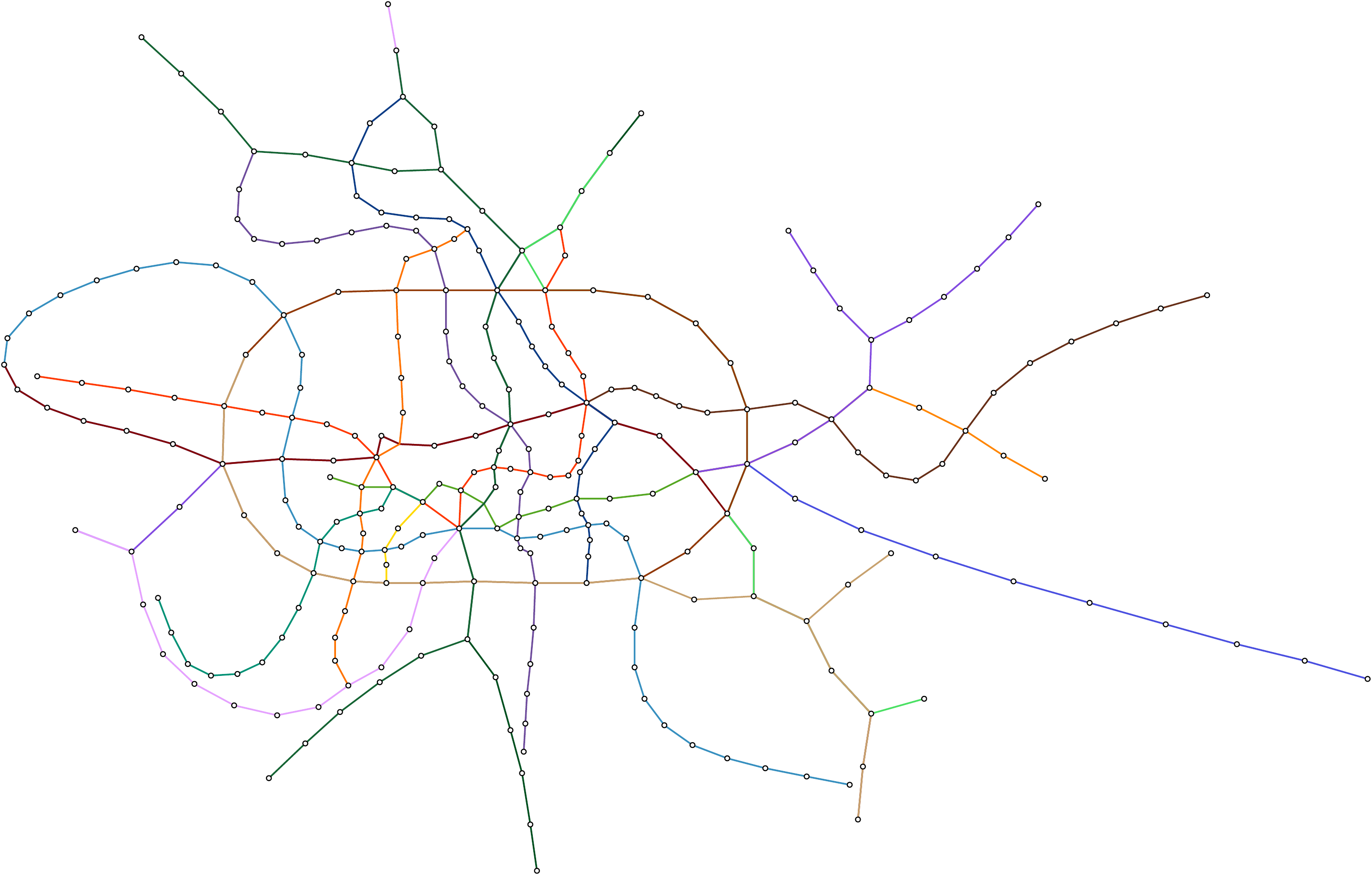} \\
        (a) & (b) \\
        \includegraphics[width=0.5\linewidth]{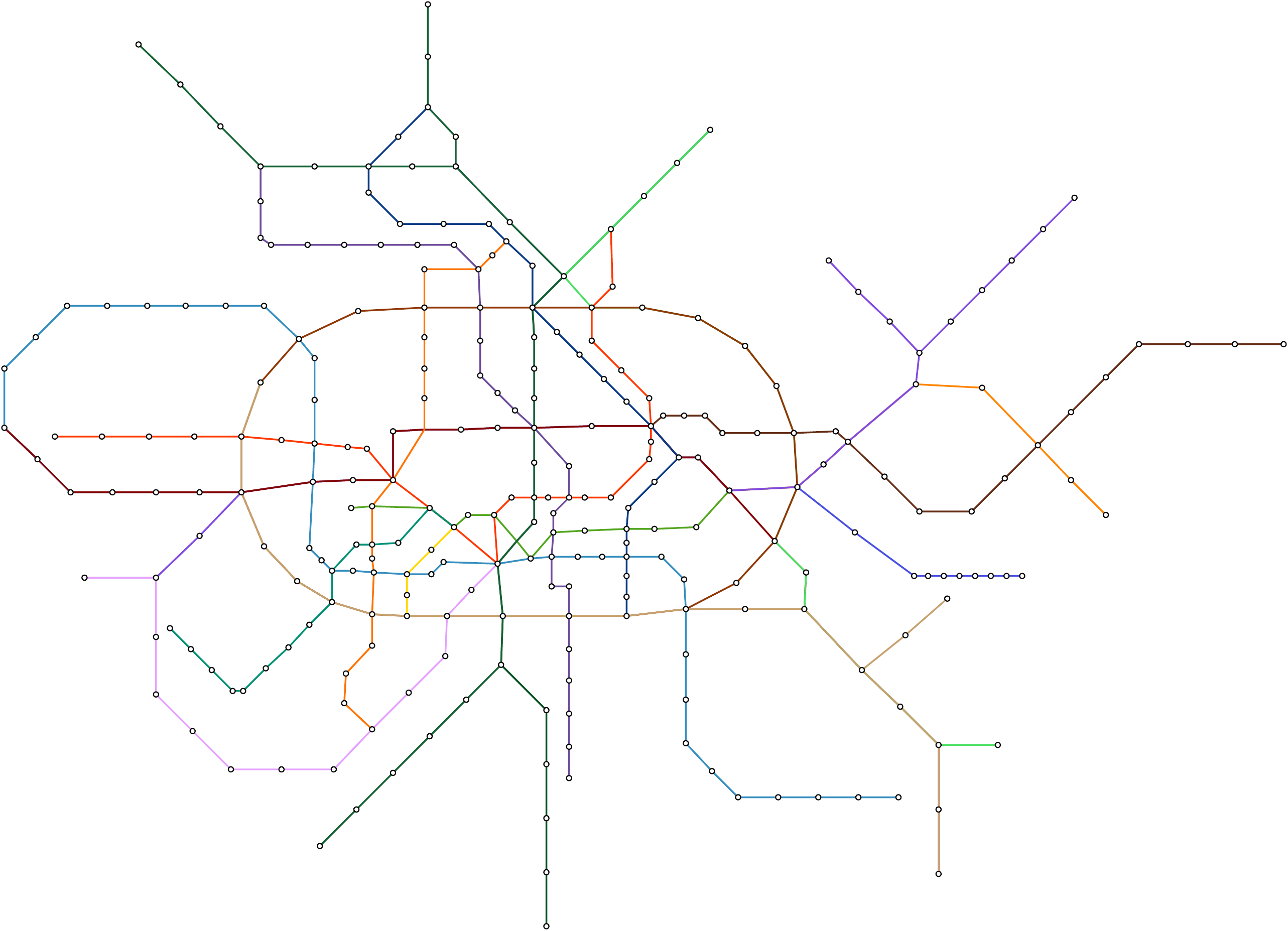} &
        \includegraphics[width=0.5\linewidth]{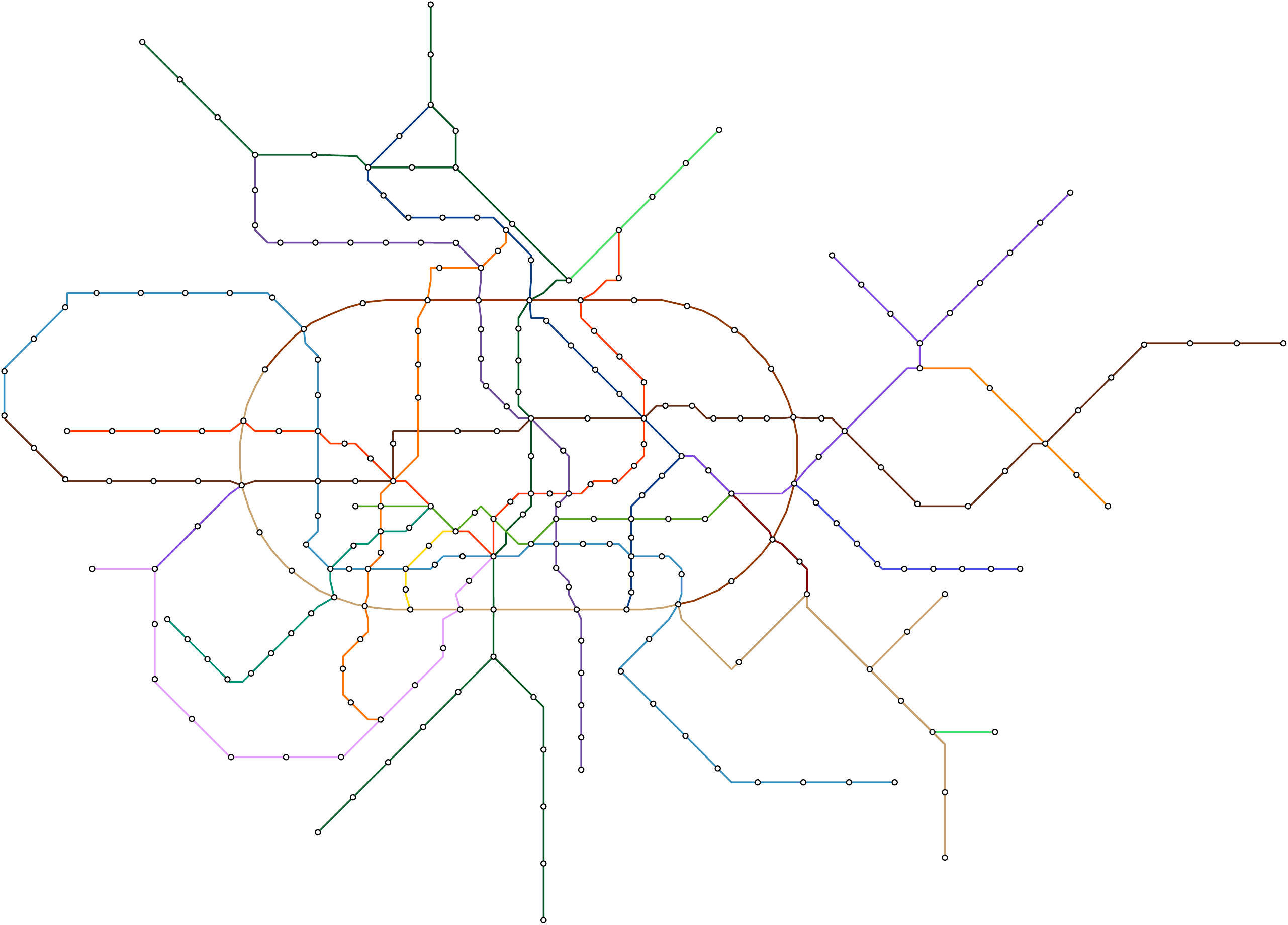} \\
        (c) & (d)\\
    \end{tabular}
    \caption{Berlin metro and S-Bahn map, consisting of 272 stations. The layout is created with a \rv{stadium}-shaped guide and a path $\connectionPath$ for the \rv{stadium} is provided. (a) Geographically accurate layout with the path $\connectionPath$ highlighted. The input guide shape is shown on the bottom left. (b) The smooth layout, (c) the mixed layout and (d) the final grid-aligned layout.}
    \label{fig:result-berlin-oval}
    }
\end{figure*}

The newly introduced edges also need to be assigned weights.
Port-to-sink connections for shape nodes are weighted the same as for normal nodes of $\grid$.
Edges between port nodes of different sinks are weighted with $c_{hop}/2$ or $c_{hop}/20$ if one or both sinks belong to $\shape$ respectively.
These cost reductions are intended to encourage the routing of edges in $N$ using edges of $\shape$.
Connections between port nodes of the same sink node are again weighted inversely proportional to the angle of their connected edges, however, since the shape is not necessarily octolinear, we need to be able to derive these weights based on any arbitrary angle $\theta$ and we therefore use $2\pi - \theta$.
Since $\theta \leq \pi$, we retain the desired property that no path of length two between two ports of the same sink is ever cheaper than their direct connecting edge.

Since the mixed layout already assigns and aligns stations and edges with the guide shape, we first and foremost want to ensure that these edges are still routed along edges of $\shape$.
Therefore we ensure that all edges of $C_{shape}$ appear in $\Sigma$ before any edge of $C_{octo}$.

When computing $V^c(s)$ for a station $s\in S_{shape}$, we drastically reduce its size. In particular, we only keep the two
nodes of $\grid$ \rv{based on our experiments}, which are closest to $s$.
\yuninline{@Soeren, please check if this footnote is necessary. CFG does not permit footnote.}
\soeren{thanks for removing}
After routing all edges, all previously removed stations are reinserted and finally all maximal sequences of stations of degree two are equally distributed.

\newcommand{\automaticYes}{computed}
\newcommand{\automaticNo}{provided}
\begin{figure}[t]
    \centering{}
    \includegraphics[width=\linewidth]{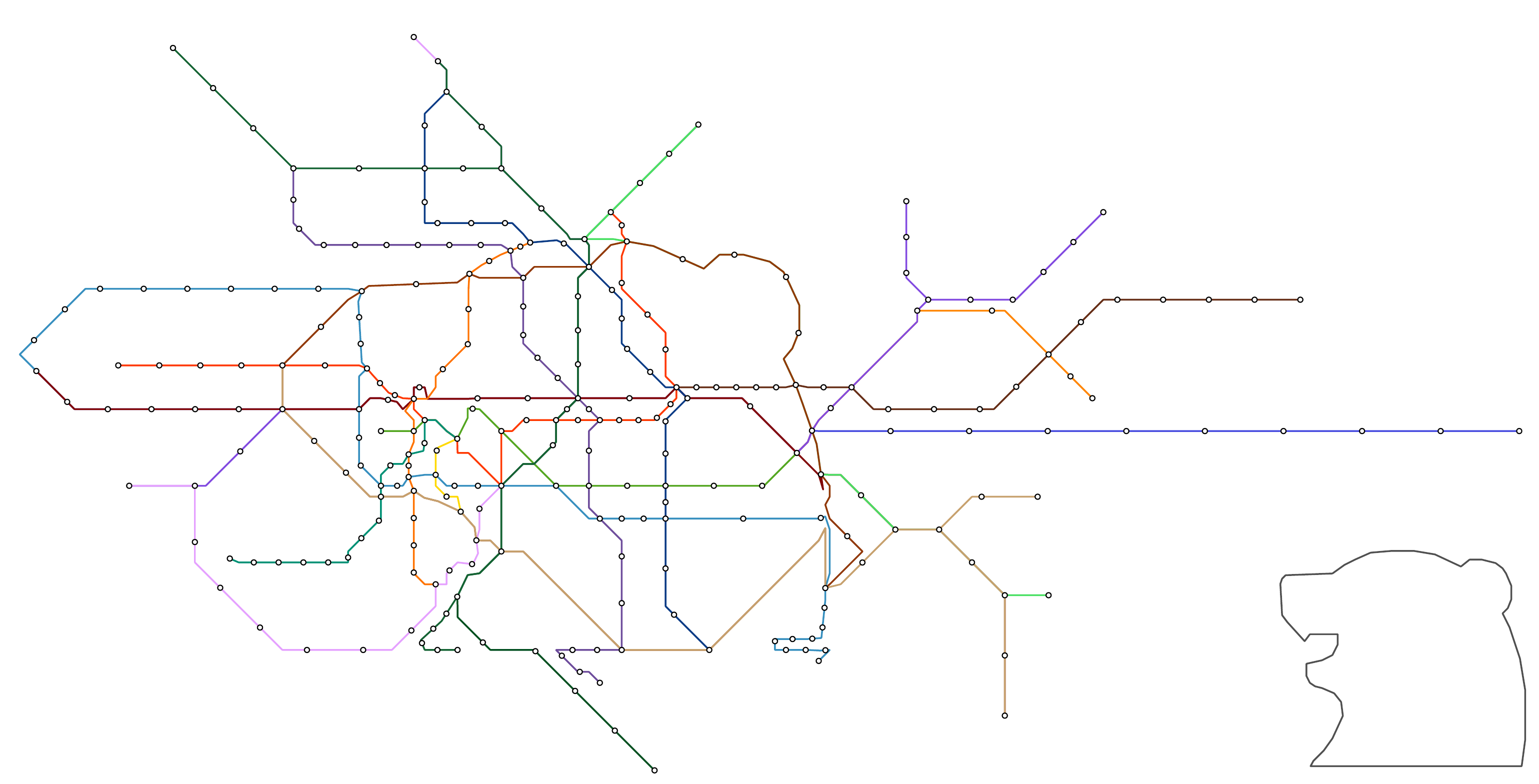}
    \caption{Berlin metro and S-Bahn map created with a bear-shaped guide, shown in the bottom right. The guide shape is inspired by the logo of Berlin.}
    \label{fig:result-berlin-bear}
    \centering{}
    \includegraphics[width=\linewidth]{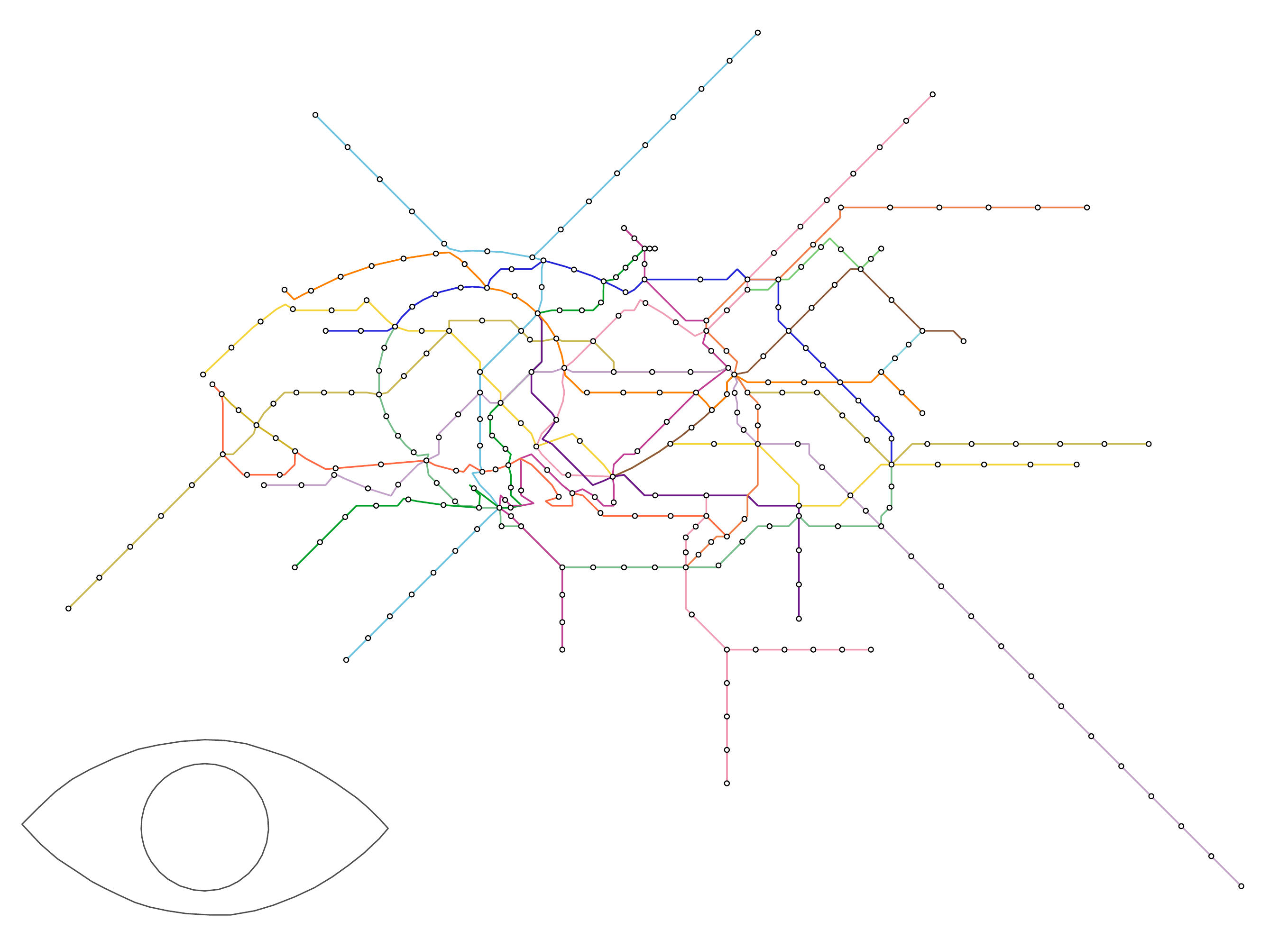}
    \caption{Paris metro map created with an eye-shaped guide consists of multiple polylines, shown in the bottom left.}
    \label{fig:result-paris-eye}
\end{figure}
\begin{figure}[t]
    \centering{}
    \includegraphics[width=0.63\linewidth]{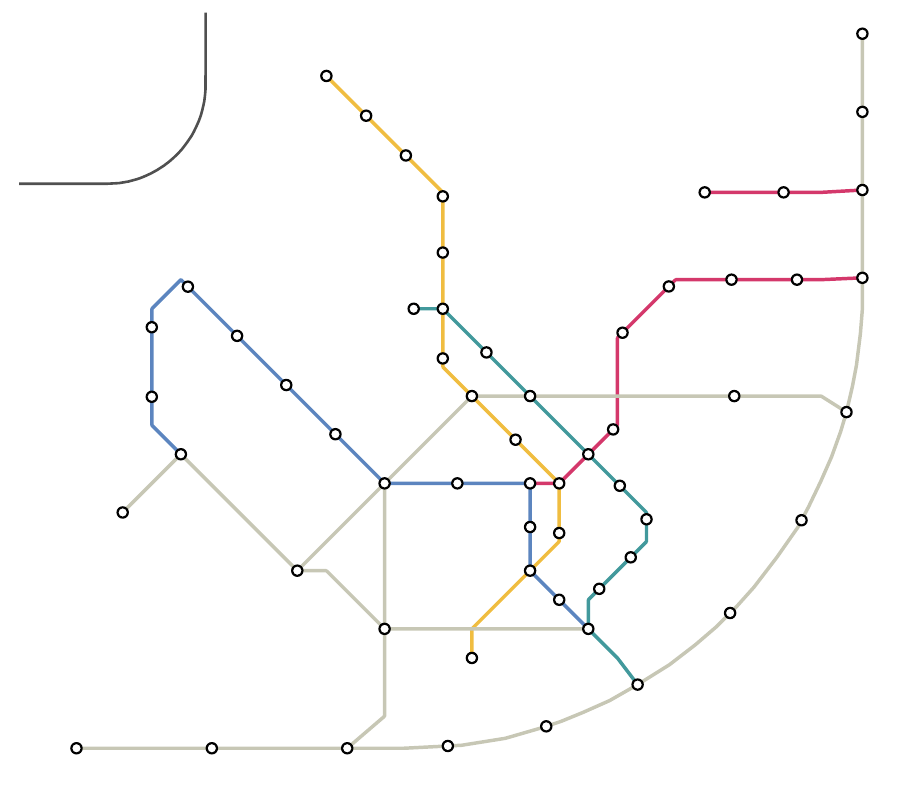}
    \caption{Lisbon metro and train map. Created with an curved guide shape, aiming to schematize the right and bottom train line that follows the coast. The guide shape is shown at the top left.}
    \label{fig:result-lisabon}
    \centering{}
    \includegraphics[width=0.8\linewidth]{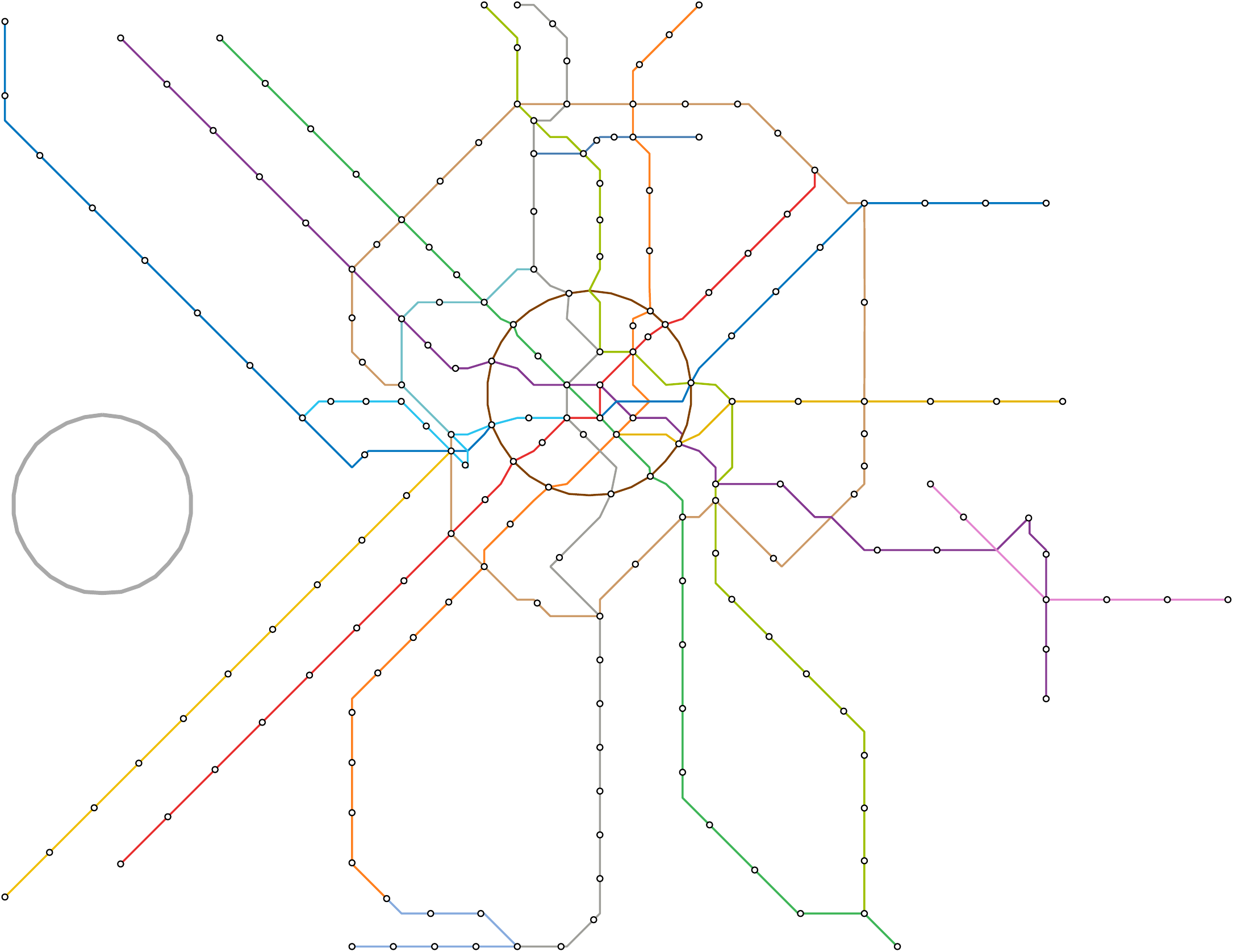}
    \caption{Moscow metro map. Created with a circular guide.}
    \label{fig:result-msocow}
\end{figure}

\section{Experimental Results}
\label{sec:result}
The route matching and the computation of the smooth and mixed layout was performed on a a MacBook Pro (2017) with a 2.9 GHz Quad-Core Intel processor.
The grid alignement was implemented in Python 3.8 using the NetworkX 2.6.3 module for shortest path and the bentley-ottman module 7.2.0 for segment crossing computation.
It was computed on a standard laptop running Ubuntu 21.20 with an Intel Core i5 processor (8 $\times$ 2.6 GHz) and 16GB of memory.
The grid size factor $f_{d}$ was $0.2$ for Moscow, Paris and Berlin and $0.3$ for all other instances.
The edge weights in $G$ were set to $c_{hop} = 20$ and $d_{min}$, $d_{low}$ and $d_{up}$ were set to $\frac{d}{5}$, $\frac{d}{2}$ and $\frac{3d}{2}$, respectively.
Our test cases can be grouped into two scenarios/potential use-cases for our approach. \rv{In the first cases} the path $\connectionPath$ was computed automatically as outlined in Section \ref{ssec:matching-automatic-scenario}.
Figure \ref{fig:teaser} shows the metro system of Taipei with a heart shape, and Figure \ref{fig:result-montreal} shows the one of Montreal with a cloverleaf shape, taken from the city's logo. Figures~\ref{fig:result-berlin-bear} and~\ref{fig:result-paris-eye} use more complex guide shapes embedded in larger metro networks, demonstrating the capabilities \rv{and limitations} of our approach.

For the second type of test case, we provide a manually defined path $\connectionPath$. In Figure~\ref{fig:result-berlin-oval} a layout of the Berlin metro and train network is created where \rv{the ring line is}
emphasized in an \rv{stadium} shape.
The provided path $\connectionPath$ is highlighted. For the Moscow system (Figure \ref{fig:result-msocow}) a similar concept is applied. For the Lisbon metro network (Figure \ref{fig:result-lisabon}) a schematic representation of the coastline \rv{guide} the train connection reaching from top right to bottom left. 
Table~\ref{tab:runtime} lists all the presented test-cases, the number of stations of the networks, the guide shape, and if $\connectionPath$ was provided or automatically computed.

\newcommand{\notInTable}{\textendash}

\section{Evaluation and Discussion}
\label{sec:discuss}

\begin{table*}[t]
    \centering{}
    \begin{tabularx}{\linewidth} {  %
    >{\raggedright\arraybackslash}X | >{\raggedleft\arraybackslash}X | >{\raggedleft\arraybackslash}X | >{\raggedleft\arraybackslash}X | >{\raggedleft\arraybackslash}X | >{\raggedleft\arraybackslash}X | >{\raggedleft\arraybackslash}X | >{\raggedleft\arraybackslash}X %
    }
         Network & Lisbon (60) Figure~\ref{fig:result-lisabon} & Montreal (68) Figure~\ref{fig:result-montreal} & Taipei (96) Figure~\ref{fig:teaser} & Moscow~(204) Figure~\ref{fig:result-msocow} & Berlin (272) Figure~\ref{fig:result-berlin-oval} & Berlin (272) Figure~\ref{fig:result-berlin-bear}  & Paris (304) Figure~ \ref{fig:result-paris-eye}\\
         \hline
         Guide Shape & Coastline~(29)$^\star$ & Cloverleaf~(91) & Heart~(52) & Circle~(30)$^\star$ & \rv{Stadium}~(53)$^\star$ & Bear~(47) & Eye~(78) \\
         \hline
         Route (s) & \notInTable   & 3.82  & 4.94   & \notInTable & \notInTable   & 80.15       & 223.52 \\
         Smooth (s)         & 1.47          & 1.67  & 4.77  & 39.44  & 133.21        & 167.47            & 189.45\\
         Mixed (s)          & 0.39          & 0.20  & 0.53  & 5.98   & 21.12         & 25.21           & 25.46\\
         \hline
         Total (s)  & 1.86          & 5.69  & 10.24 & 45.42 & 154.33        & 272.84            & 438.43\\ 
         \end{tabularx}
    \caption{
    Test instances %
    and guide shapes with number of vertices (after planarization and before subsampling, respectively) in parenthesis with their corresponding runtimes for the route matching and deformation processes.
    All times are given in seconds.
    Shapes marked with $^\star$ were manually defined, and therefore have no route matching time.
    }
    \label{tab:runtime}
    
\end{table*}

In this section, we examine the performance of our approach, and the visual quality of the maps generated using our system.

\subsection{Performance Evaluation}

The measured computation times for the figures of the paper are summarized in Table~\ref{tab:runtime}.
We observe that \rv{the running time} increases based on the network sizes and route sizes.
\rv{As the grid alignment is built based on the shortest path algorithm with complexity $N(|V|^2)$, the runtimes are not yet competitive for real-time applications (a minute for smaller instances like Montreal and Lisbon, multiple minutes for medium instances like Singapore and Moscow and up to 15 minutes for the large instances Berlin, Paris, since we did not parallelize our code).
}
\yuninline{@Soeren, please check!}
\soereninline{I added 'in our implementation', since faster implementations exist. Significantly so.}
\yuninline{@Soeren, do you mean algorithm-wise or implementation-wise? parallelization for example}

\subsection{Visual Quality Evaluation}

To examine the visual quality of our layouts, we conducted an online user study. \yun{supplementary materials?} 
\rv{
As previously summarized in Section~\ref{ssec:design}, one goal of our approach is to make the embedded shapes recognizable on maps (D6) so that they become helpful for general route-finding tasks that can be done on the maps.
Therefore, we are interested in two research questions, including (1) \emph{Are shapes in maps significant enough to be identified?} and (2) \emph{Are embedded shapes helpful or harmful for route planning tasks in comparison to classical design?}
} 
For the \emph{shape recognizability} (E1), we asked the participants to mark a shape if they saw one using a polyline drawing tool via mouse clicking.
To evaluate the \emph{map usability} (E2), we explicitly asked participants to trace and count the number of stations between a start station and a destination station. The tasks are done on classical octolinear maps and \rv{our mixed metro maps}. 
To analyze the result of (E1), we overlay the polyline coordinates with the map and annotate each result manually. As for (E2), we analyze \rv{task accuracy and time performance} on both map styles.
The study \rv{begins with example training}, followed by formal questions on small (approx. $100$ stations), medium (approx. $200$ stations), and large (approx. $300$ stations) networks.

\begin{figure}[t]
    \centering{
    \setlength{\tabcolsep}{0pt}
    \begin{tabular}{ccc}
        \includegraphics[width=0.4\linewidth]{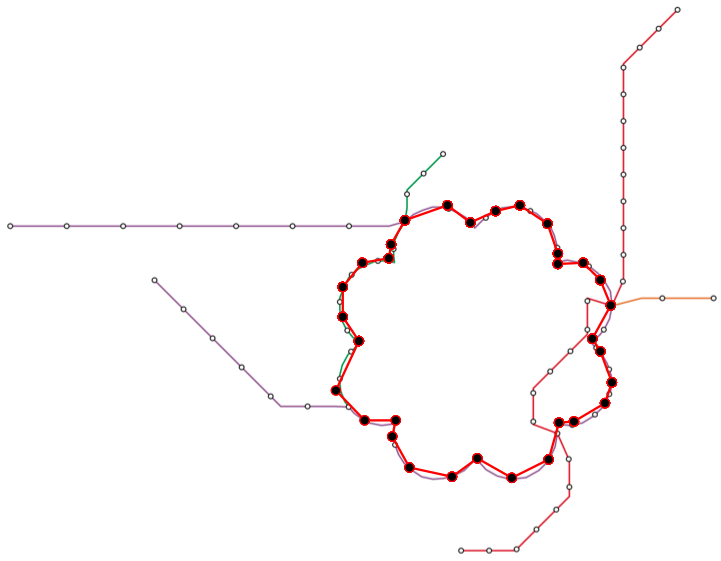} &
        \includegraphics[width=0.3\linewidth]{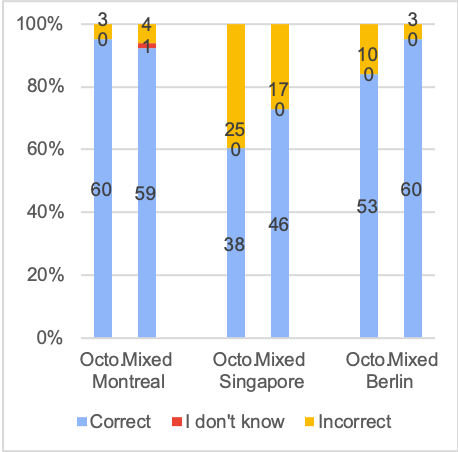} &
        \includegraphics[width=0.3\linewidth]{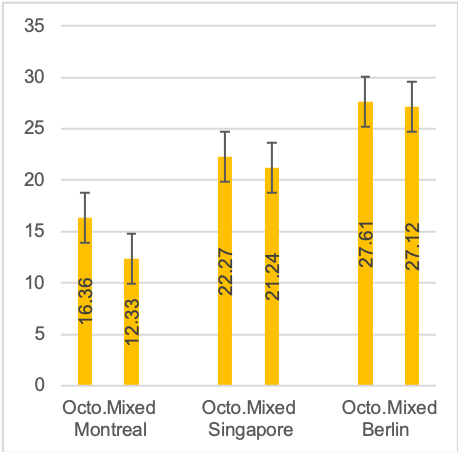} \\
        (a) & (b) & (c)
    \end{tabular}
\caption{(a) An example of the shape recognition task, and (b) \rv{accuracy and (c) time} of the route finding task. \rv{The time is recorded in seconds and the error bars represent the standard errors.}}
\label{fig:evaluation}
    }
\end{figure}

We received in total \rv{$65$} results, while we removed \rv{one incomplete answer} and one which did not pass our training exercises. We analyzed \rv{$63$} results from the participants ($\rv{16}$ females, $\rv{40}$ males, and $\rv{7}$ \rv{not disclosed}) with ages ranging from $20$ to \rv{$59$}. Five are professional transit map users (e.g., cartographers, designers, transport company employees, etc). \rv{$15$} are everyday users, \rv{$25$} are frequent users, and \rv{$15$} are occasional users. One participant rarely uses a transit map and two participants never used it.
Figure~\ref{fig:evaluation} (a) gives an example of a route marking result from a participant, and Figure~\ref{fig:evaluation} (b) shows a summary of the route finding task accuracy. For (E1), \rv{we prepared maps from cities in several countries and with different guide shapes}, including Montreal (flower), Singapore (heart), Paris (circle), Berlin (bear), and Paris (eye).
\rv{More than $95\%$} of the participants marked the shapes correctly in the the Montreal (flower), Singapore (heart), and Paris (circle) maps. \rv{The eye in the Paris (eye) map was recognized by $89\%$ of the participants.} The coarse shape of the bear was recognized by \rv{$60\%$} of the participants, while none of them identified the teeth of the bear. \rv{We assume this} could be because the teeth are a small portion of the entire shape and the stations are not sufficient enough to represent it.
Except for the Berlin (bear) map \rv{($58\%$), more than $97\%$ participants} considered the shapes in the study significant. 
We also asked participants if the symbolic shapes helped them recognize the maps.
This is done trough a 5 Point Likert Scale Analysis (yes, partially yes, neutral, partially no, and no).
\rv{When the shapes are simple (flower ($3.75$), heart ($3.51$), and circle ($3.71$)), the answers lean toward positive,  while if the shapes are complicated (bear ($2.85$) and eye ($3.31$)), the preference is close to neutral or negative.}

For (E2), we compare \rv{task performance on a classical} octolinear layout and a layout with embedded shapes on Montreal (flower), Singapore (circle), and Berlin (stadium). With simple shapes (e.g., circle, stadium, etc.), \rv{we see a slight improvement in task accuracy on Singapore and Berlin mixed maps (Figure~\ref{fig:evaluation}(b)), and the time used to accomplish the route planning tasks is reduced in general (Figure~\ref{fig:evaluation}(c)). The Berlin (stadium) has better accuracy because the stadium is a nice shape for representing a circular route of this network.}
We also received some explicit feedback from the participants.
\rv{Ten participants mentioned in the feedback form that they found the shape design interesting and potentially helpful, especially when the shapes are simple, while two participants} disliked embedding artificial shapes due to the inaccuracy introduced in comparison to more topographically correct maps.
Two participants mentioned that corners on the shapes were important features that were expected to be mapped to stations in the final result. It is an interesting observation, while we did not come up with an intuitive extension to achieve this.
One participant suggested that embedding a shape covering a larger area would increase the map's memorability. 
Another two participants mentioned that the embedded shapes might influence \rv{their} planning decisions since they introduced additional bends in the map. 

\subsection{Limitations \rv{and Discussion}} \label{ssec:limitations}

Although the route matching generally found a good alignment between the guide shape and the metro network in our test cases, we cannot guarantee their global optimality. %
In addition, formalizing and predicting the recognizability of shapes is a hard task, especially on shapes with multiple polylines. Likely in such a scenario not all features and subsections of a shape contribute equally to the recognizability. Our approach does not differentiate sub-regions of the guide shape and treats the entire guide shape equally.

\rv{O}ur deformation process creates the layout by minimizing sets of local constraints, which may cause unexpected results. 
For example, when deforming the transit network, the guide shape can prevent areas with a higher station density to expand, like in the Moscow map (Figure~\ref{fig:result-msocow}), where a larger circular route would be beneficial. 
Note that we also do not claim that any shape will be appropriate for a transit map.%
However, we see other potentials in the marketing, such as logo (Figure~\ref{fig:example}) and poster design~\cite{zerozero}.
Another limitation is inherited from the octolinear transit map design, 
where we do not accept a station degree that is larger than eight.
In our implementation, we split high-degree stations into multiple stations with degrees less than eight in a pre-processing step.

\begin{figure}[t]
    \centering{}
    \includegraphics[width=\linewidth]{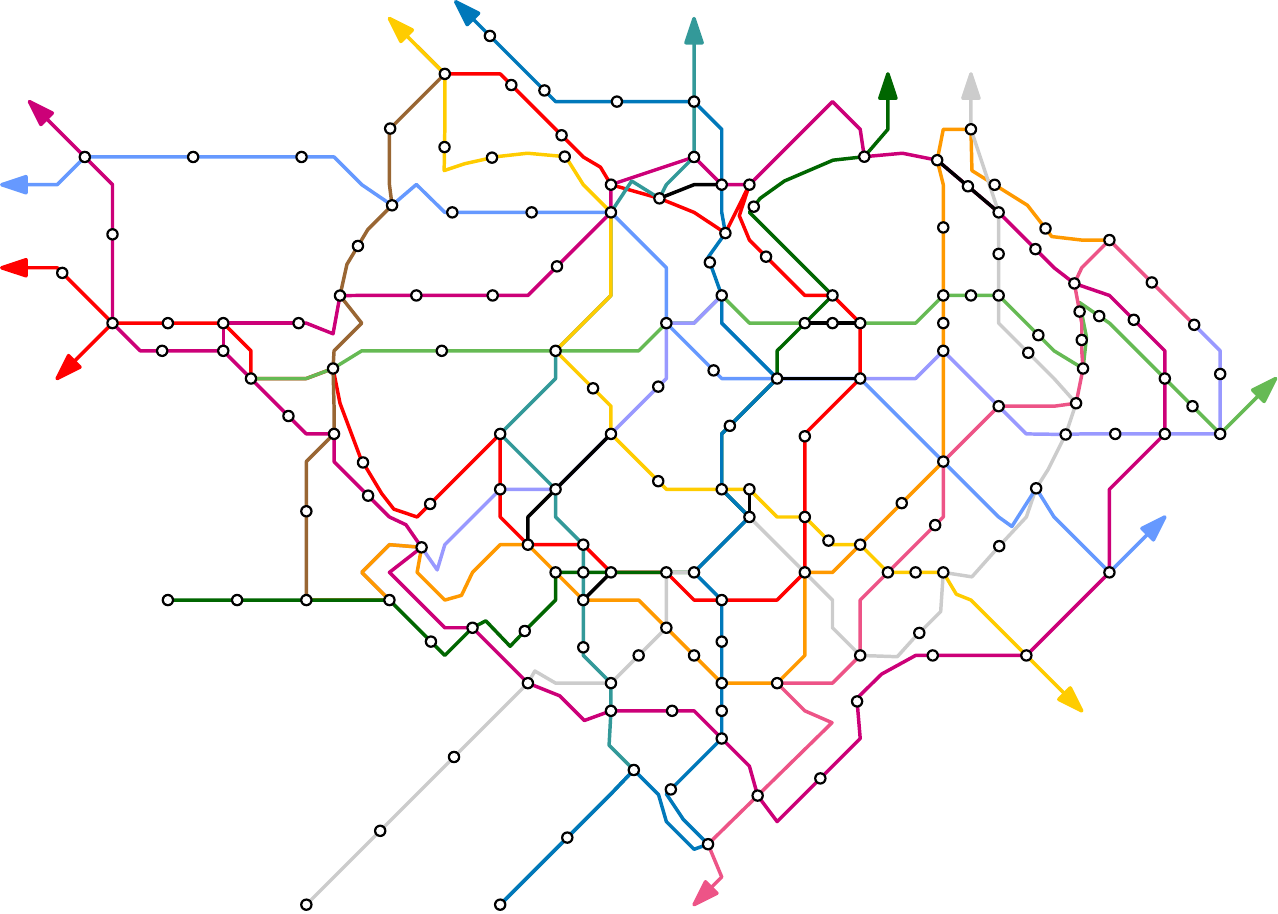}
    \caption{Tokyo metro map with a heart shaped guide guide. Five edges could not be routed. The grid alignment was computed based on the smooth layout and only the dense center of the map is shown.}
    \label{fig:result-failed_tokyo}
\end{figure}

The grid alignment 
requires that the vertices of $\shape$ are connected with a suitable set of vertices of the grid,
\rv{which} results in non-octolinear edges at the interface between the grid and $\shape$.
The iterative computation \rv{might not} always find a routing for all edges, although even for \rv{large} networks, we \rv{found} layouts where \rv{only few} number of edges ($<5$) fail.
An example is the layout of Tokyo, shown in Figure~\ref{fig:result-failed_tokyo}.
\rv{The running time bottleneck of the grid alignment is the grid size and the shortest path algorithm.}
\yuninline{@Soeren, please check!}
\soereninline{Yes that is true, though a big part is computation of the corssings in the grid, which coul be sped up. But that is kind of subsumed in grid size.}
However, the underlying method of using repeated shortest path computations has been shown to be highly efficient on both octolinear~\cite{BastBS20} and flexible grids~\cite{BastBS21} and has recently been made openly available as part of a toolchain for transit map generation~\cite{brositoolchain}.
Finally, our current approach applies a post-processing labeling technique~\cite{niedermann2018algorithmic}, which can potentially produce rather small text labels.
\rv{The implementation is available at \href{https://github.com/TobiasBat/Shape-Guided-Mixed-Metro-Map-Layout}{https://github.com/TobiasBat/Shape-Guided-Mixed-Metro-Map-Layout}}.
\yuninline{@Soeren, please check. move the link to citation. CFG does not permit footnote.}
\soereninline{Thanks for moving it}

\section{Conclusion and Future Work}
\label{sec:conclude}

In this paper, we introduced a new layout approach for synthesizing more engaging, mixed-style metro maps. This is achieved by embedding recognizable shapes into a classical metro map. The presented algorithm handles automatic and interactive route matching, shape-aware deformation, and finally, grid alignment sub-problems. With our results and evaluation, we show that the synthesized maps are of good quality and that the embedded shapes are intuitively recognizable.
As a primary future research direction, we aim to investigate embedding more complex shapes, but also in combination with non-octolinear grids. 
One initial idea here is to decompose the network together with the shape hierarchically.
We also plan to investigate more scalable methods to enable a real-time workflow for larger networks with proper labeling as well as to develop systematic quality metrics on such representations that are heavily linked to human shape recognition.
\rv{Finally, we aim to conduct a usability test that covers the aforementioned spectrum to examine the function of shapes in a layout design.}

\section*{Acknowledgments}
\rv{
We thank Dr. Benjamin Niedermann from yWorks GmbH, T{\"u}bingen, Germany, who shared his metro map labeling code with us.
MN acknowledges funding from the Austrian Science Fund (FWF) under grant P31119.
}

\bibliographystyle{eg-alpha-doi}
\bibliography{paper,bibfiles/book,bibfiles/design,bibfiles/metaphor,bibfiles/survey,bibfiles/machine,bibfiles/user}

\clearpage
\appendix

\section{Path Invariant Testing Experiment} \label{sup:path}

\begin{figure*}[t]
    \centering{
    \setlength{\tabcolsep}{0pt}
    \begin{tabular}{cccc}
        \includegraphics[width=0.25\linewidth]{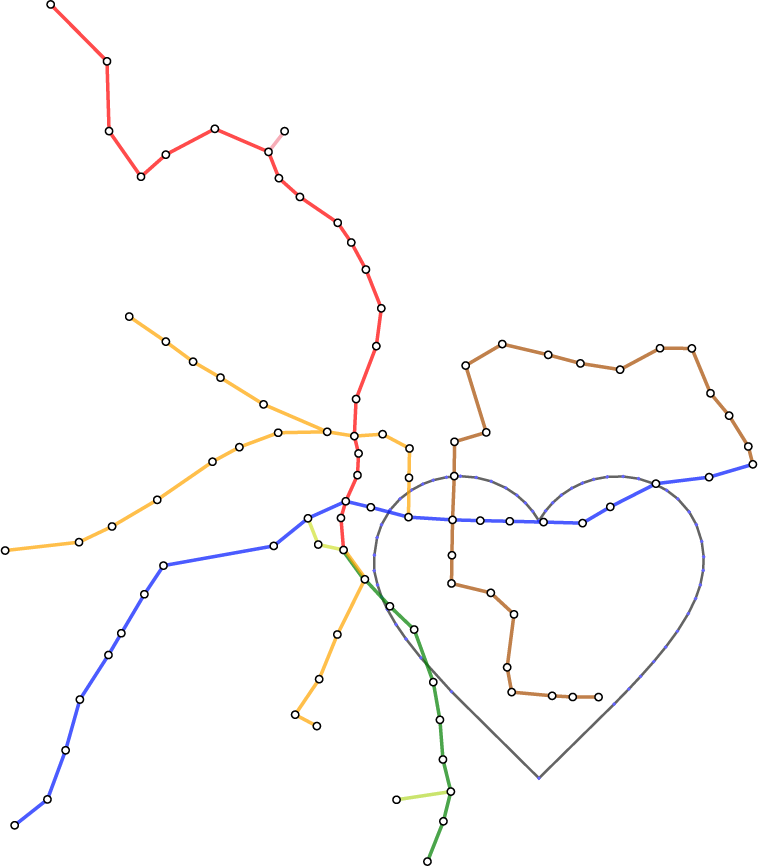} &
        \includegraphics[width=0.25\linewidth]{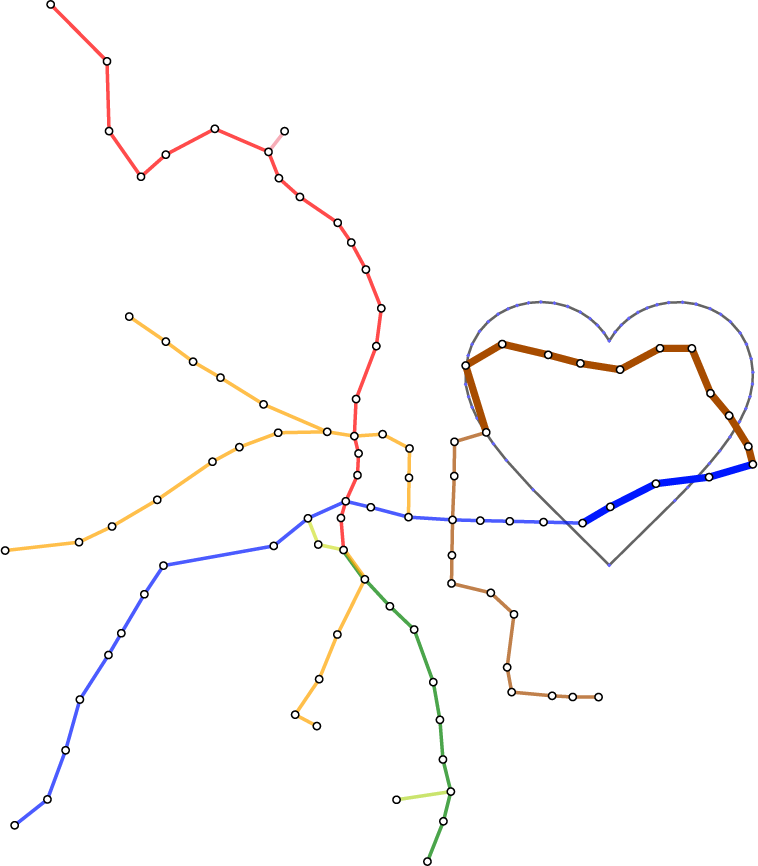} &
        \includegraphics[width=0.25\linewidth]{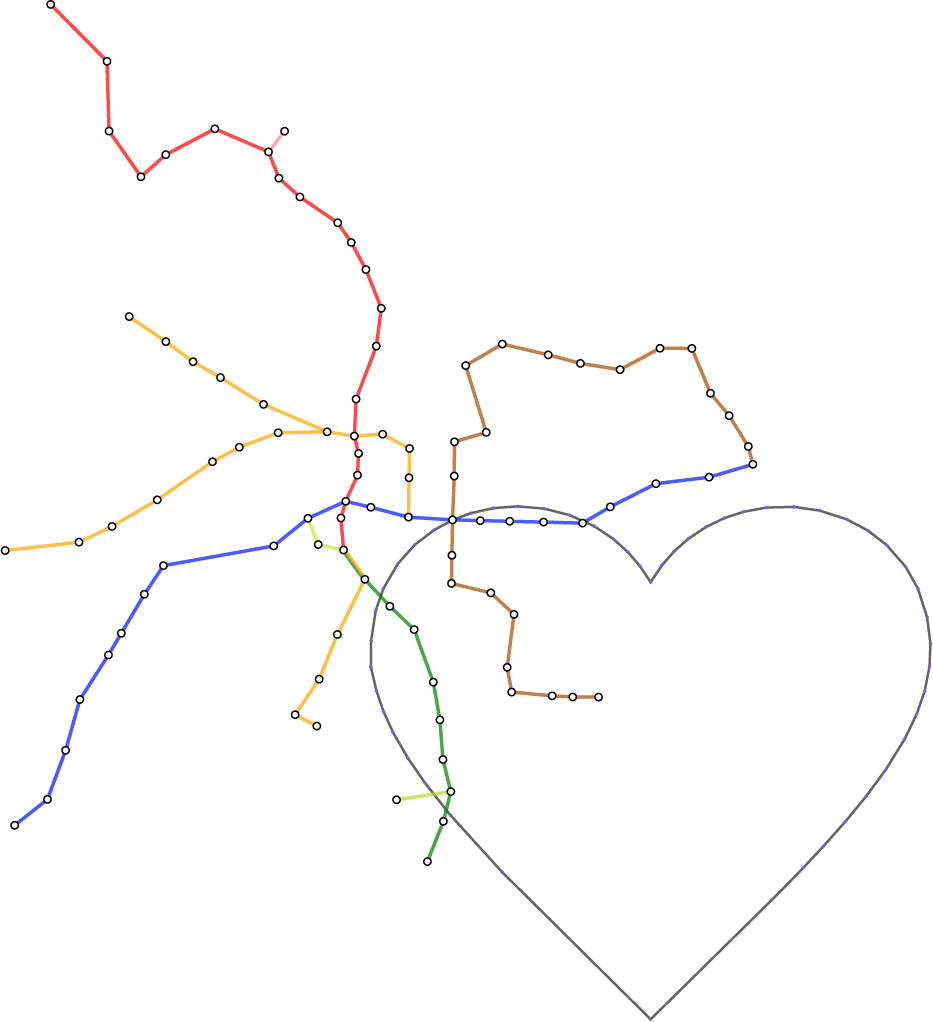} & 
        \includegraphics[width=0.25\linewidth]{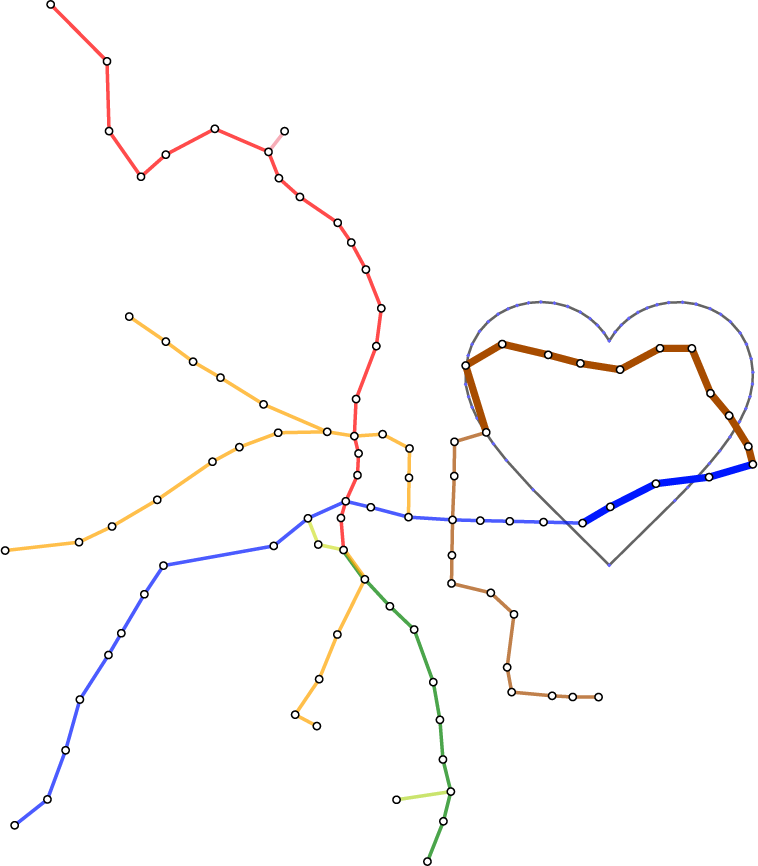} \\
        (a) Input & (a) Output & (b) Input & (b) Output  \\
        \includegraphics[width=0.25\linewidth]{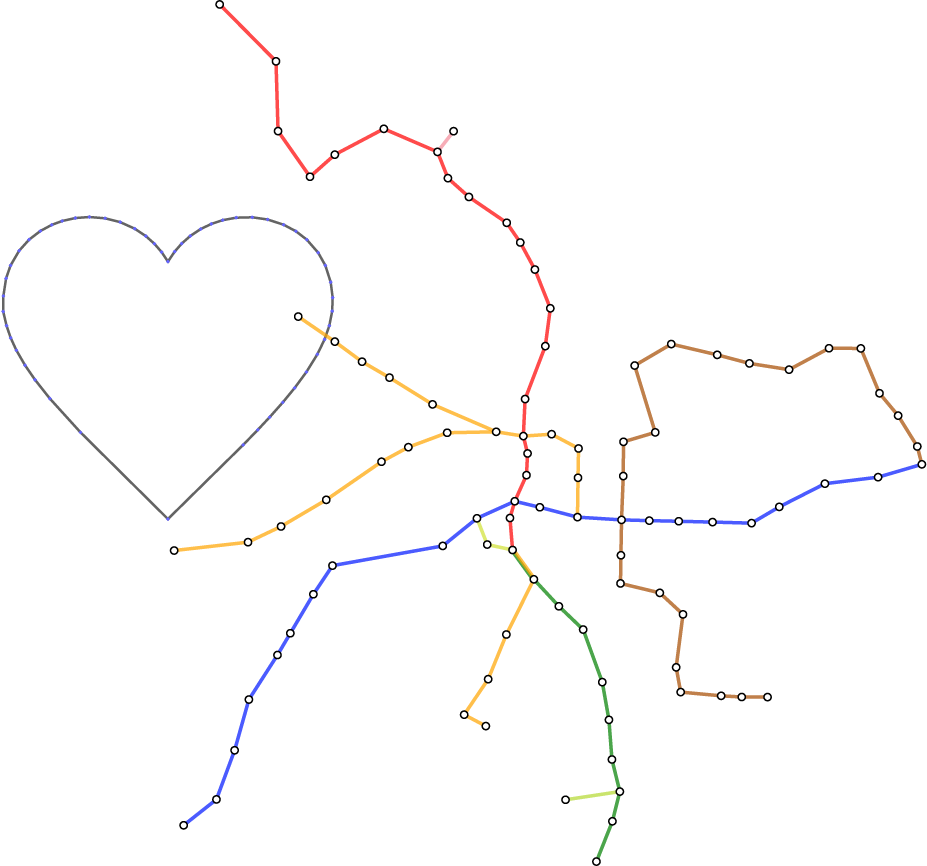} &
        \includegraphics[width=0.25\linewidth]{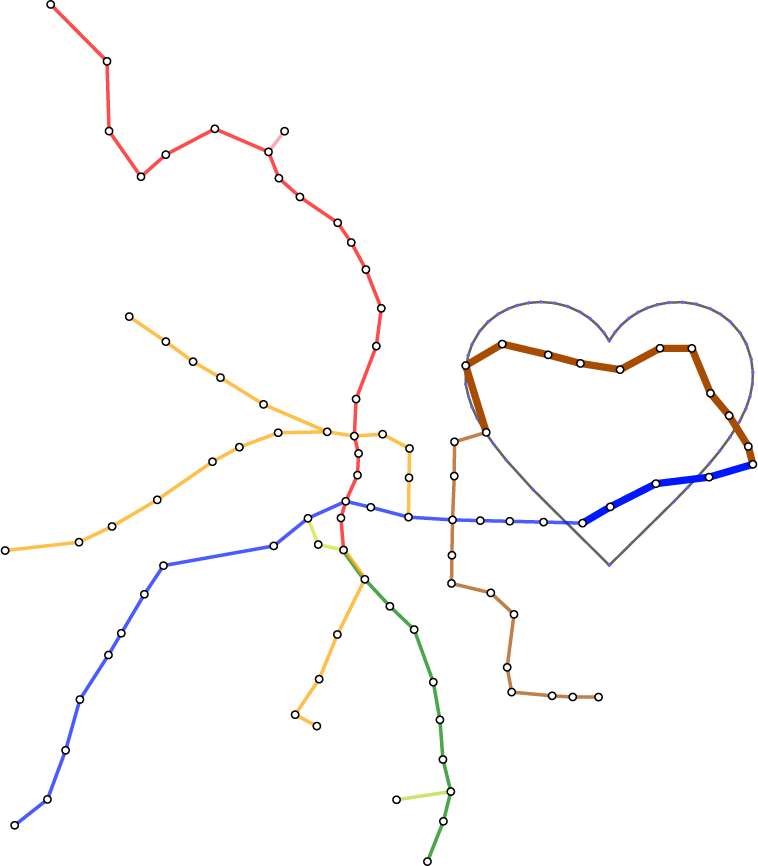} &
        \includegraphics[width=0.25\linewidth]{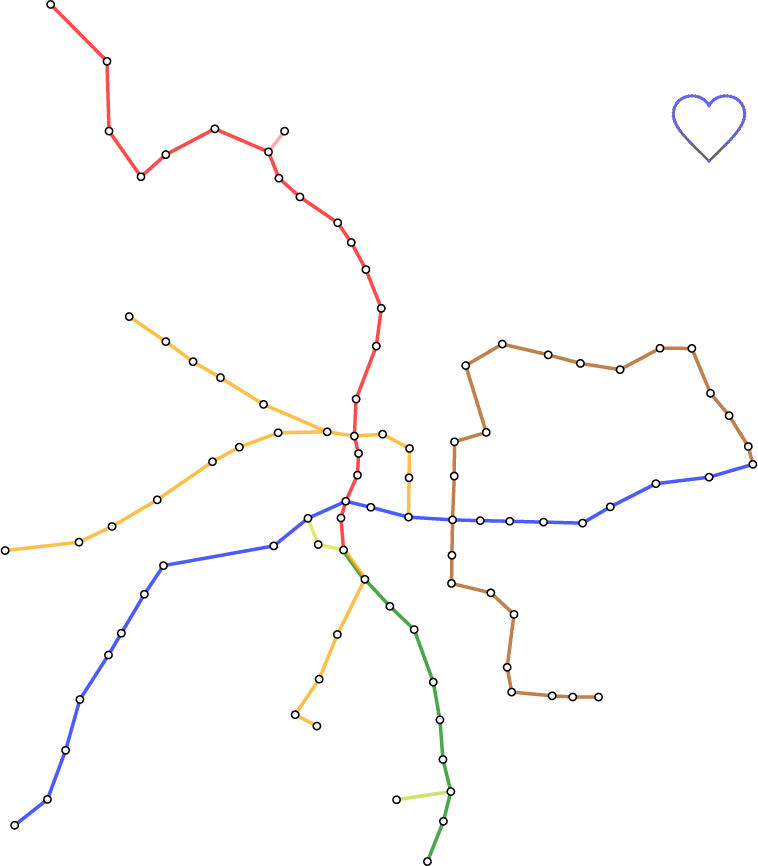} & 
        \includegraphics[width=0.25\linewidth]{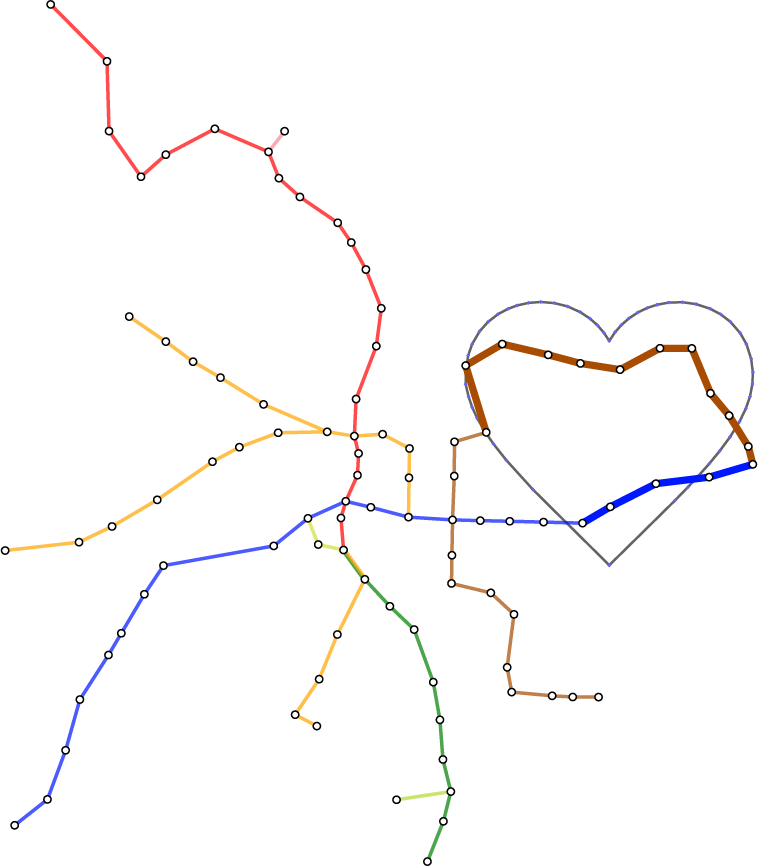} \\
        (c) Input) & (c) Output & (d) Input & (d) Output  \\
        \includegraphics[width=0.25\linewidth]{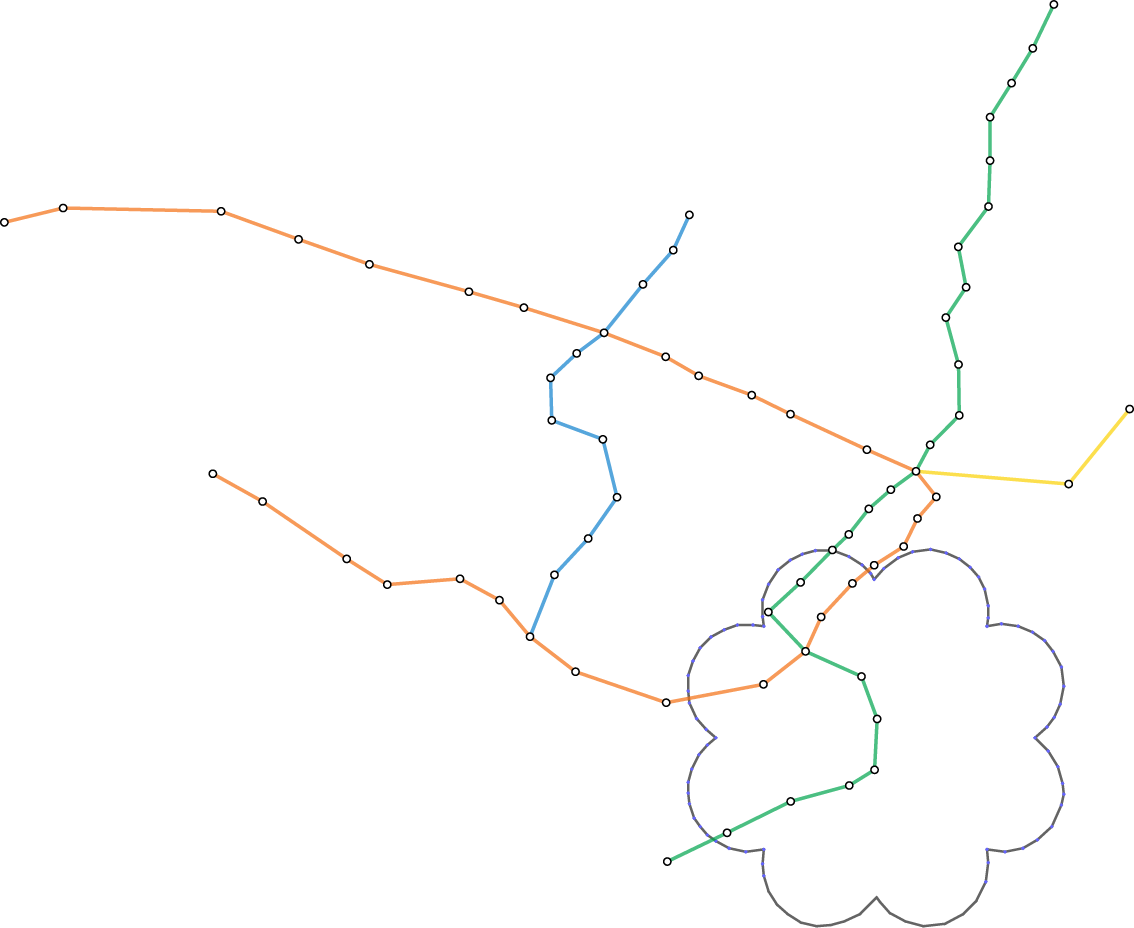} &
        \includegraphics[width=0.25\linewidth]{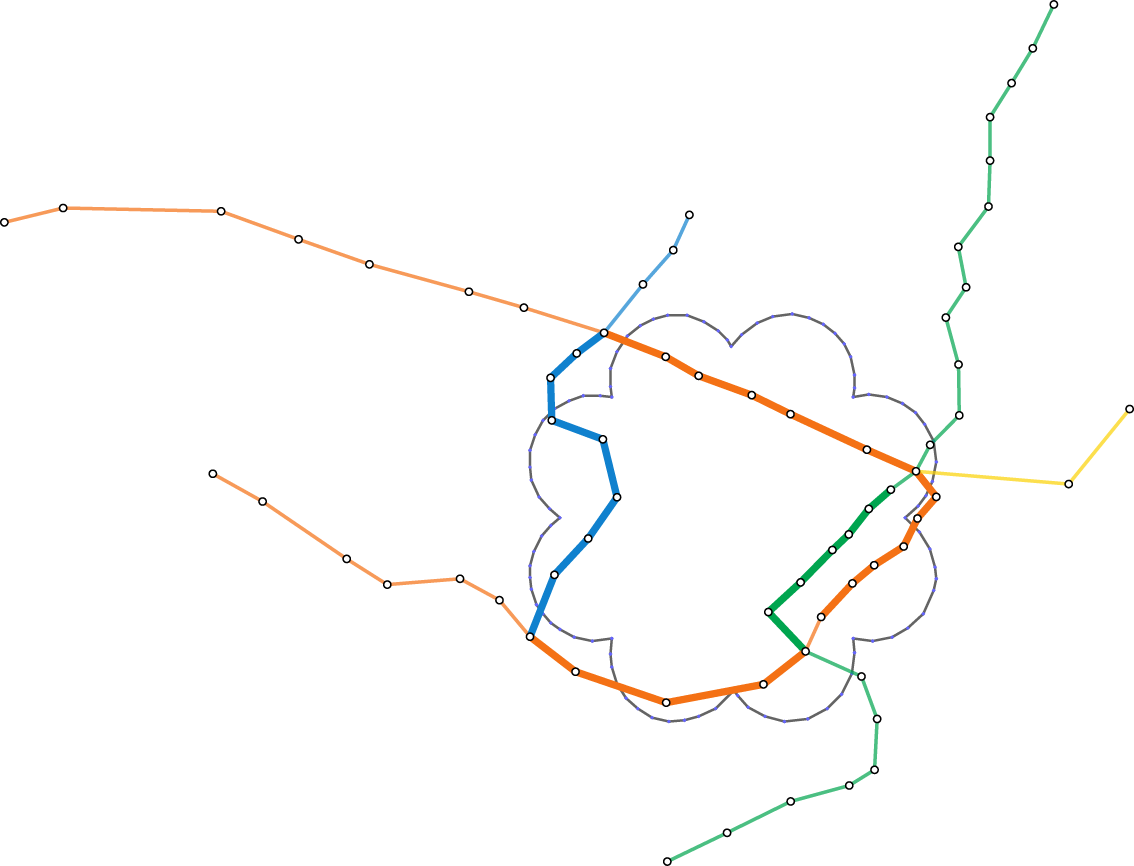} &
        \includegraphics[width=0.25\linewidth]{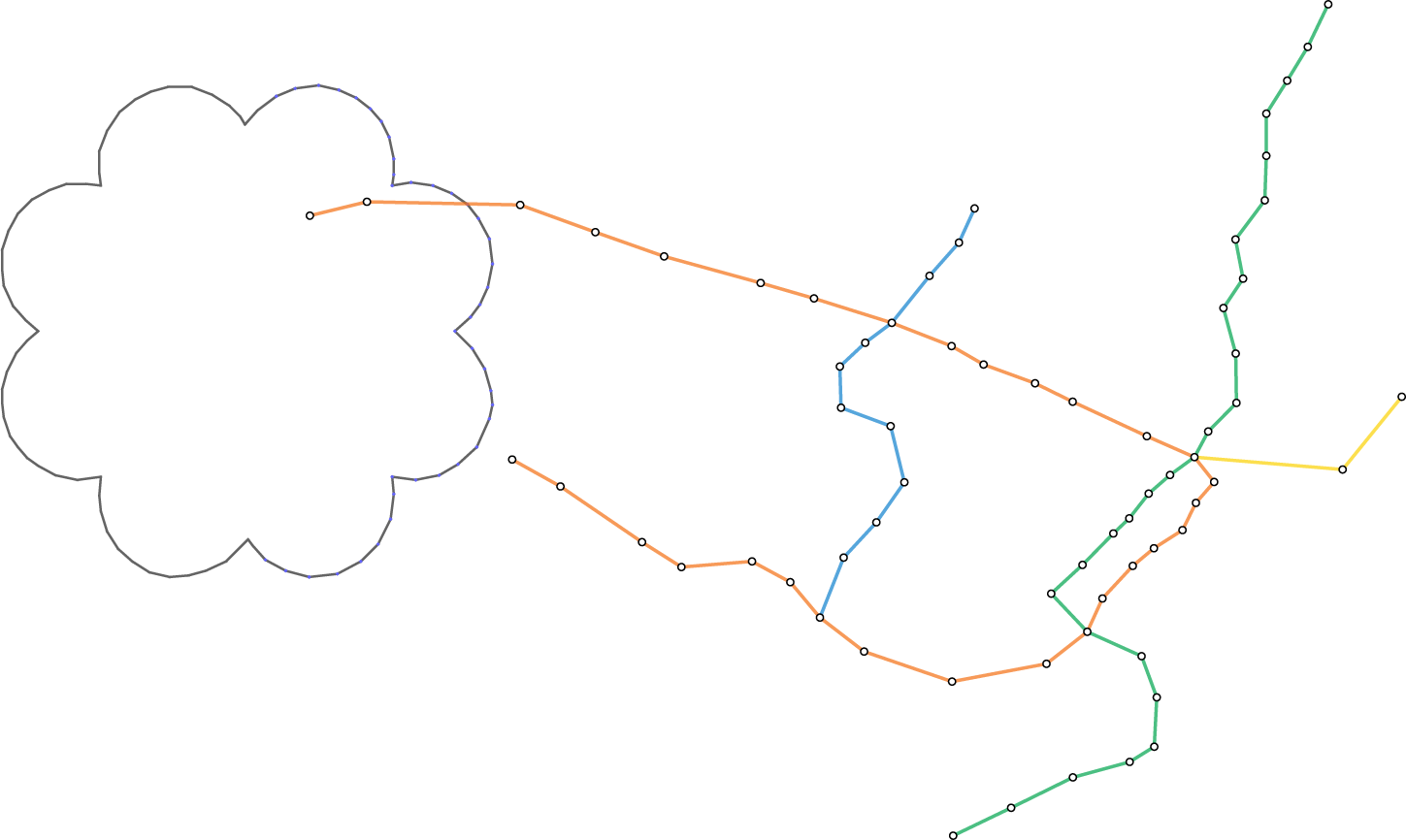} & 
        \includegraphics[width=0.25\linewidth]{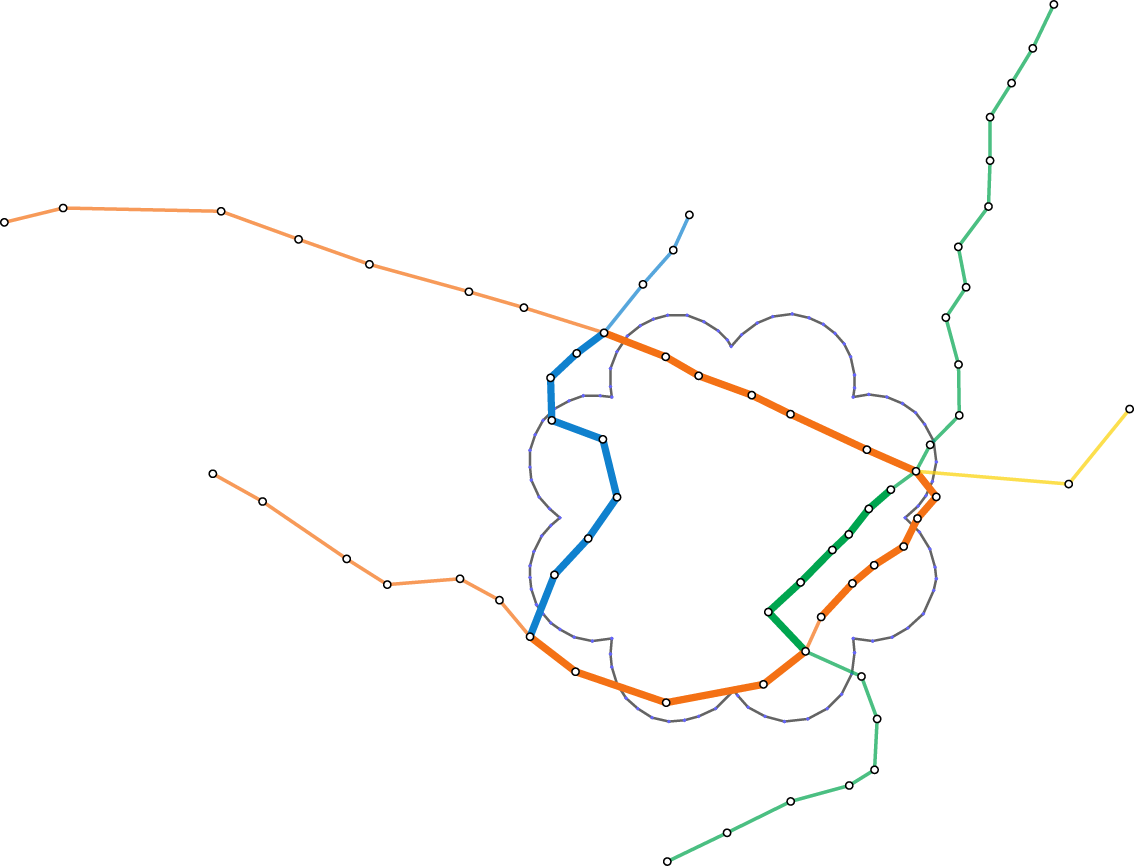} \\
        (e) Input & (e) output & (f) Input & (f) Output  \\
        \includegraphics[width=0.25\linewidth]{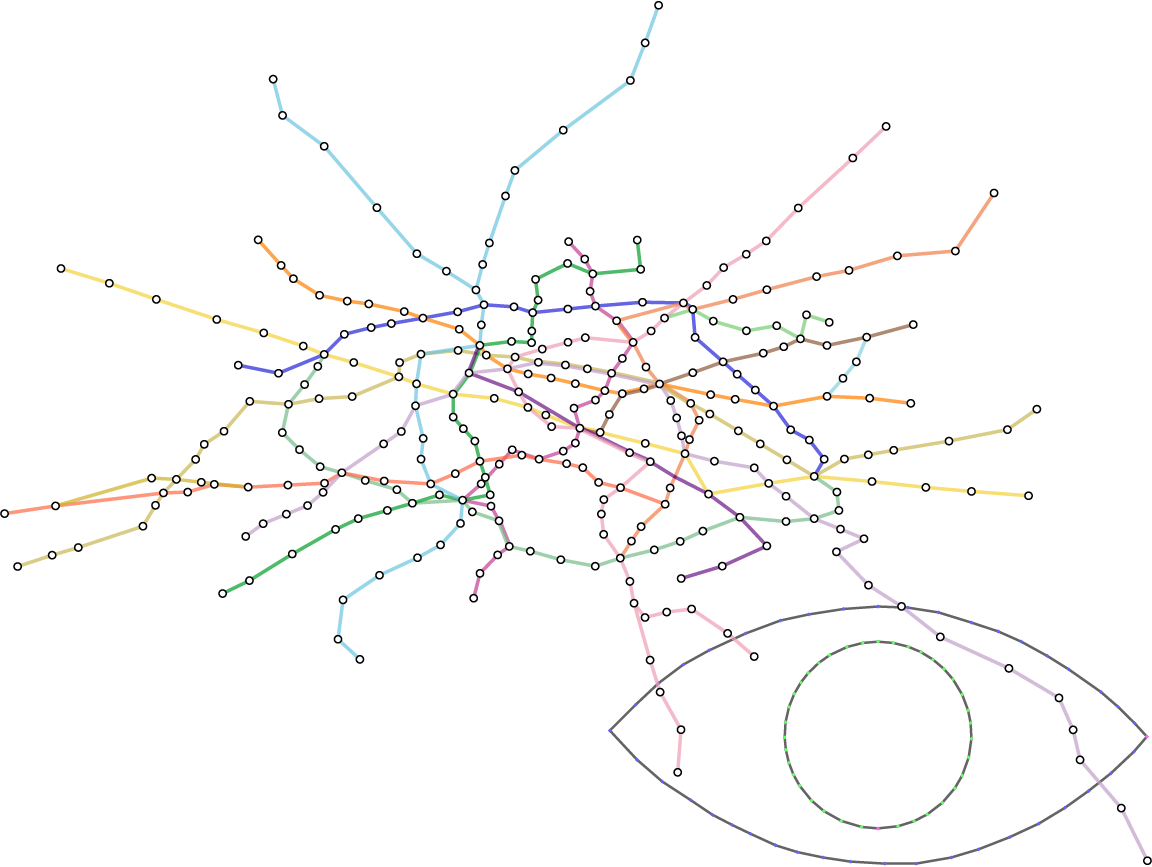} &
        \includegraphics[width=0.25\linewidth]{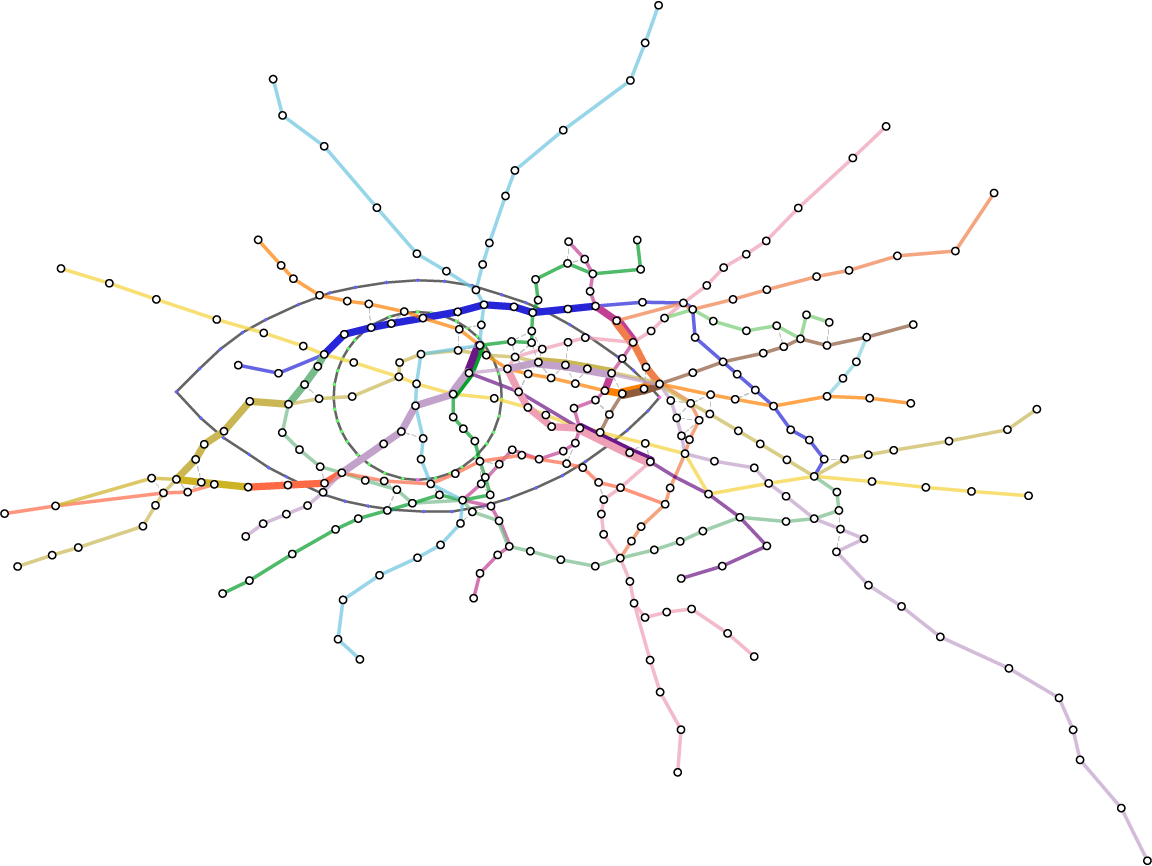} &
        \includegraphics[width=0.25\linewidth]{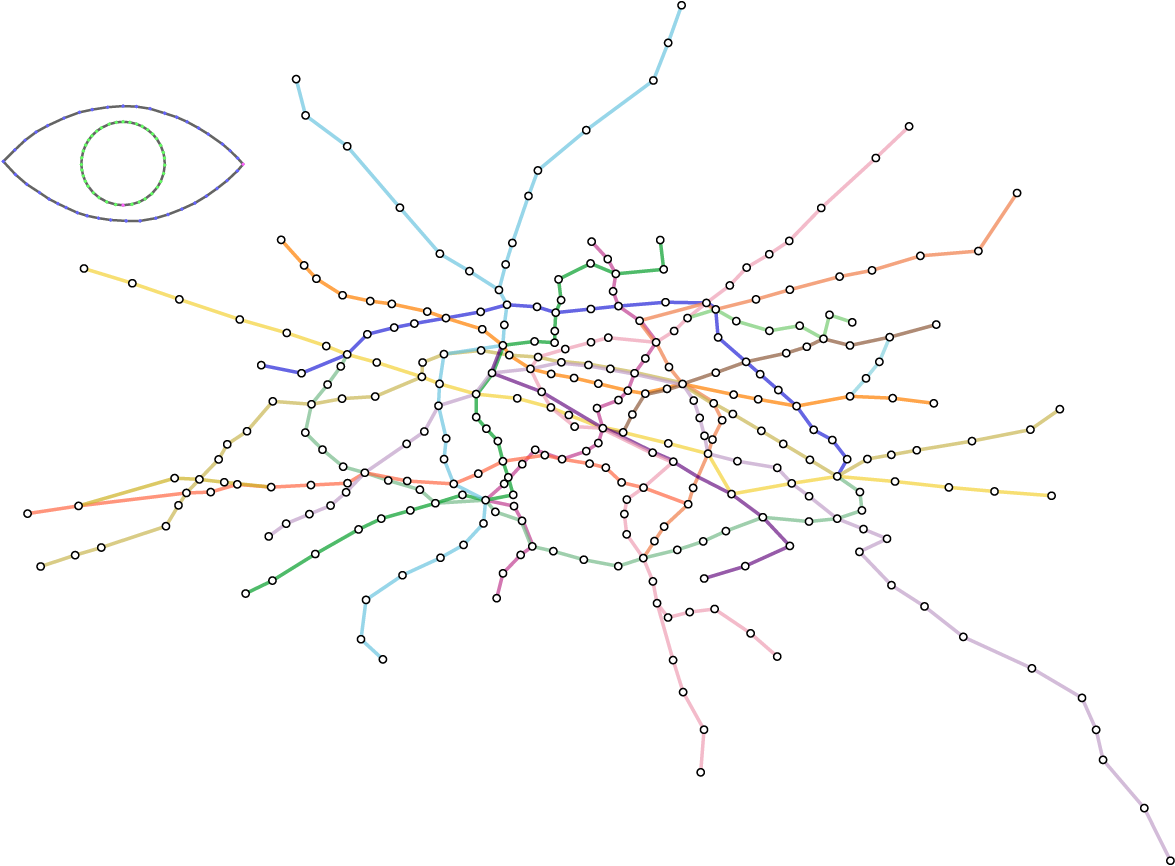} & 
        \includegraphics[width=0.25\linewidth]{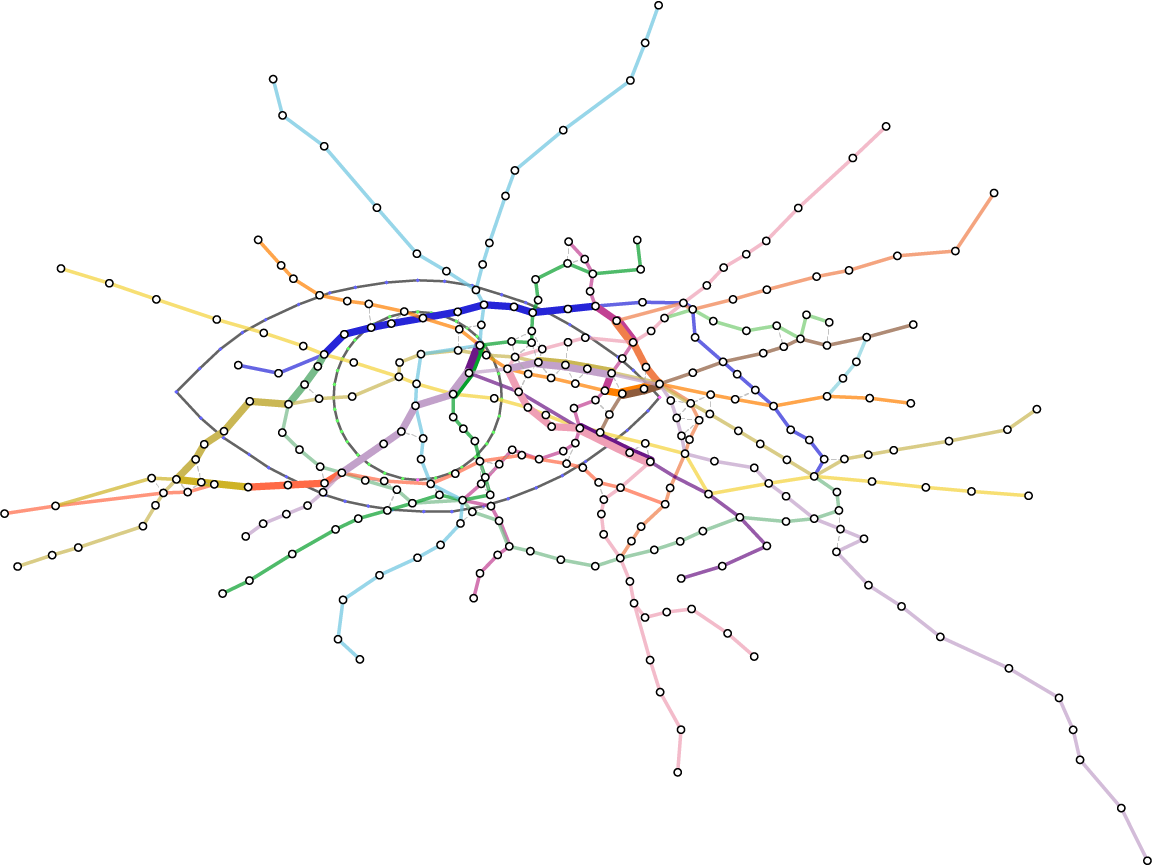} \\
        (g) Input & (g) Output & (h) Input & (h) Output \\
    \end{tabular}
    \caption{\rv{Results of the translation and scale invariant test. For each test-case, we show the input guide shape (gray) in the input image and the resulting path (highlighted with a bold line) in the output image.}}
    \label{fig:supp-path}
    }
\end{figure*}

\rv{In this section, we would like to show several examples to demonstrate the proposed approach is translation and scale invariant, as explained in Section~\ref{sec:align}. For each test case, we generate an arbitrary guide shape with random size and position. The results are shown in Figure~\ref{fig:supp-path}, where we show the position and scale of the guide shape in the input image and the resulting path in the corresponding output image. Figures~\ref{fig:supp-path}(a)-(d) show the examples of Taipei metro system, Figures~\ref{fig:supp-path}(e)-(f) are the examples of Montreal metro system, and Figures~\ref{fig:supp-path}(g)-(h) depict the examples of Paris metro system. The guide shapes include a heart, a flower, and an eye, respectively. Based on this experiment, shape size and position are invariant since the found paths are identical even though the shape size and position are different.}

\clearpage

\section{Parameter Testing Experiment} \label{sup:parameter}

\begin{figure*}[t]
    \centering{
    \setlength{\tabcolsep}{0pt}
    \begin{tabular}{ccc}
        \includegraphics[width=0.25\linewidth]{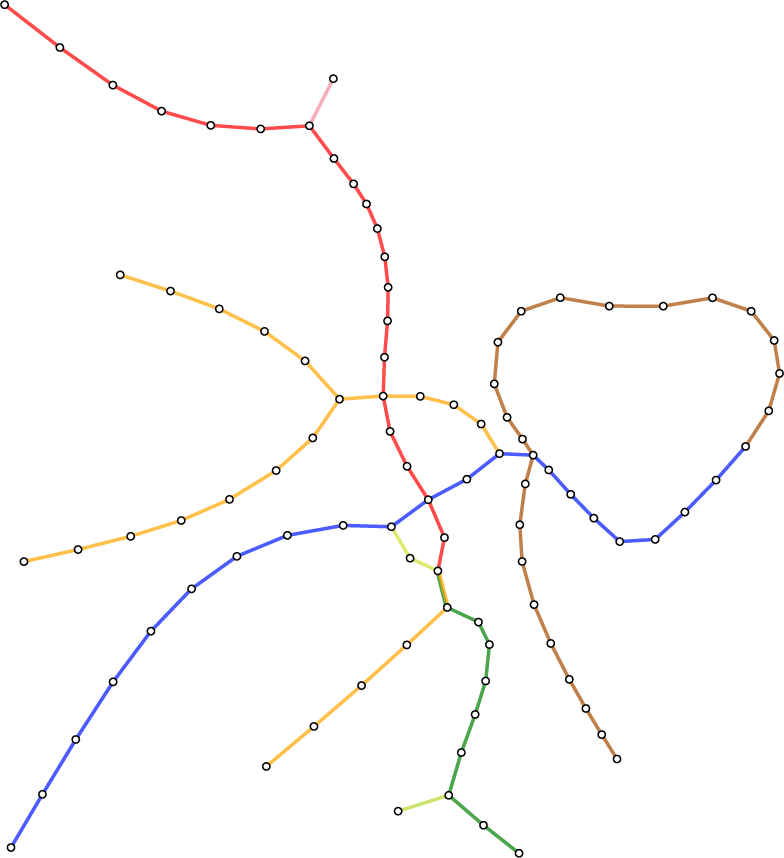} &
        \includegraphics[width=0.25\linewidth]{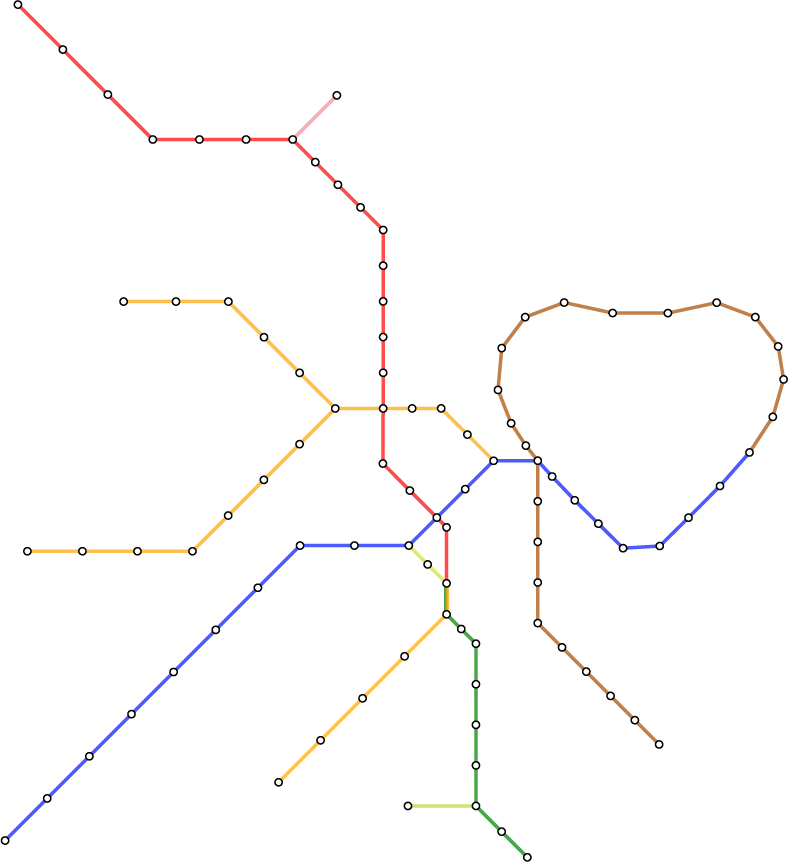} &
        \includegraphics[width=0.25\linewidth]{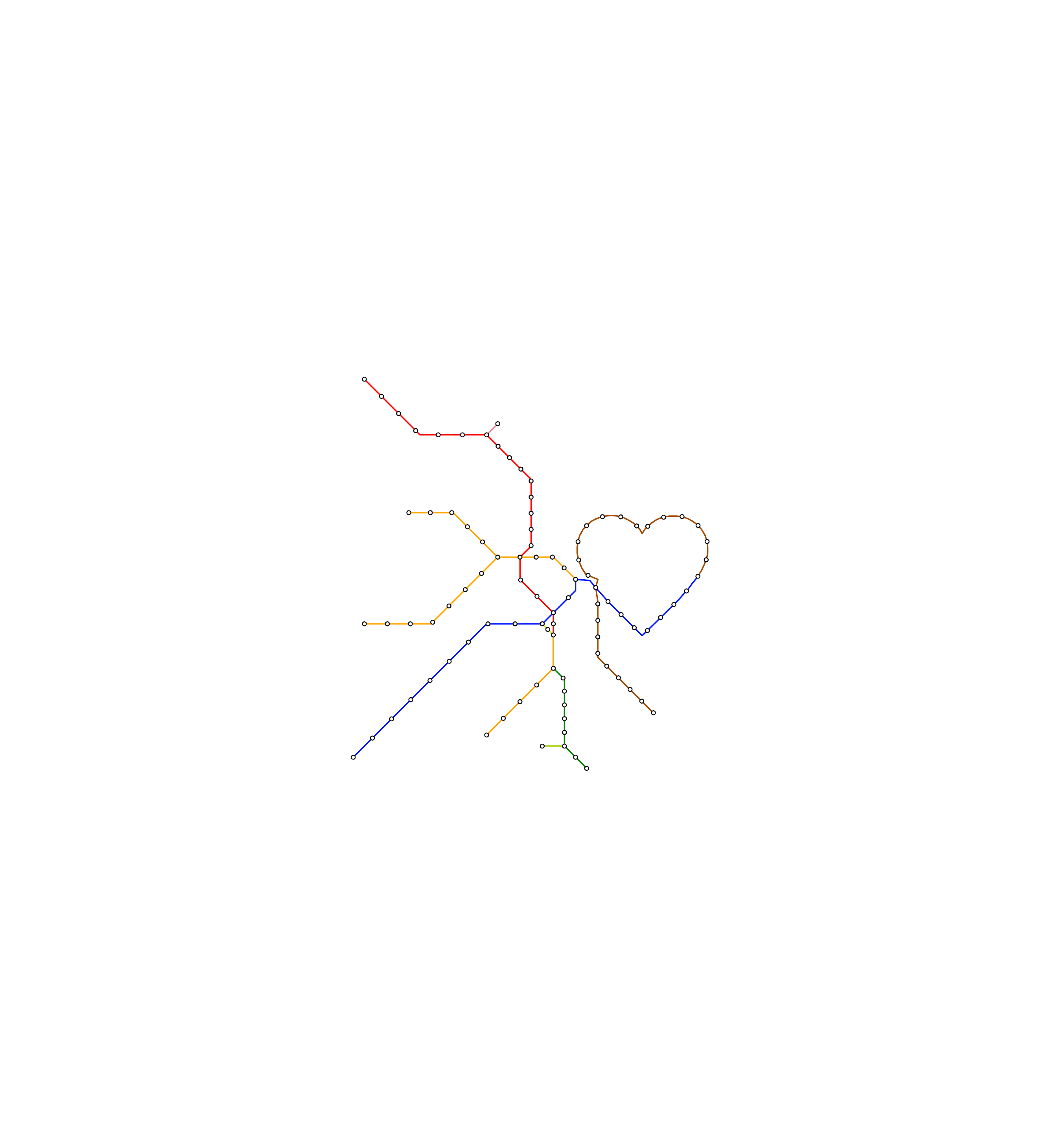} \\
         \multicolumn{3}{c}{\rv{(a) Default Test-case; Parameters for the smooth stage: $w_c = 4$, $w_l = 1$, $w_a = 2$, $w_p = 0.16$.}} \\
        \includegraphics[width=0.25\linewidth]{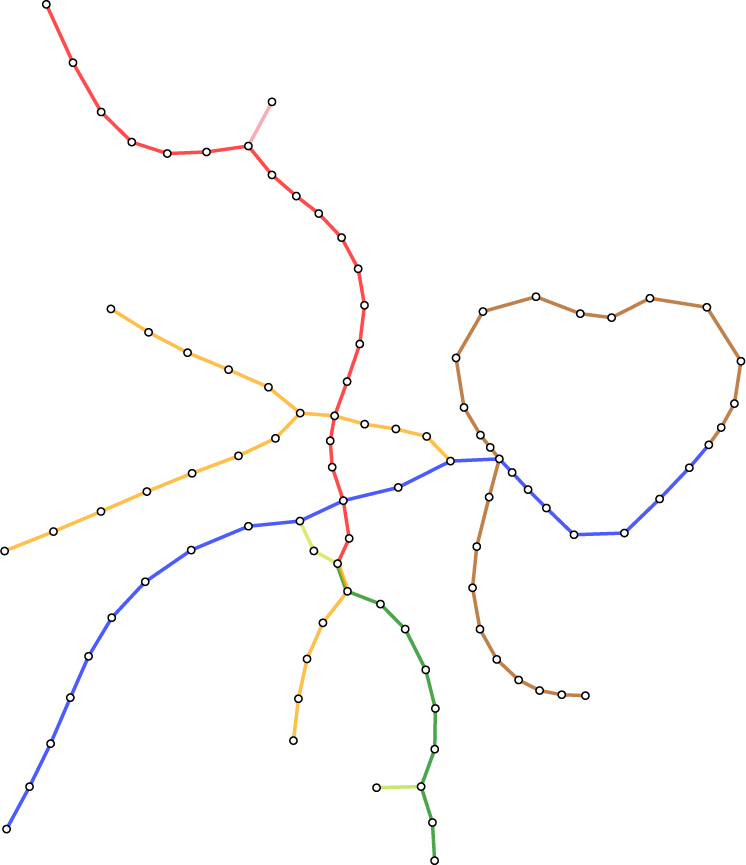} &
        \includegraphics[width=0.25\linewidth]{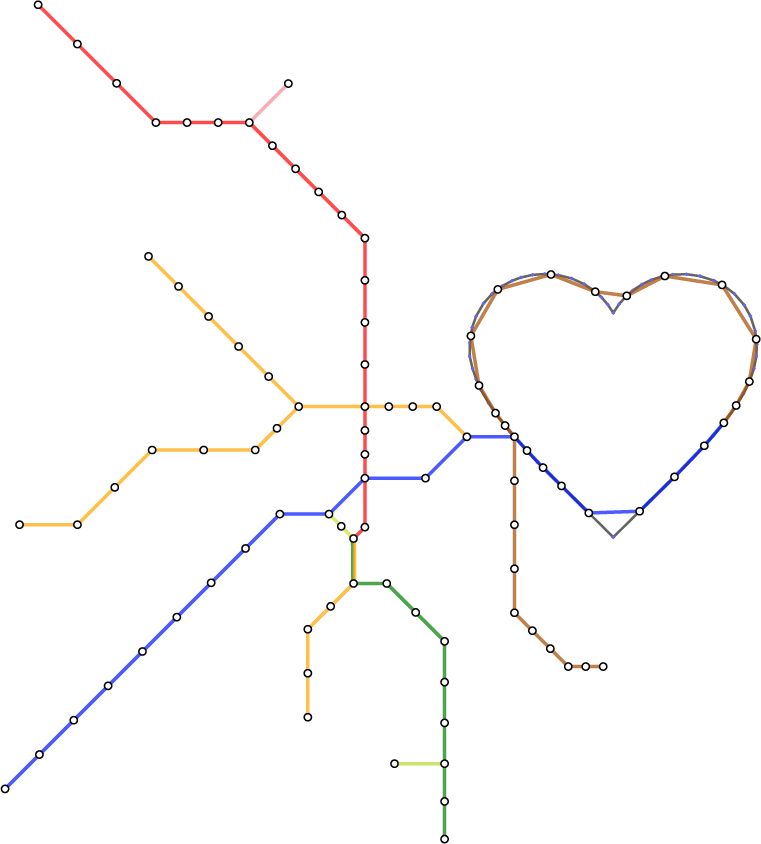} &
        \includegraphics[width=0.25\linewidth]{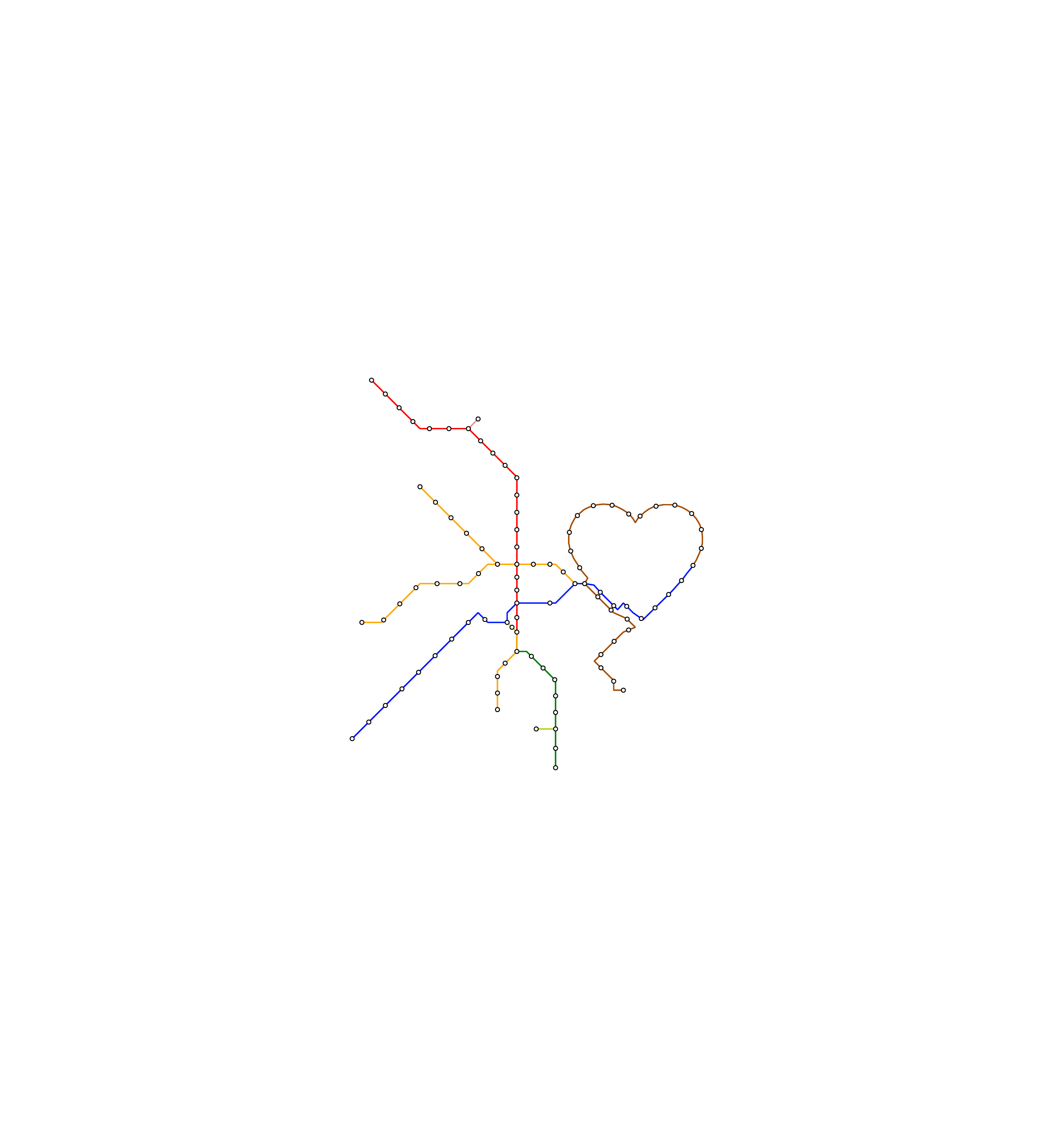} \\
        \multicolumn{3}{c}{\rv{(b) Test-case 1; Parameters for the smooth stage: $w_c = 10$, $w_l = 1$, $w_a = 2$, $w_p = 0.16$.}} \\ 
        \includegraphics[width=0.25\linewidth]{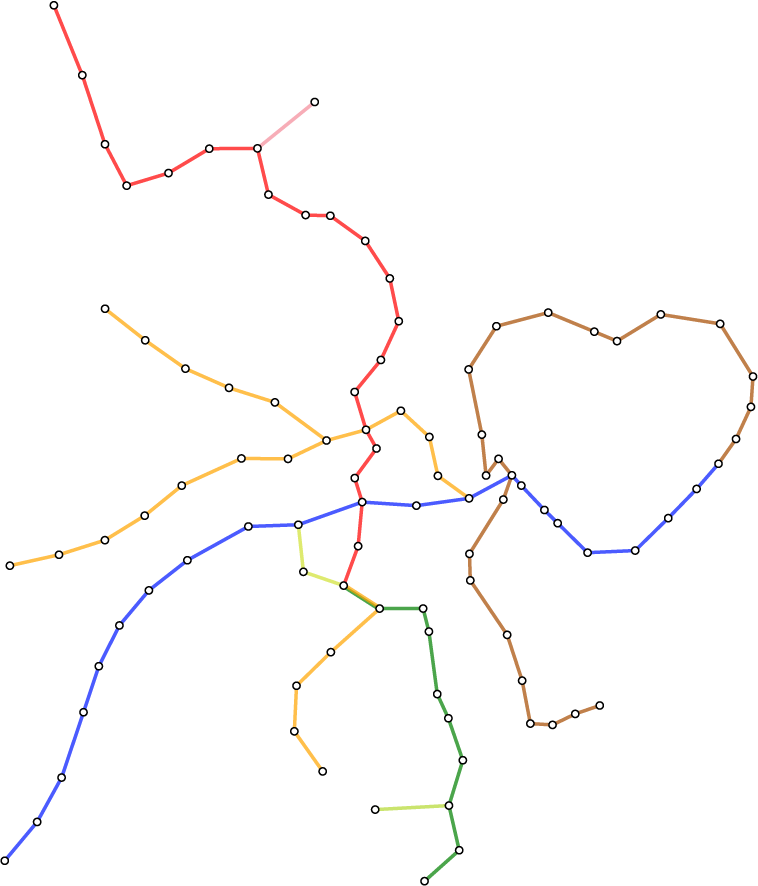} &
        \includegraphics[width=0.25\linewidth]{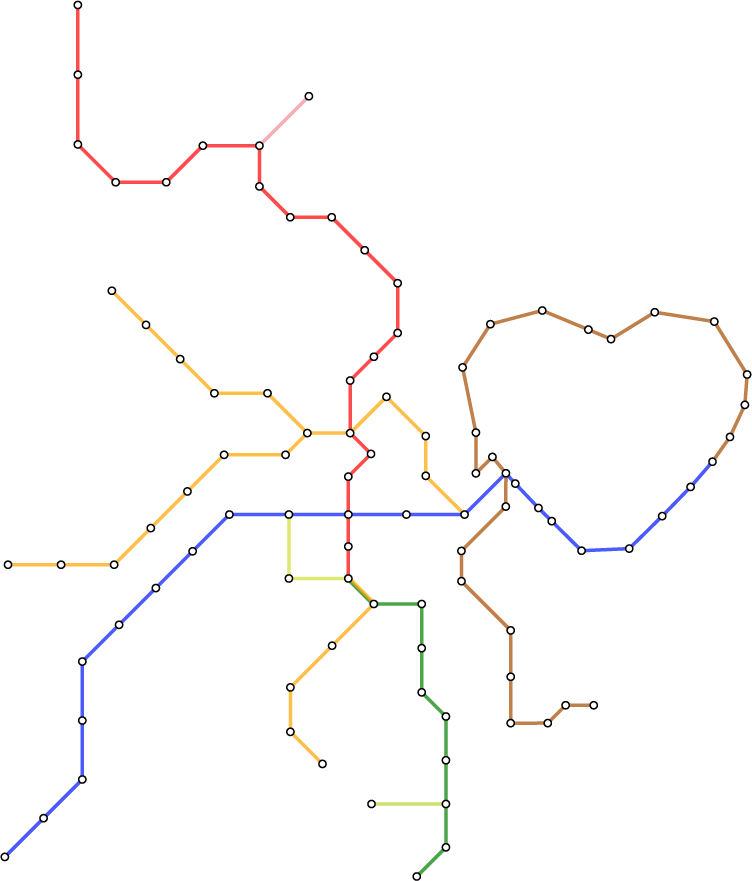} &
        \includegraphics[width=0.25\linewidth]{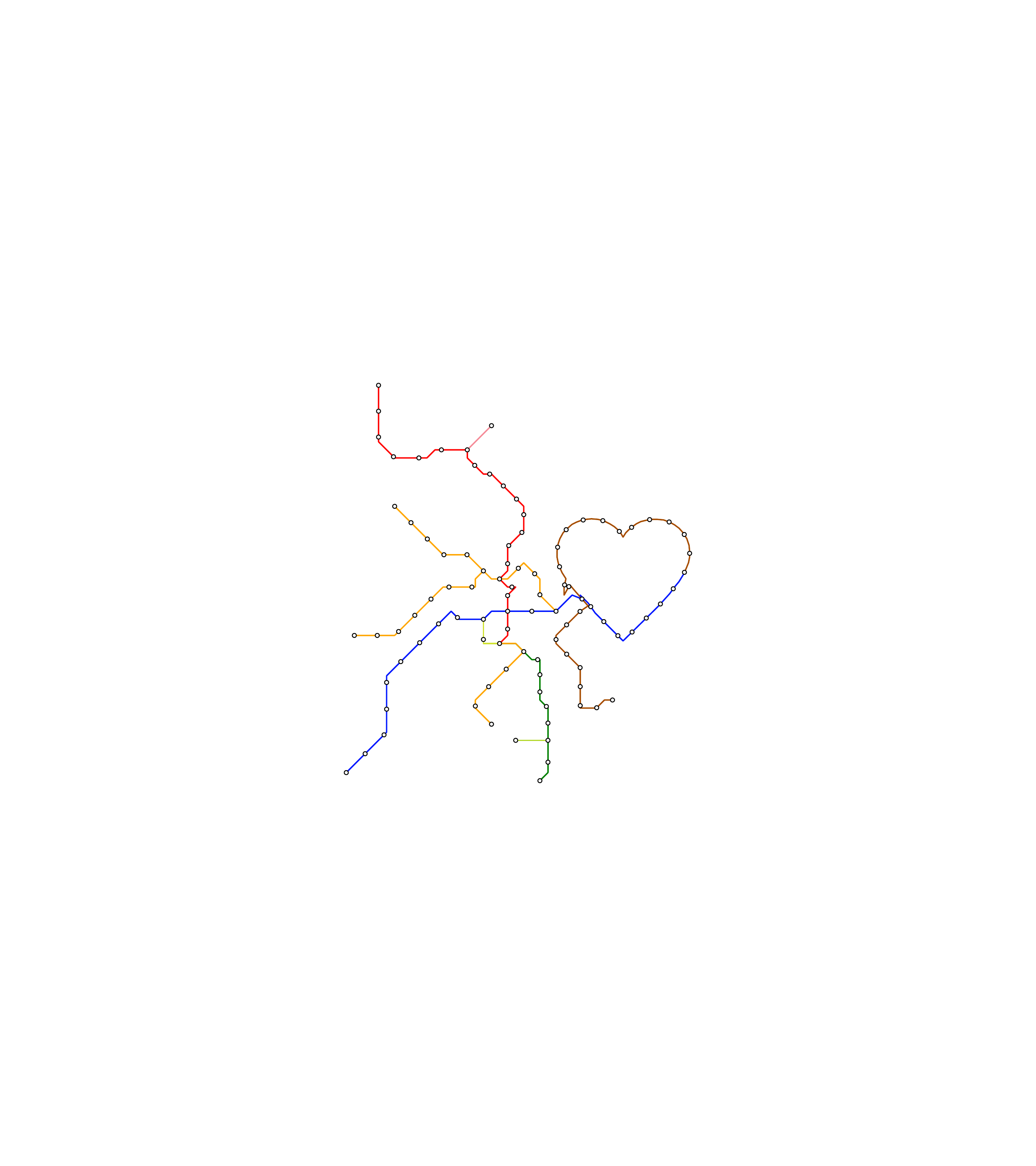} \\
        \multicolumn{3}{c}{\rv{(c) Test-case 2; Parameters for the smooth stage: $w_c = 4$, $w_l = 1$, $w_a = 0.5$, $w_p = 0.16$.}} \\ 
        \includegraphics[width=0.25\linewidth]{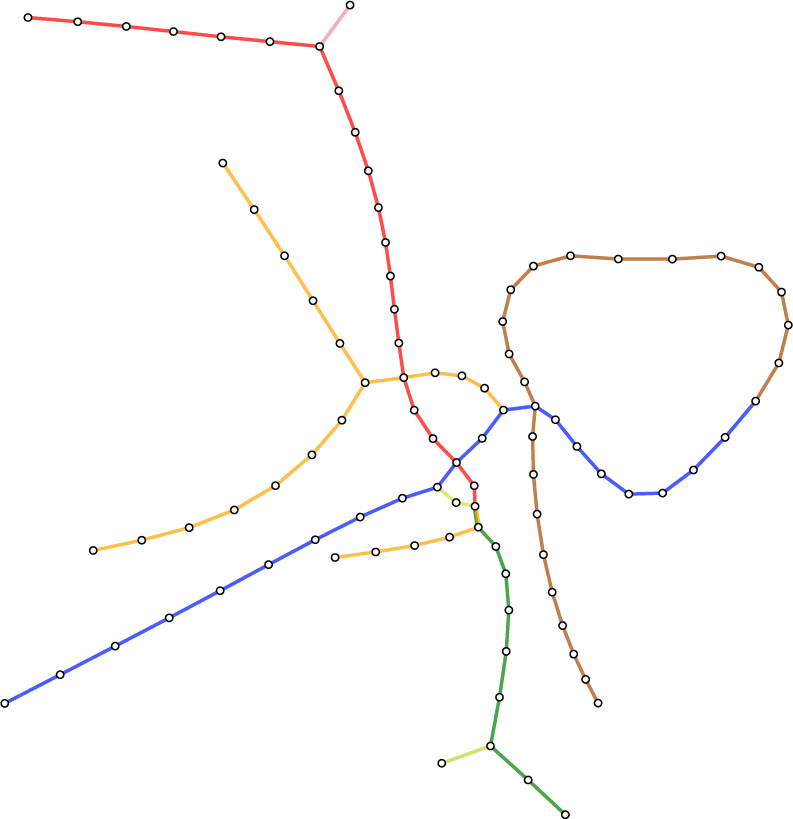} &
        \includegraphics[width=0.25\linewidth]{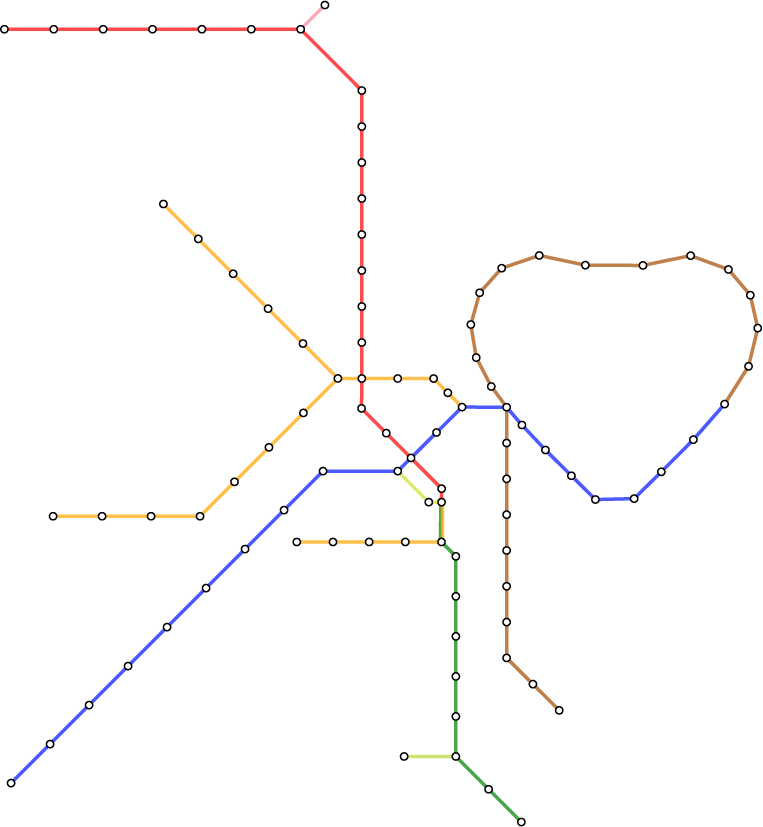} &
        \includegraphics[width=0.25\linewidth]{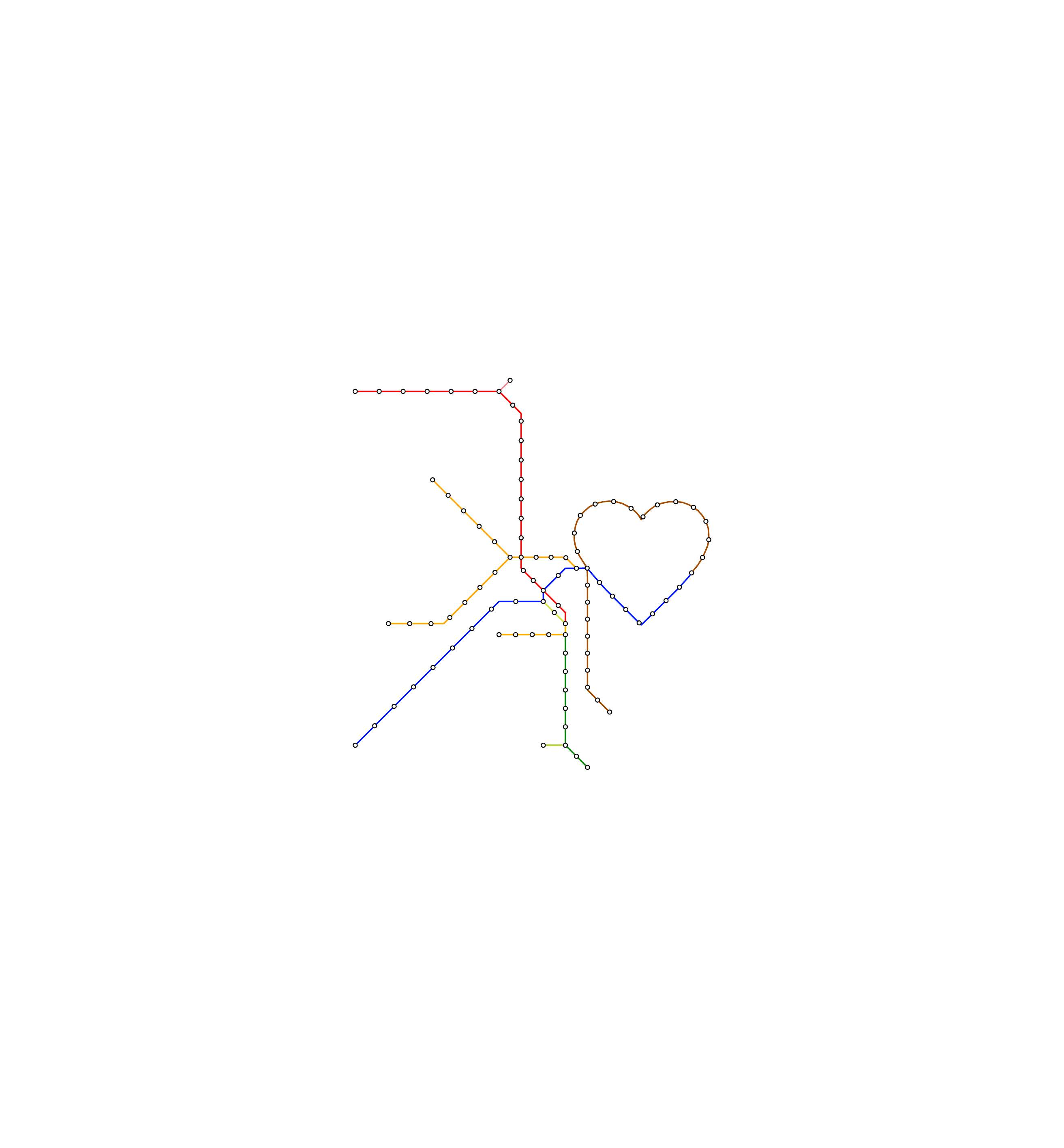} \\
        \multicolumn{3}{c}{\rv{(d) Test-case 3; Parameters for the smooth stage: $w_c = 4$, $w_l = 1$, $w_a = 5.0$, $w_p = 0.16$.}} \\
    \end{tabular}
    \caption{\rv{Taipei metro system computed with different weights for the smooth optimization process.
    The mixed stage was computed with the parameters $w_o= 2$, $w_p = 0.1$ and $w_c = 10$.
    Left: smooth layout, center: mixed layout and right: grid aligned layout.
    }}
    \label{fig:taipei-smooth}
    }
\end{figure*}

\begin{figure*}[t]
    \centering{
    \setlength{\tabcolsep}{0pt}
    \begin{tabular}{ccc}
        \includegraphics[width=0.25\linewidth]{figures/parameters/taipei/default/smooth/Metro.png} &
        \includegraphics[width=0.25\linewidth]{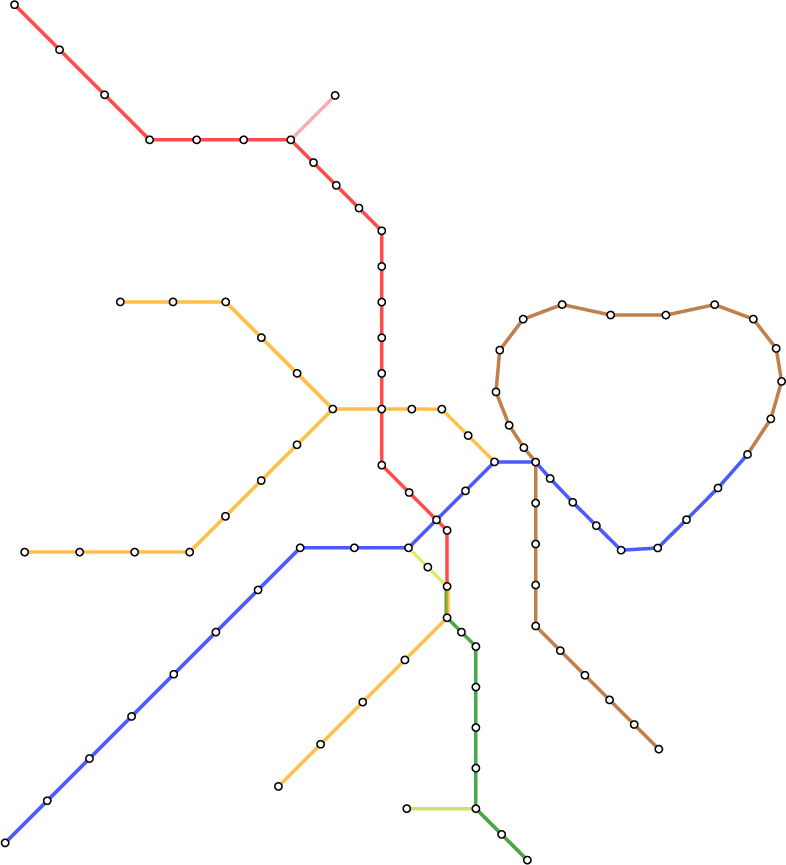} &
        \includegraphics[width=0.25\linewidth]{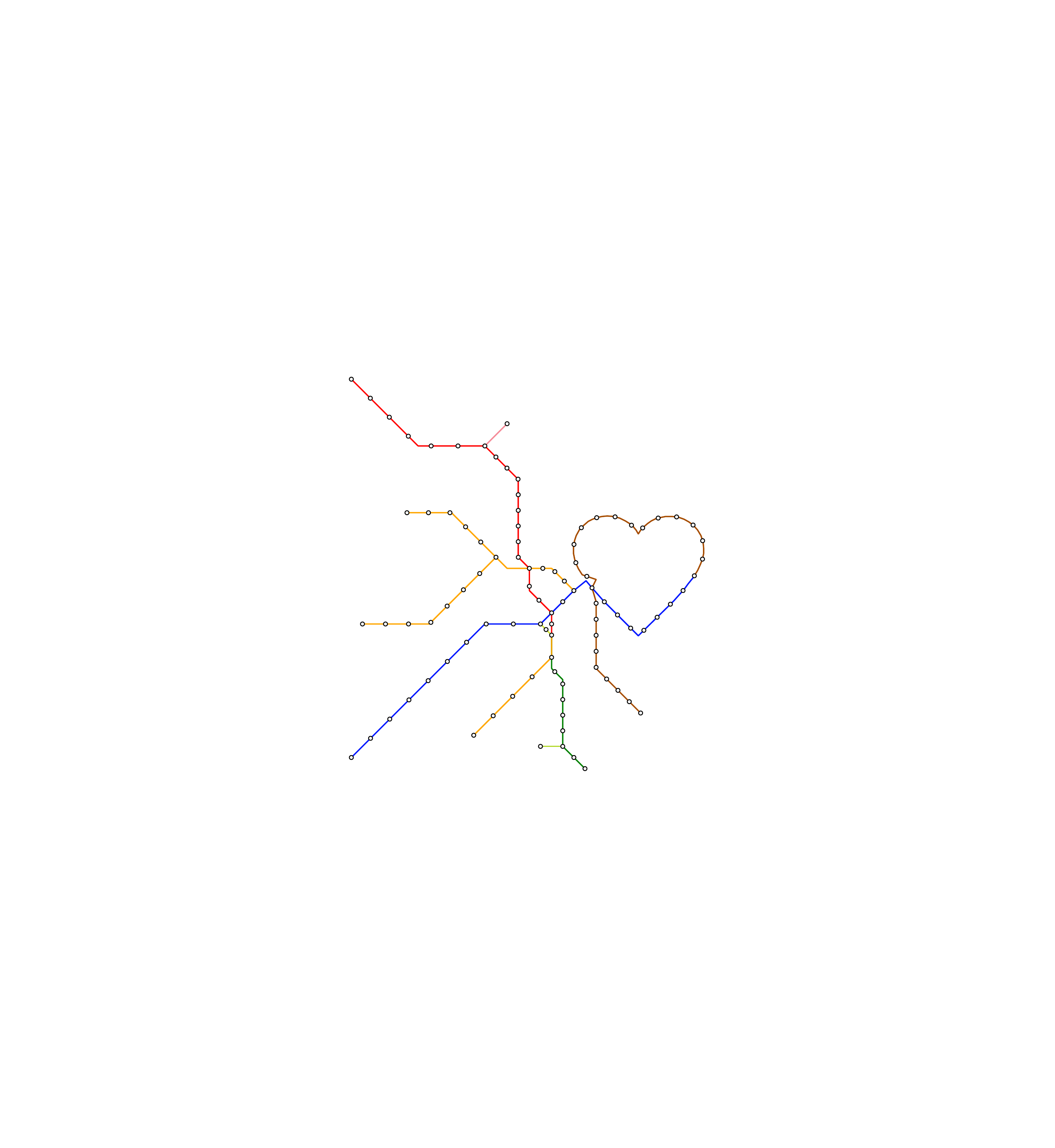} \\
        \multicolumn{3}{c}{\rv{(e) Test-case 4; Parameters for the mixed stage: $w_o = 5$, $w_p = 0.1$, $w_c = 5$.}} \\
        \includegraphics[width=0.25\linewidth]{figures/parameters/taipei/default/smooth/Metro.png} &
        \includegraphics[width=0.25\linewidth]{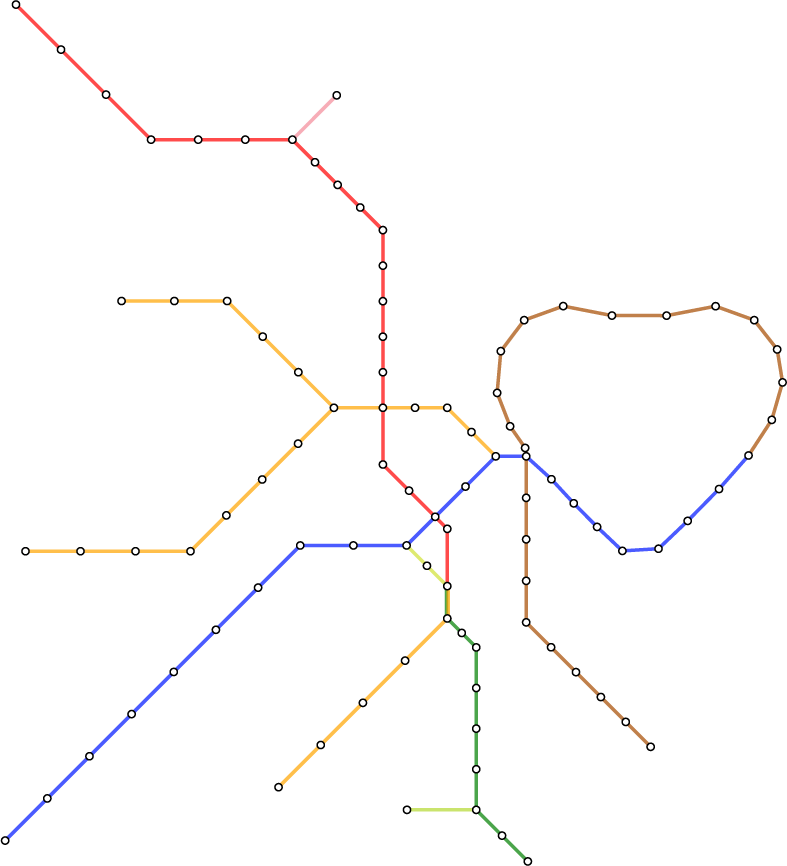} &
        \includegraphics[width=0.25\linewidth]{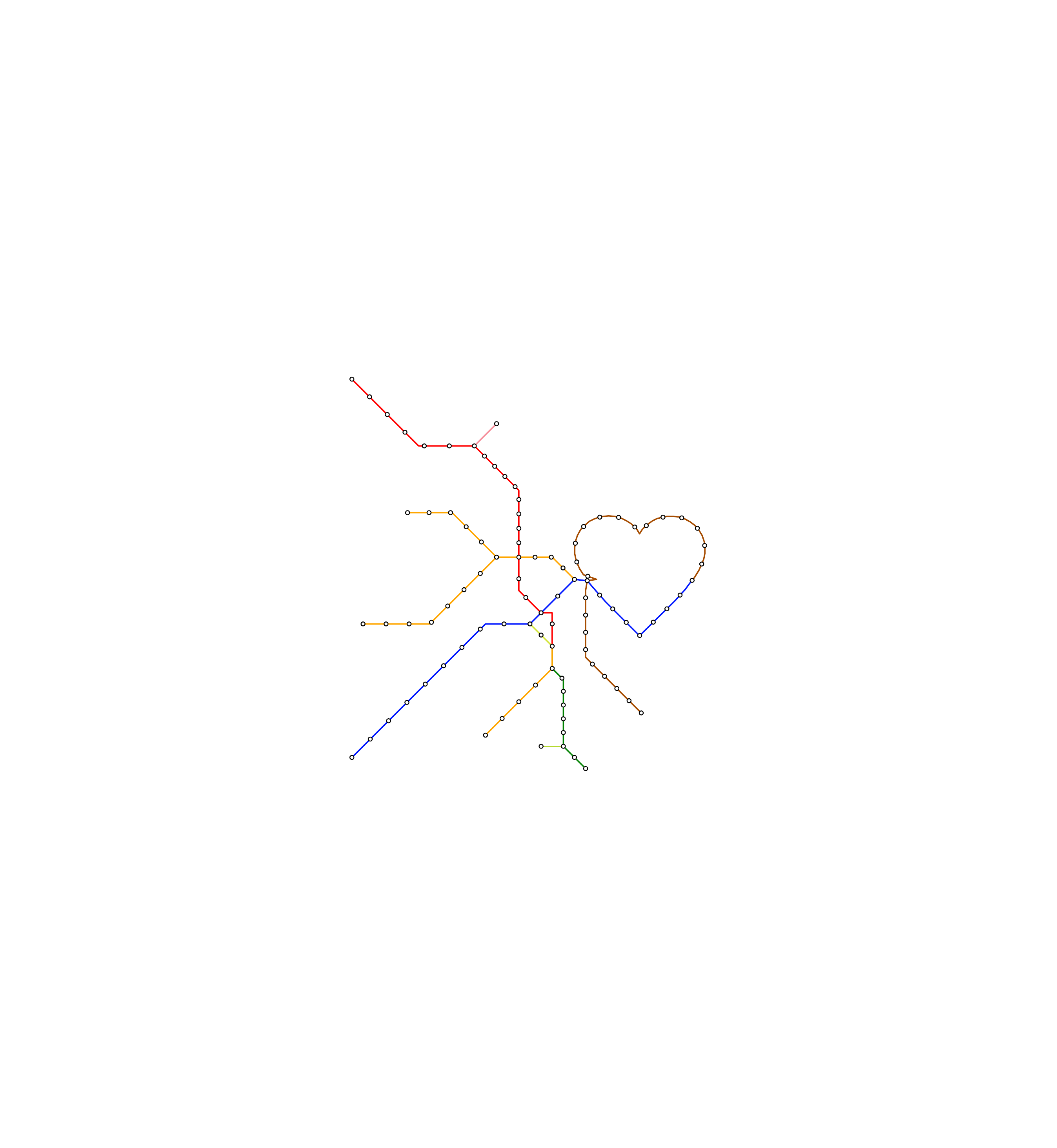} \\
        \multicolumn{3}{c}{\rv{(f) Test-case 5; Parameters for the mixed stage: $w_o = 10$, $w_p = 0.1$, $w_c = 2$.}} \\
        \includegraphics[width=0.25\linewidth]{figures/parameters/taipei/default/smooth/Metro.png} &
        \includegraphics[width=0.25\linewidth]{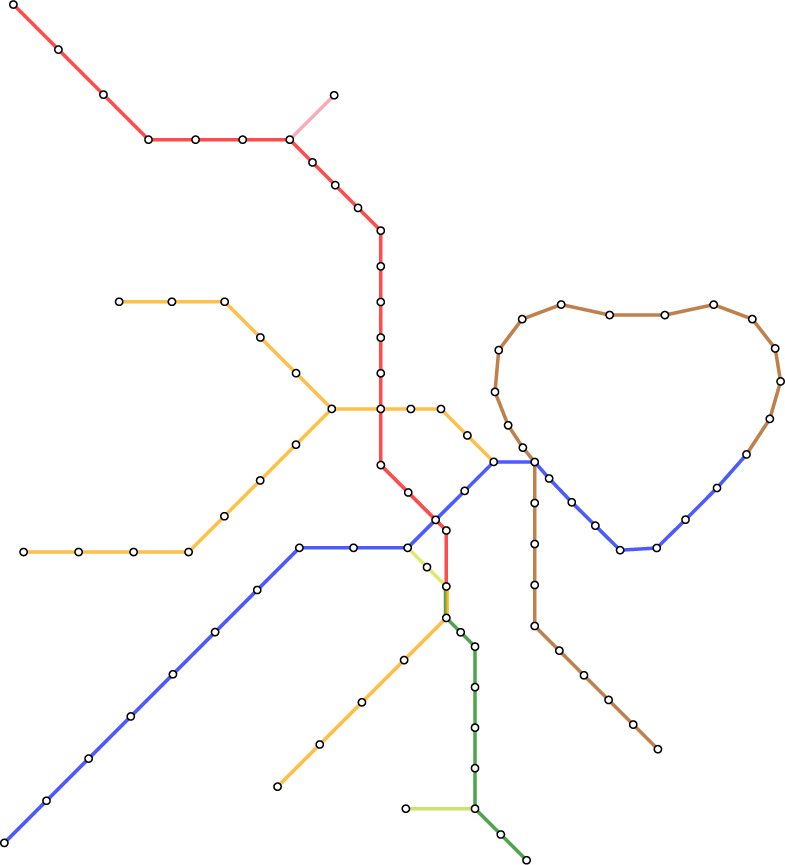} &
        \includegraphics[width=0.25\linewidth]{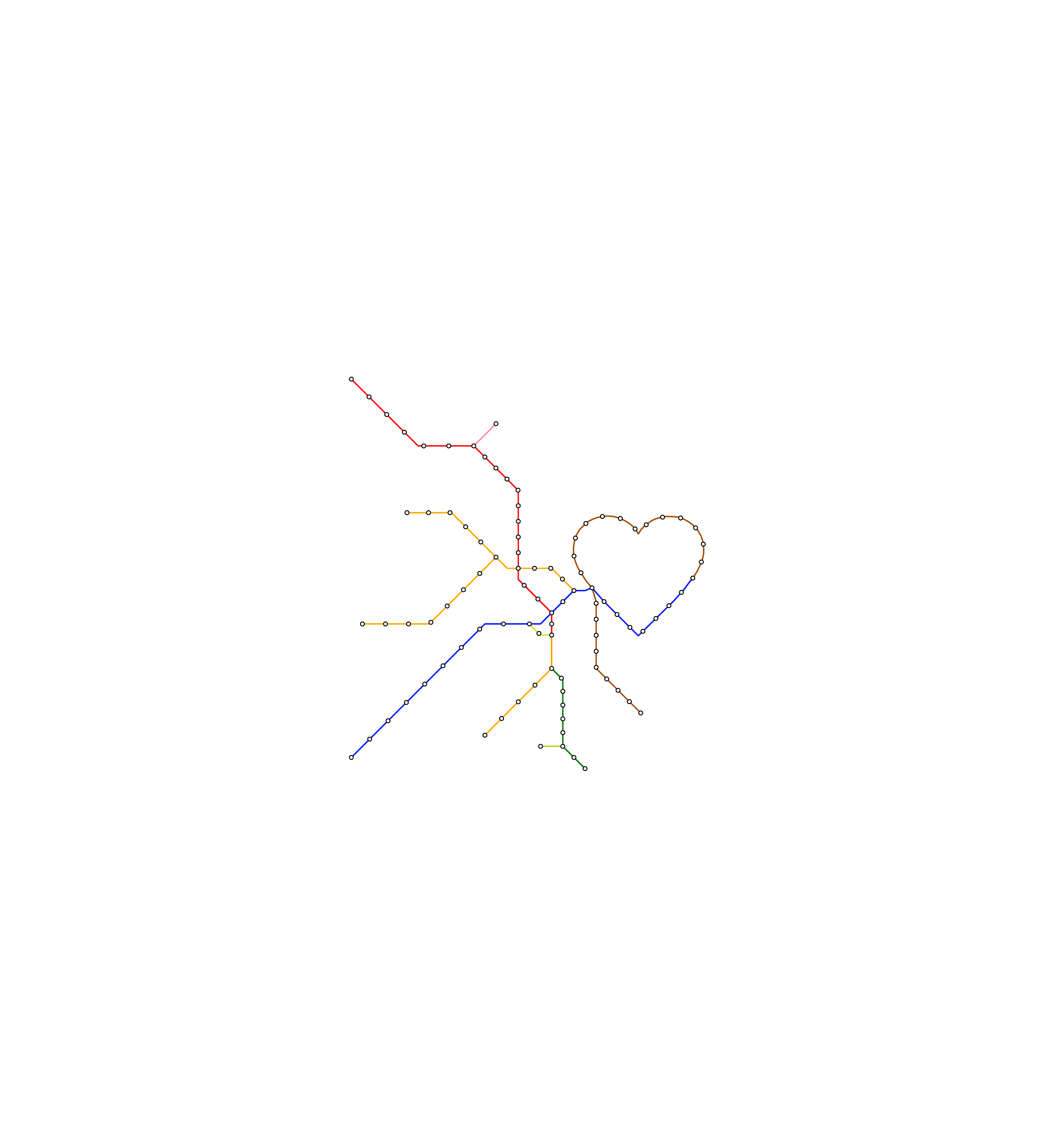} \\
        \multicolumn{3}{c}{\rv{(g) Test-case 6; Parameters for the mixed stage: $w_o = 10$, $w_p = 0.1$, $w_c = 10$.}} \\
        \includegraphics[width=0.25\linewidth]{figures/parameters/taipei/default/smooth/Metro.png} &
        \includegraphics[width=0.25\linewidth]{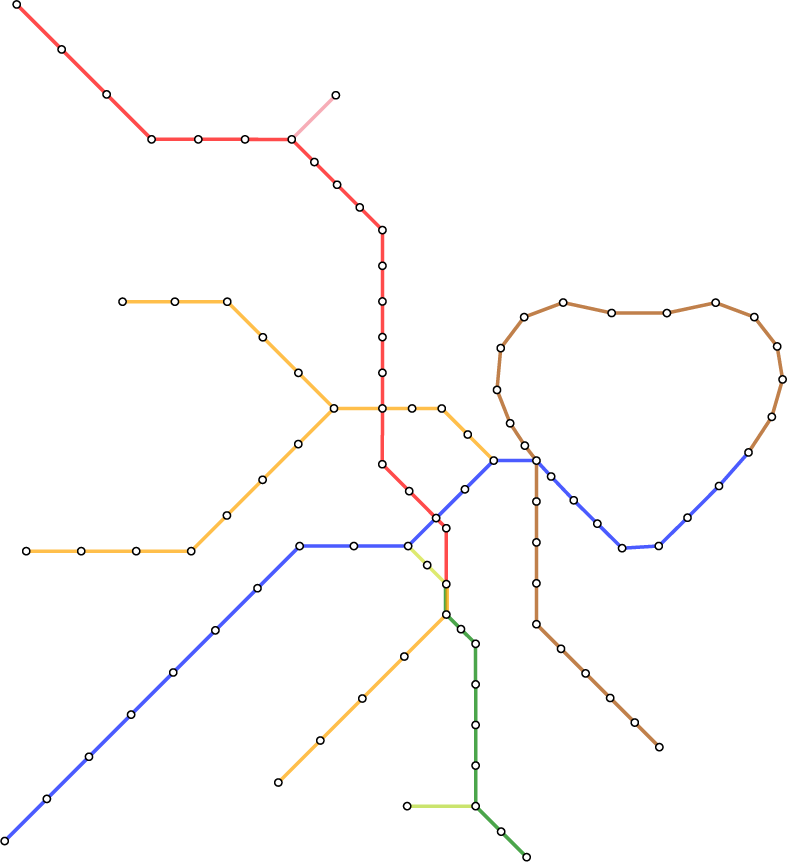} &
        \includegraphics[width=0.25\linewidth]{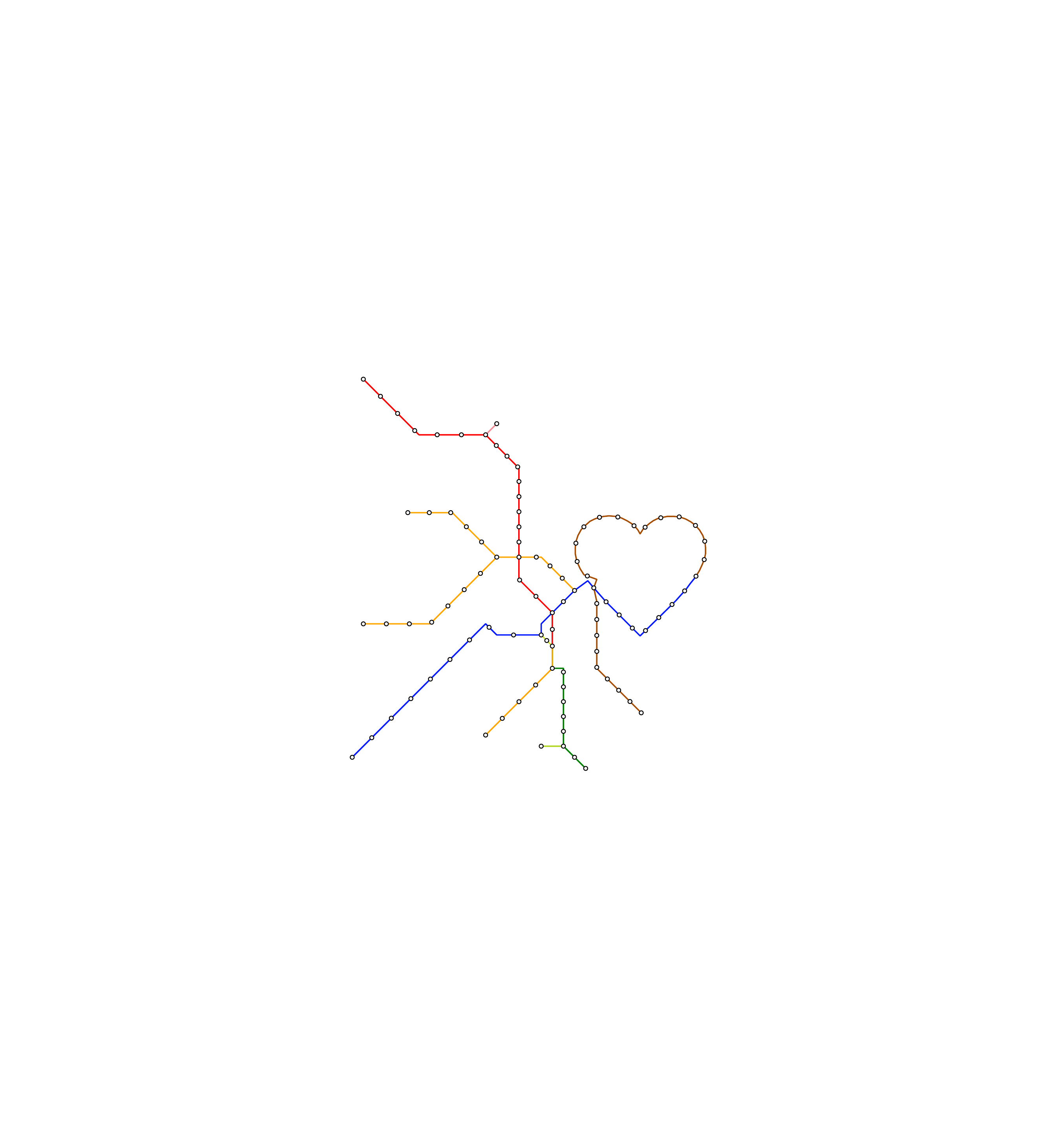} \\
        \multicolumn{3}{c}{\rv{(h) Test-case 7; Parameters for the mixed stage: $w_o = 2$, $w_p = 0.1$, $w_c = 2$.}} \\
    \end{tabular}
    \caption{\rv{Taipei metro system computed with different weights for the mixed optimization process. The smooth stage was computed with the parameters $w_c = 10$, $w_l = 1$, $w_a = 2$ and $w_p = 0.16$.
    Left: smooth layout, center: mixed layout and right: grid aligned layout.
    }}
    \label{fig:taipei-mixed}
    }
\end{figure*}

\begin{figure*}[t]
    \centering{
    \setlength{\tabcolsep}{0pt}
    \begin{tabular}{ccc}
        \includegraphics[width=0.33\linewidth]{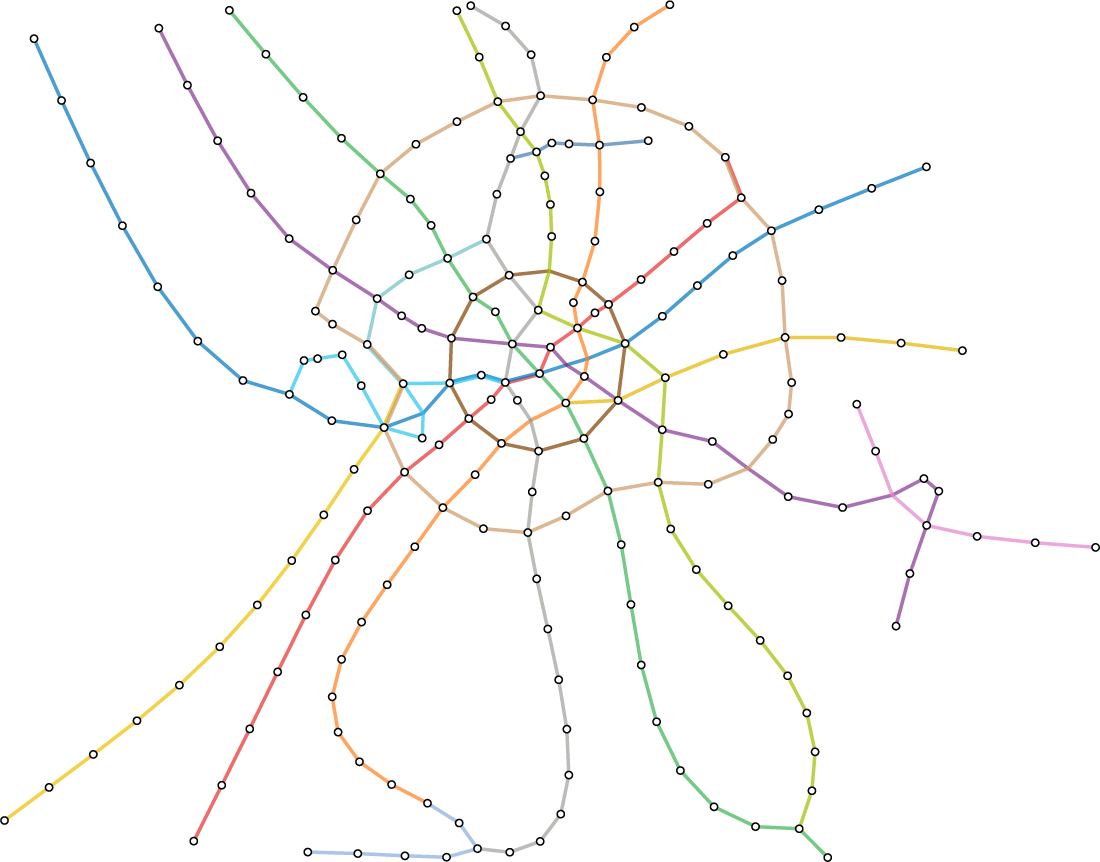} &
        \includegraphics[width=0.33\linewidth]{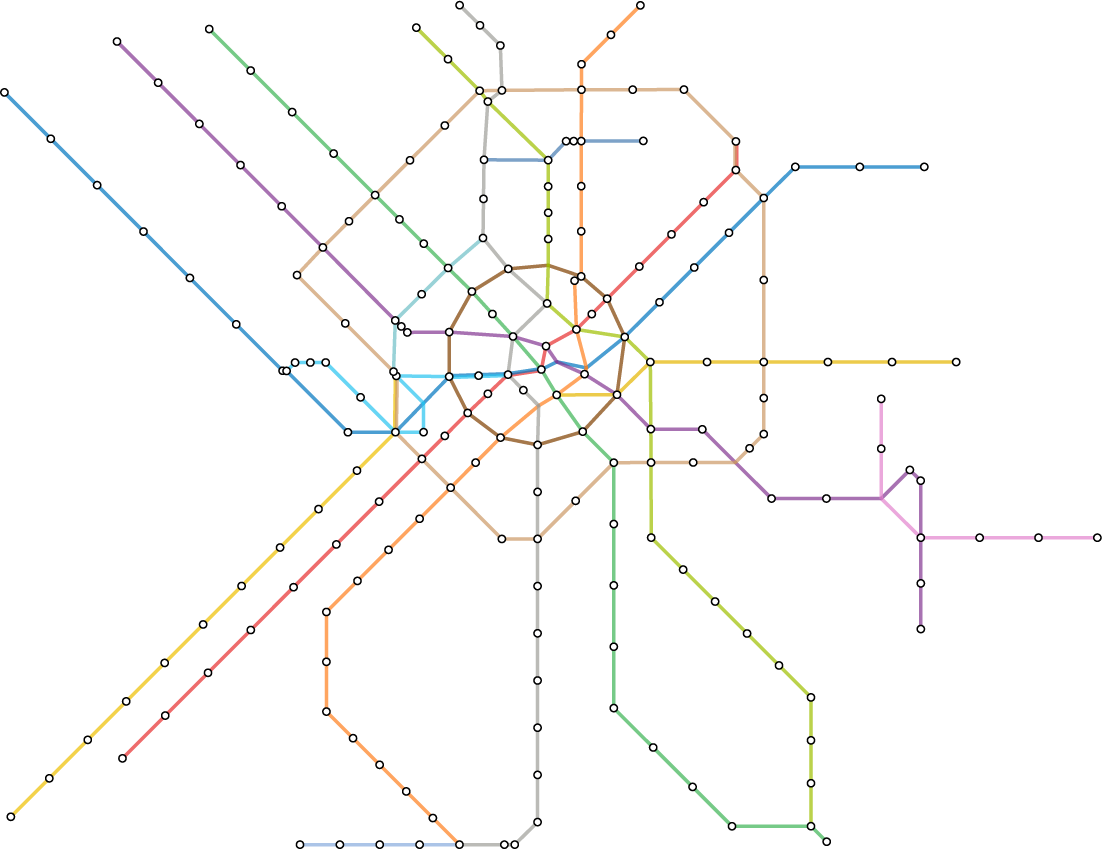} &
        \includegraphics[width=0.33\linewidth]{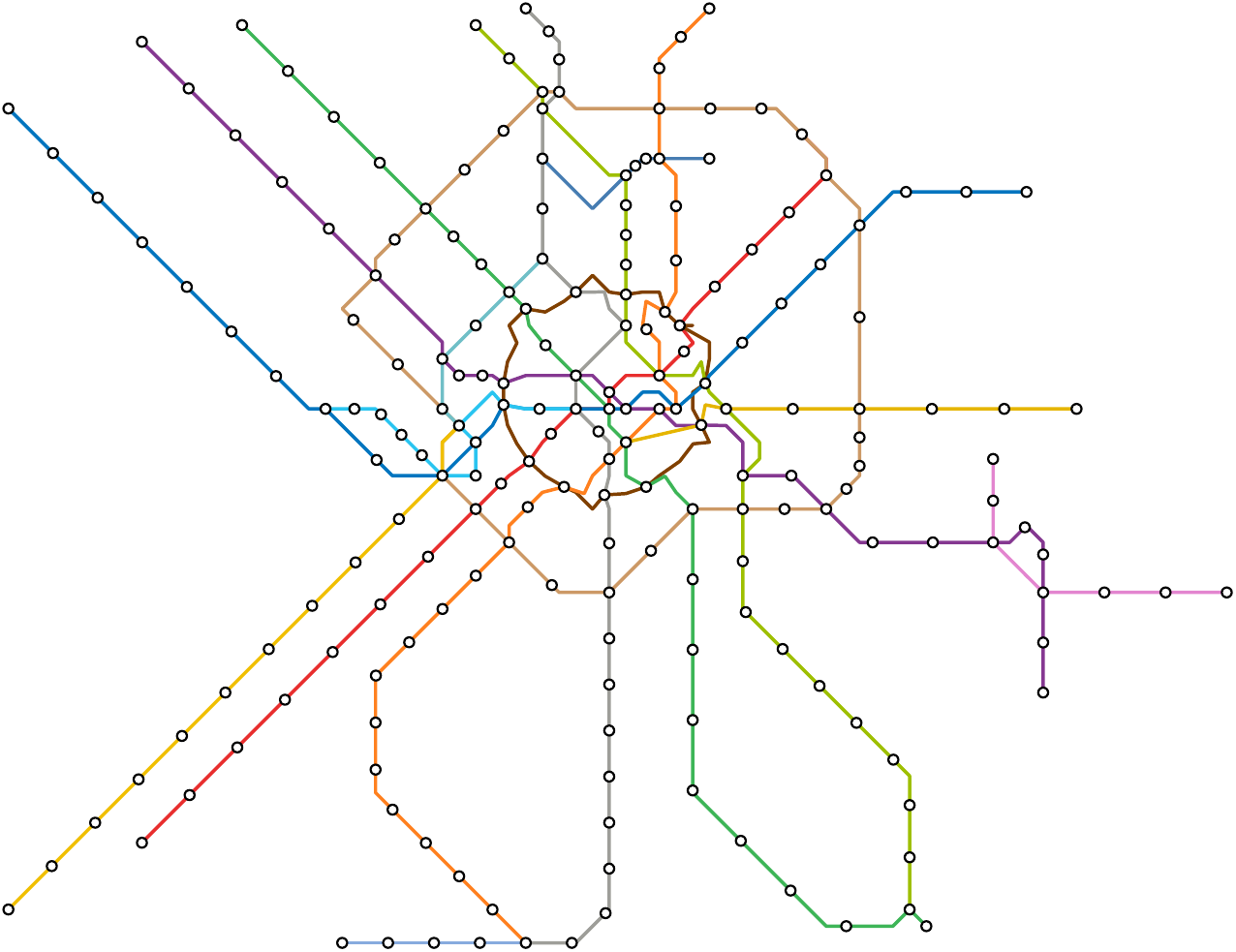} \\
         \multicolumn{3}{c}{\rv{(a) Default Test-case; Parameters for the smooth stage: $w_c = 4$, $w_l = 1$, $w_a = 2$, $w_p = 0.16$.}} \\
        \includegraphics[width=0.33\linewidth]{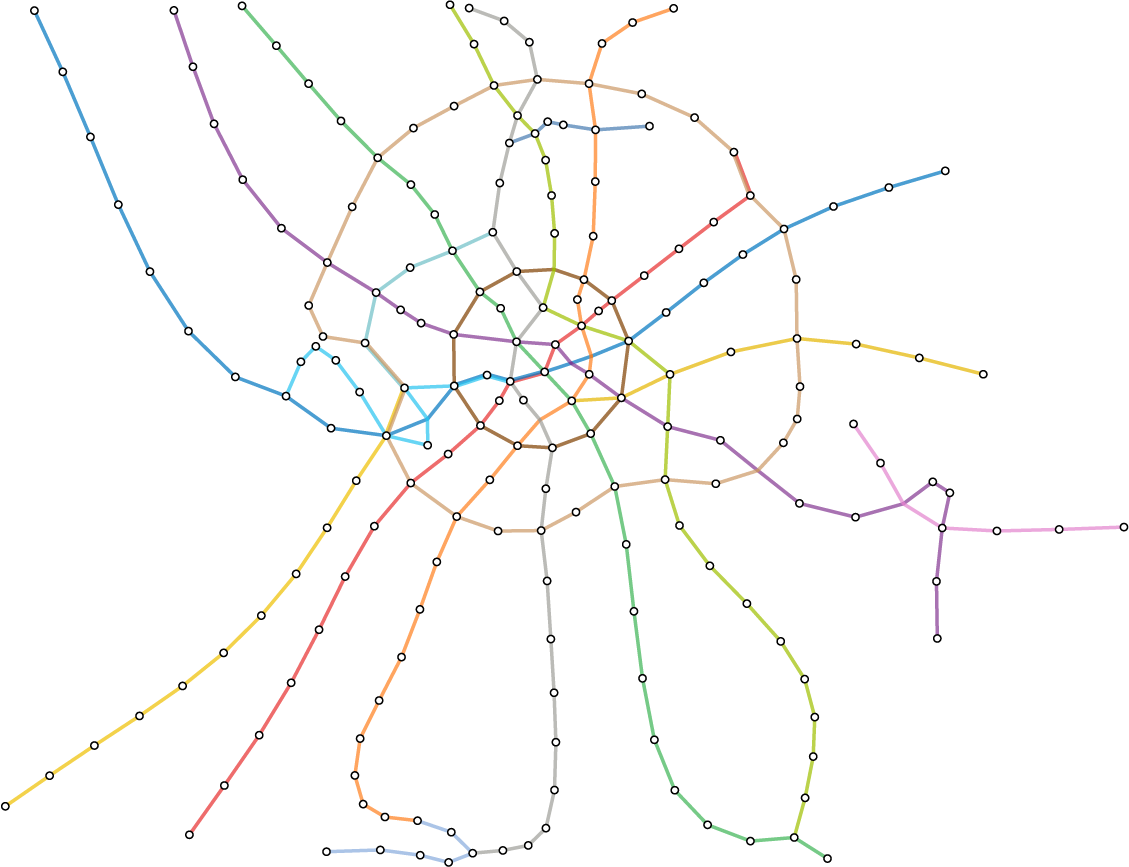} &
        \includegraphics[width=0.33\linewidth]{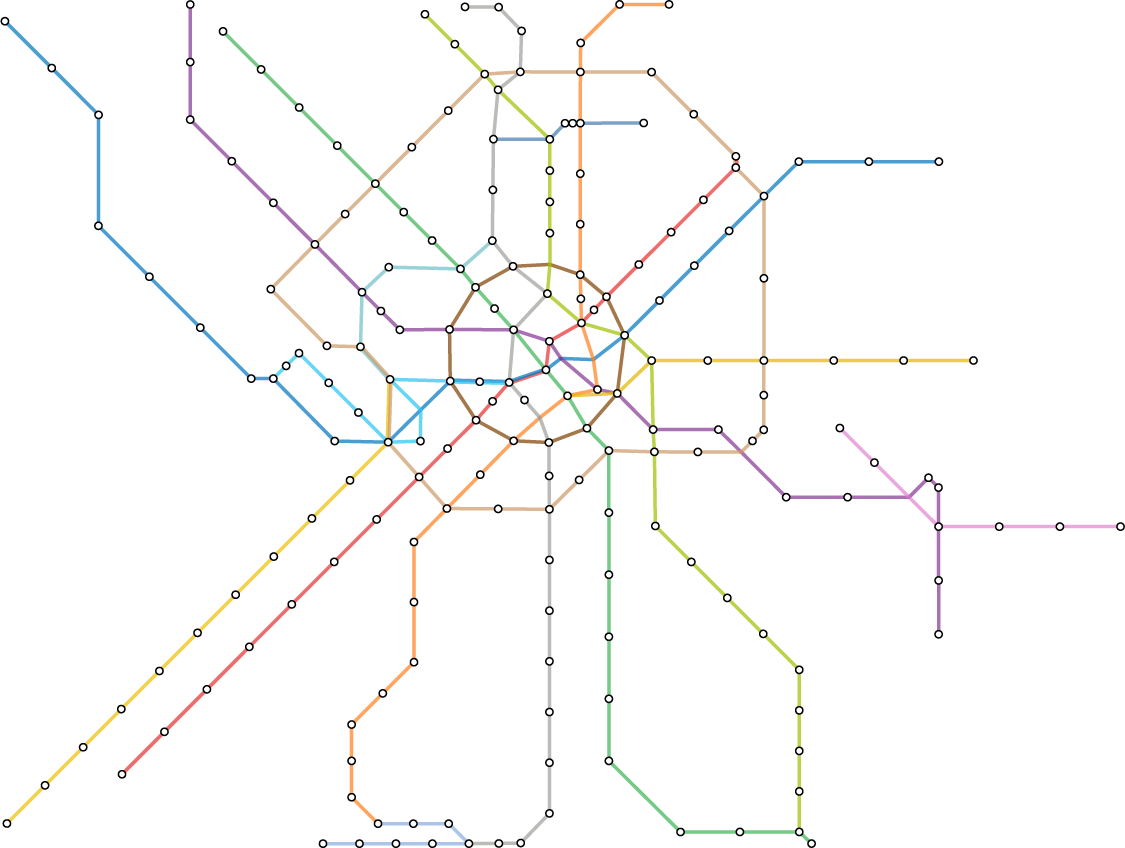} &
        \includegraphics[width=0.33\linewidth]{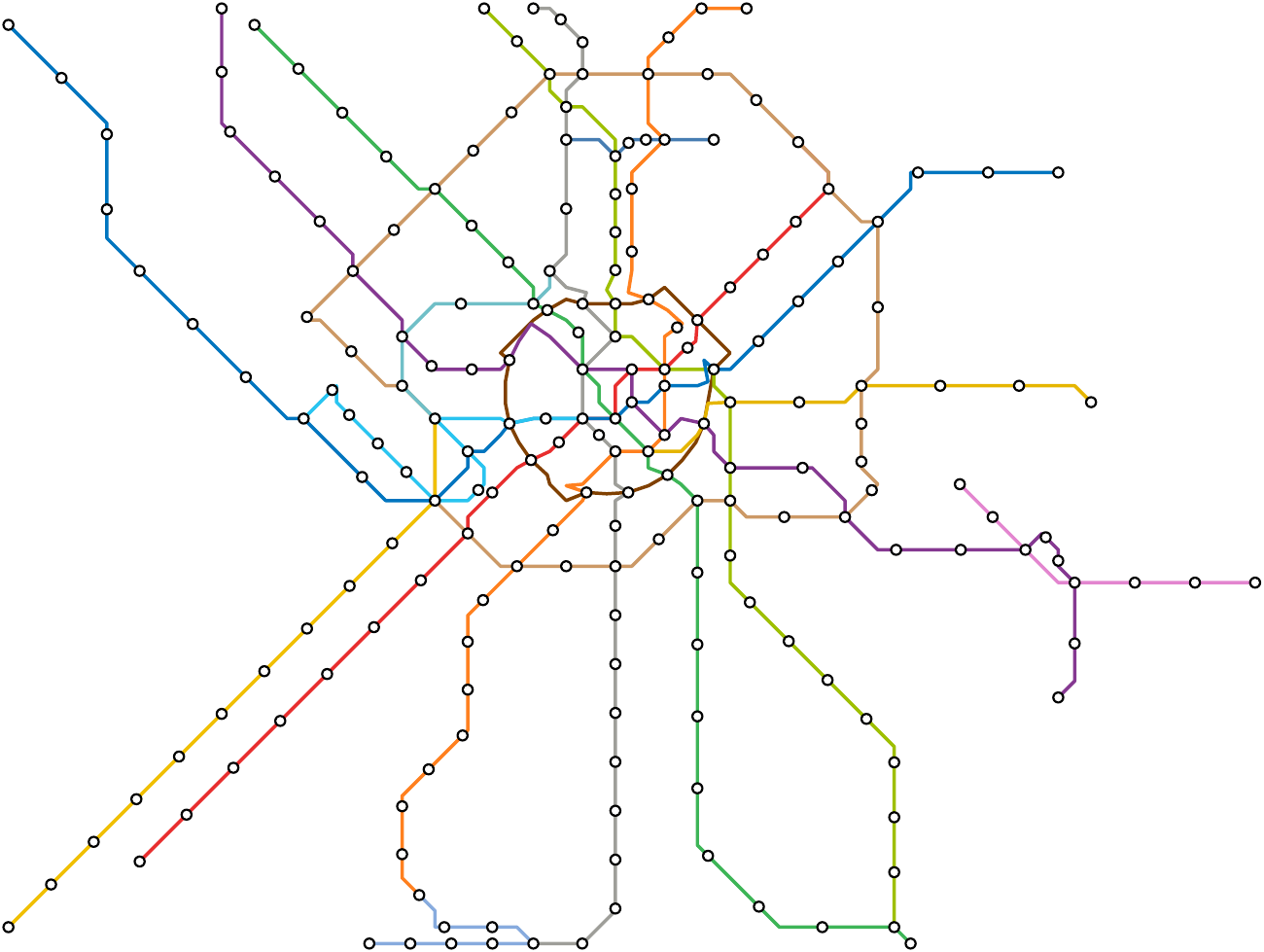} \\
       \multicolumn{3}{c}{\rv{(b) Test-case 1; Parameters for the smooth stage: $w_c = 10$, $w_l = 1$, $w_a = 2$, $w_p = 0.16$.}} \\ 
        \includegraphics[width=0.33\linewidth]{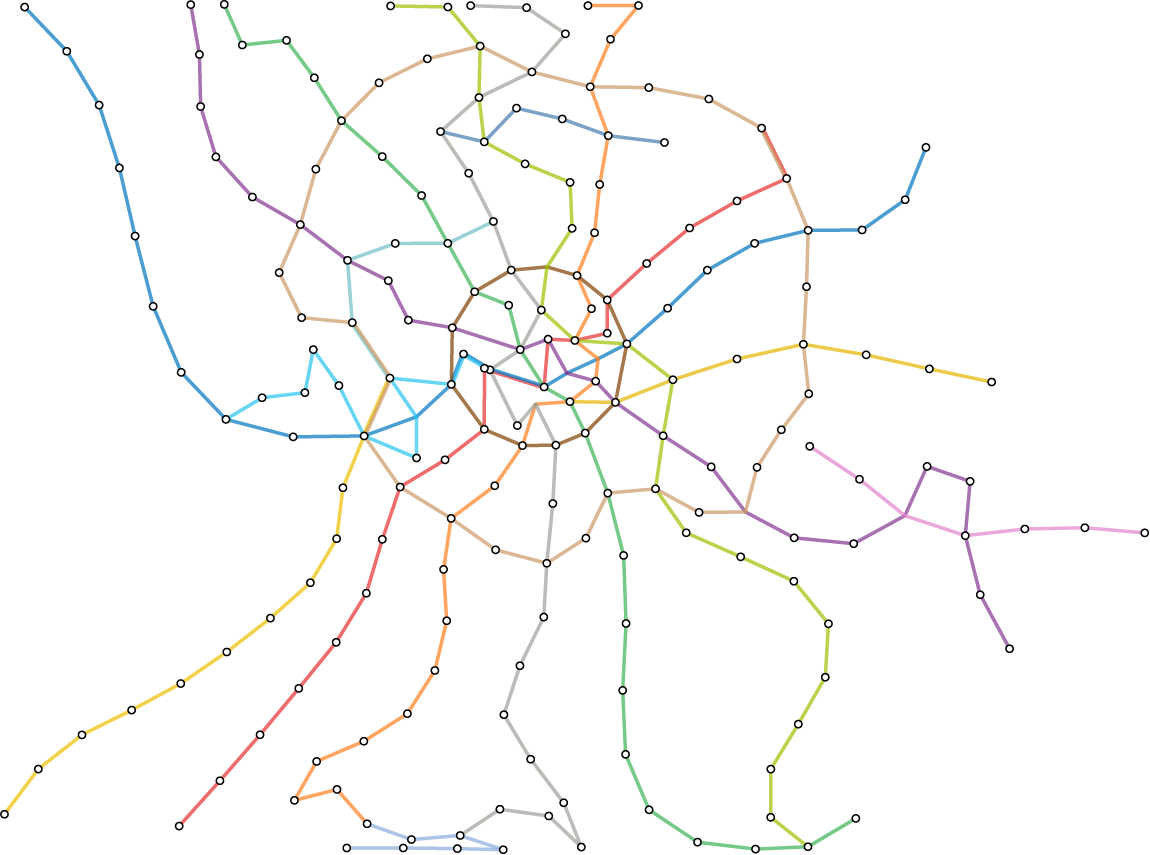} &
        \includegraphics[width=0.33\linewidth]{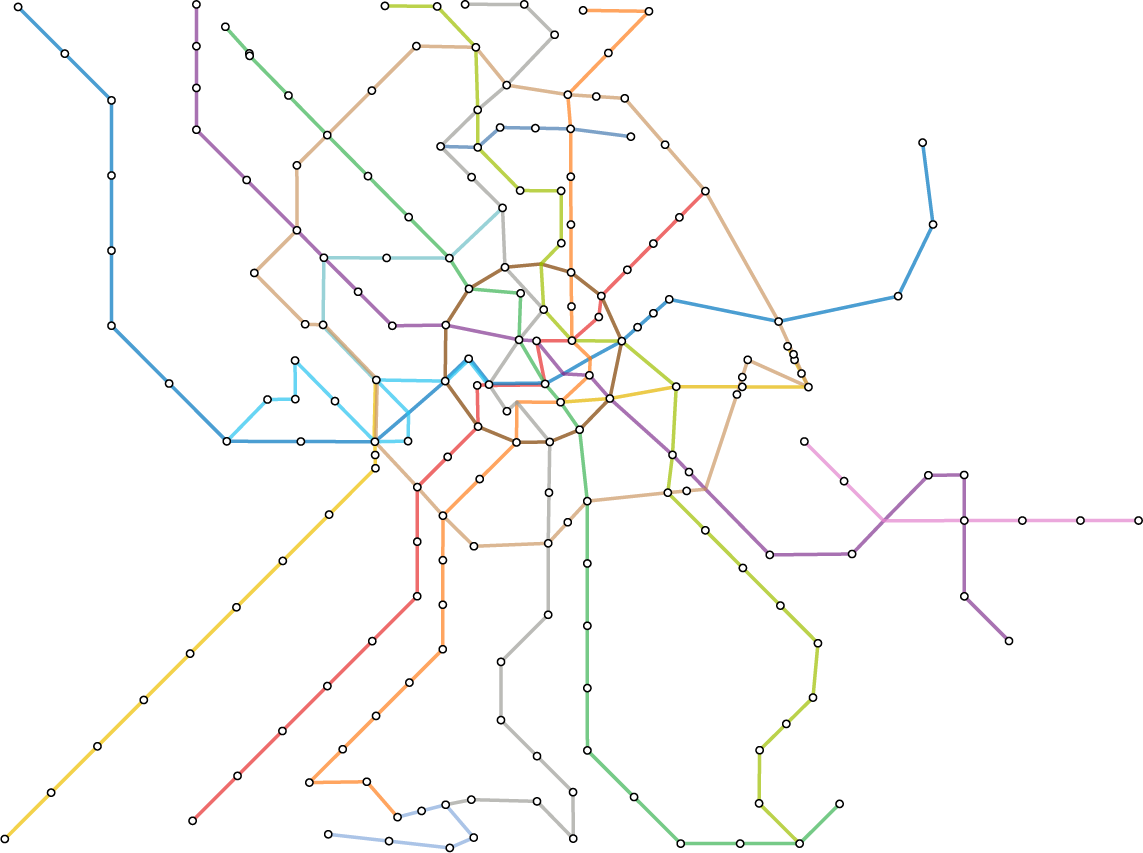} &
        \includegraphics[width=0.33\linewidth]{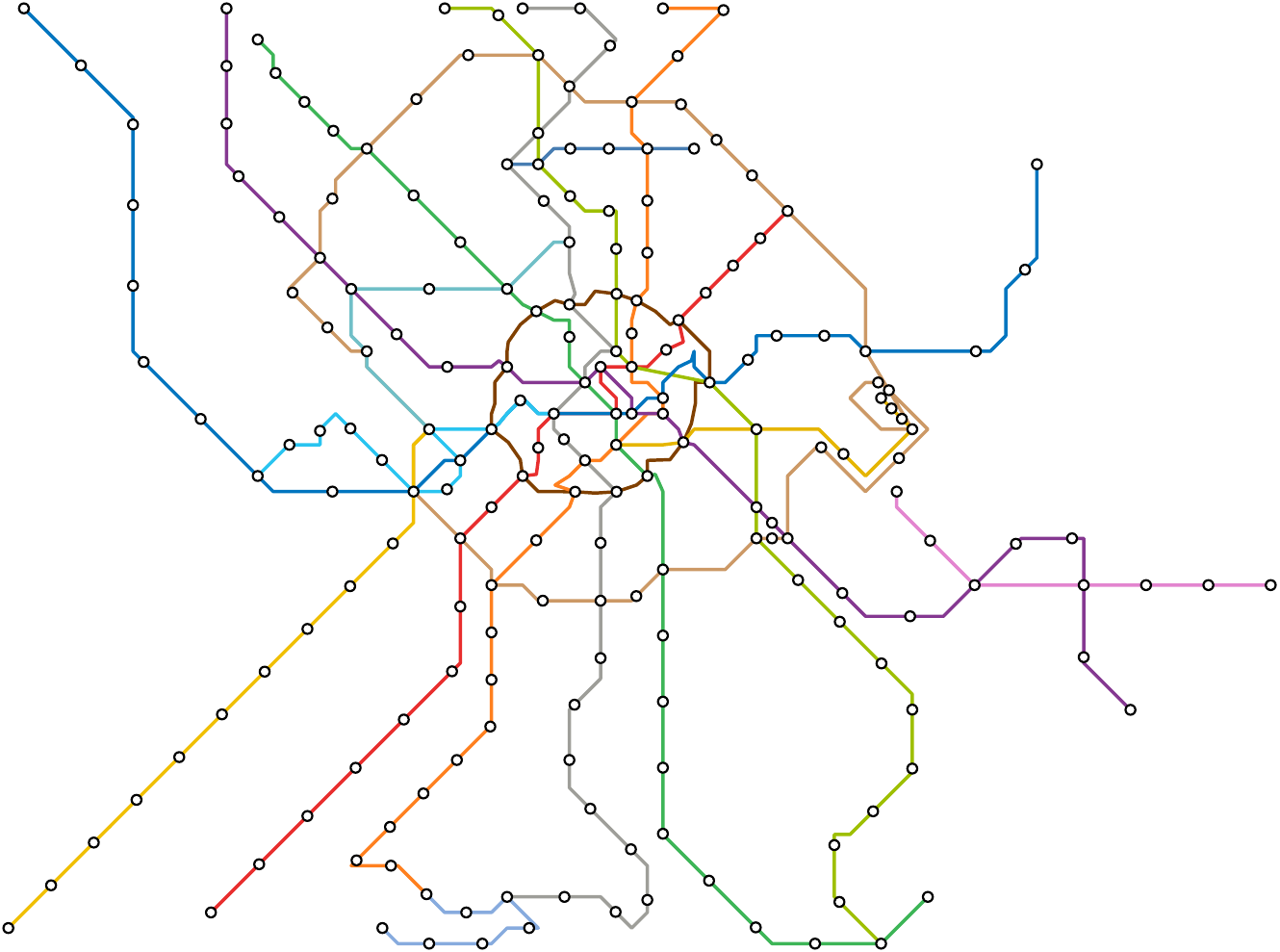} \\
        \multicolumn{3}{c}{\rv{(c) Test-case 2; Parameters for the smooth stage: $w_c = 4$, $w_l = 1$, $w_a = 0.5$, $w_p = 0.16$.}} \\
        \includegraphics[width=0.33\linewidth]{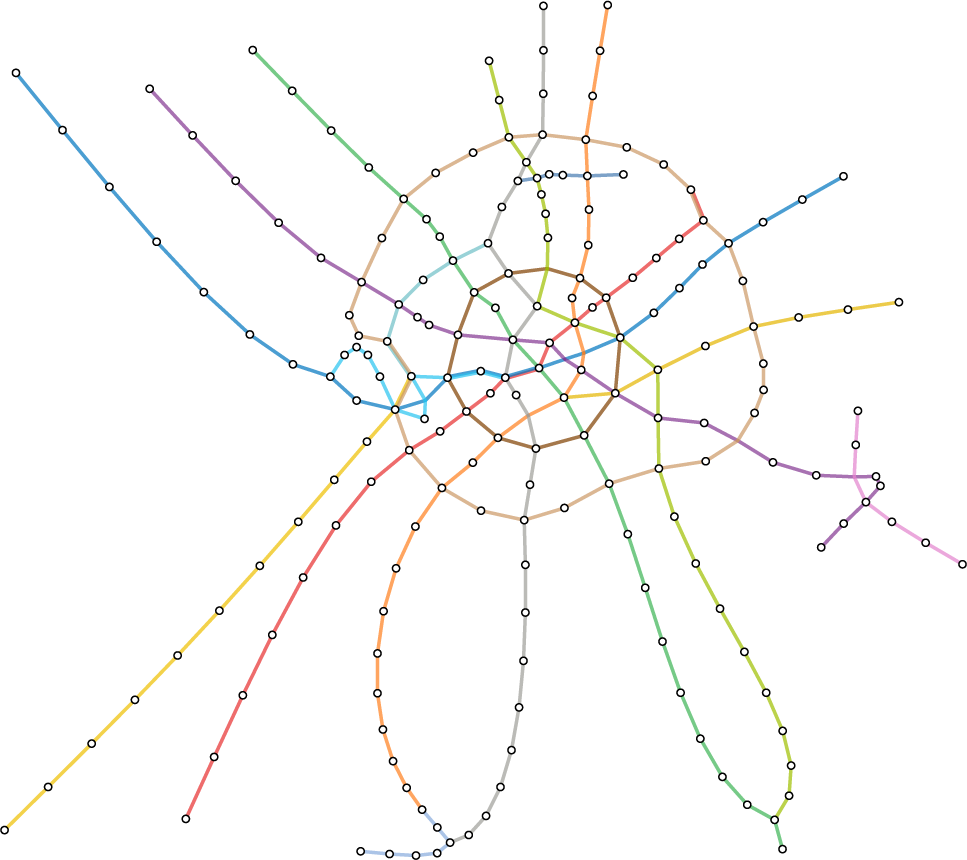} &
        \includegraphics[width=0.33\linewidth]{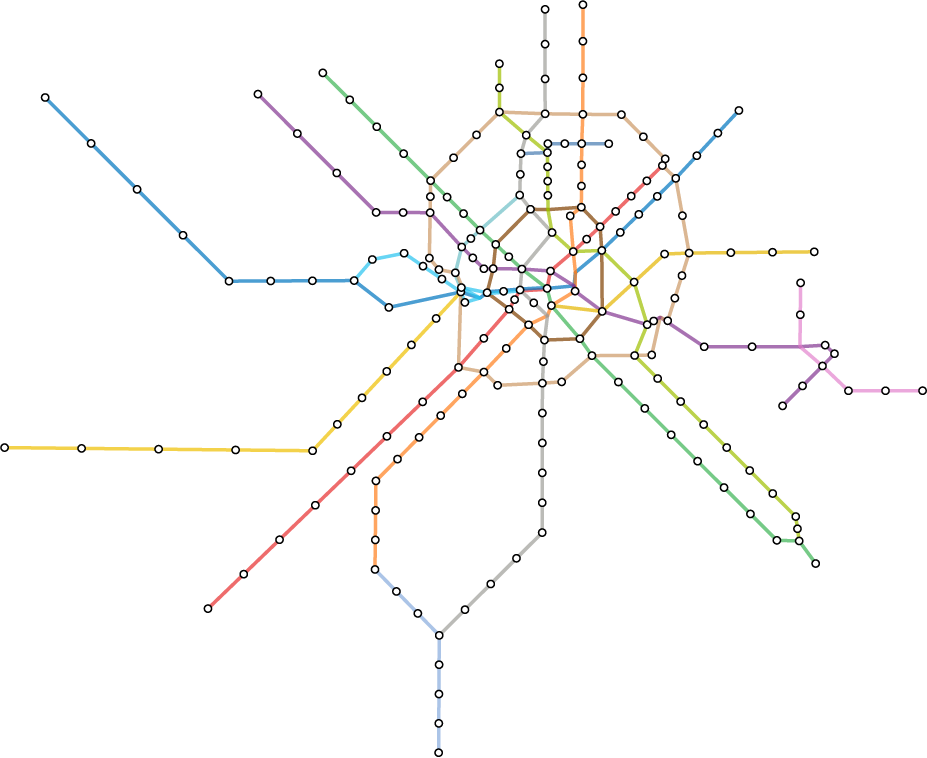} &
        \includegraphics[width=0.33\linewidth]{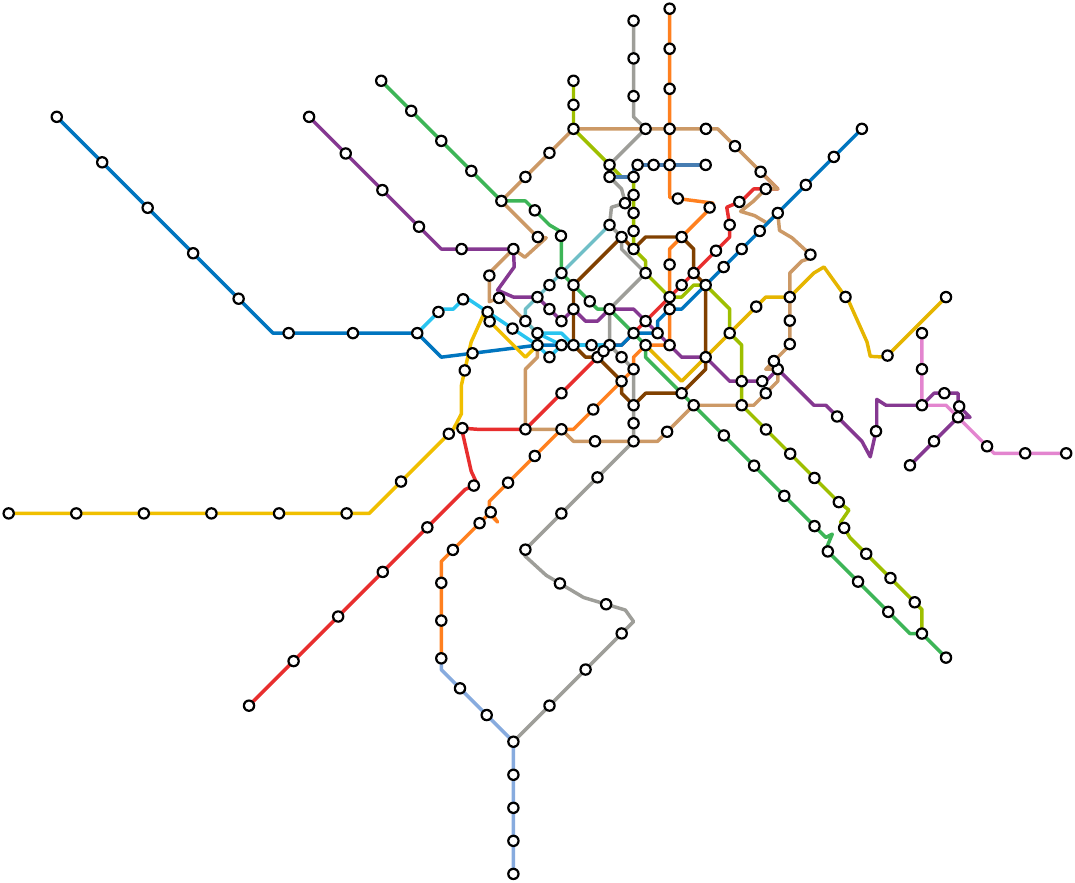} \\
        \multicolumn{3}{c}{\rv{(d) Test-case 3; Parameters for the smooth stage: $w_c = 4$, $w_l = 1$, $w_a = 5.0$, $w_p = 0.16$.
        }} \\
    \end{tabular}
    \caption{\rv{Moscow metro system computed with different weights for the smooth optimization process. The mixed stage was computed with the parameters $w_o= 2$, $w_p = 0.1$ and $w_c = 10$.
    Left: smooth layout, center: mixed layout and right: grid aligned layout.
    }}
    \label{fig:moscow-smooth}
    }
\end{figure*}

\begin{figure*}[t]
    \centering{
    \setlength{\tabcolsep}{0pt}
    \begin{tabular}{ccc}
        \includegraphics[width=0.33\linewidth]{figures/parameters/moscow/default/smooth/Metro.png} &
        \includegraphics[width=0.33\linewidth]{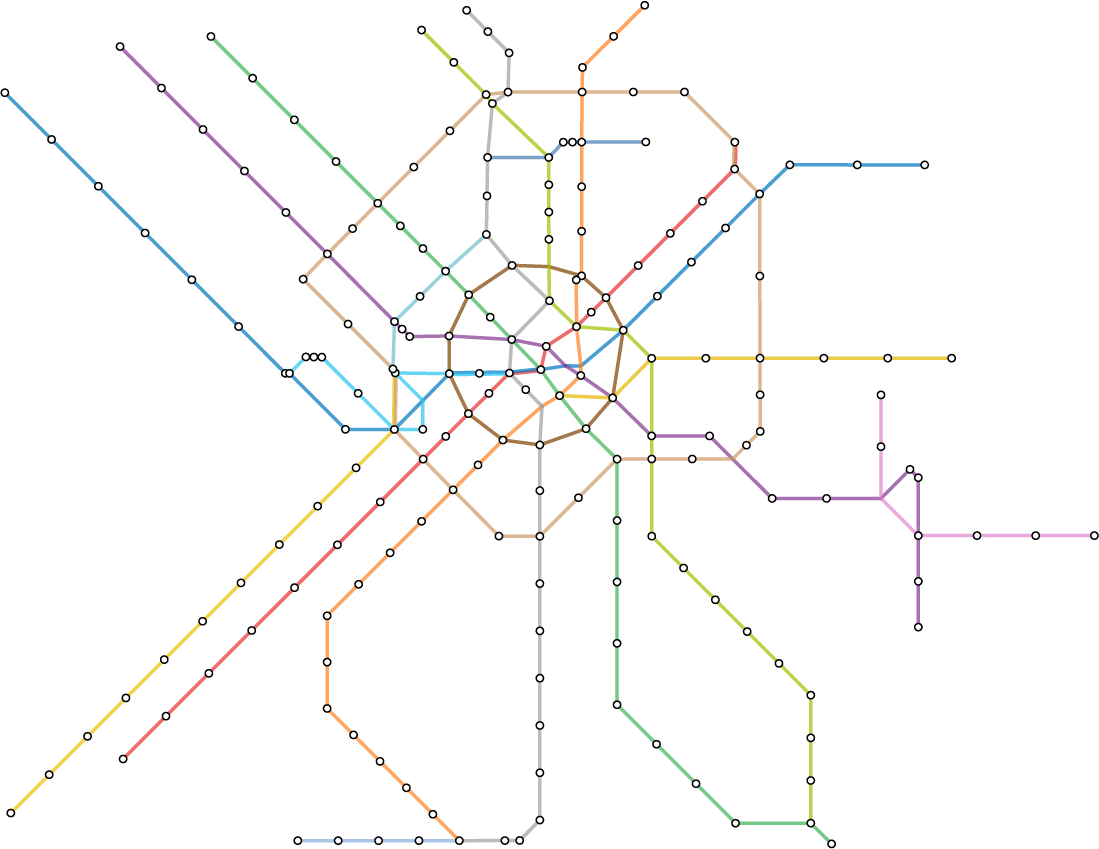} &
        \includegraphics[width=0.33\linewidth]{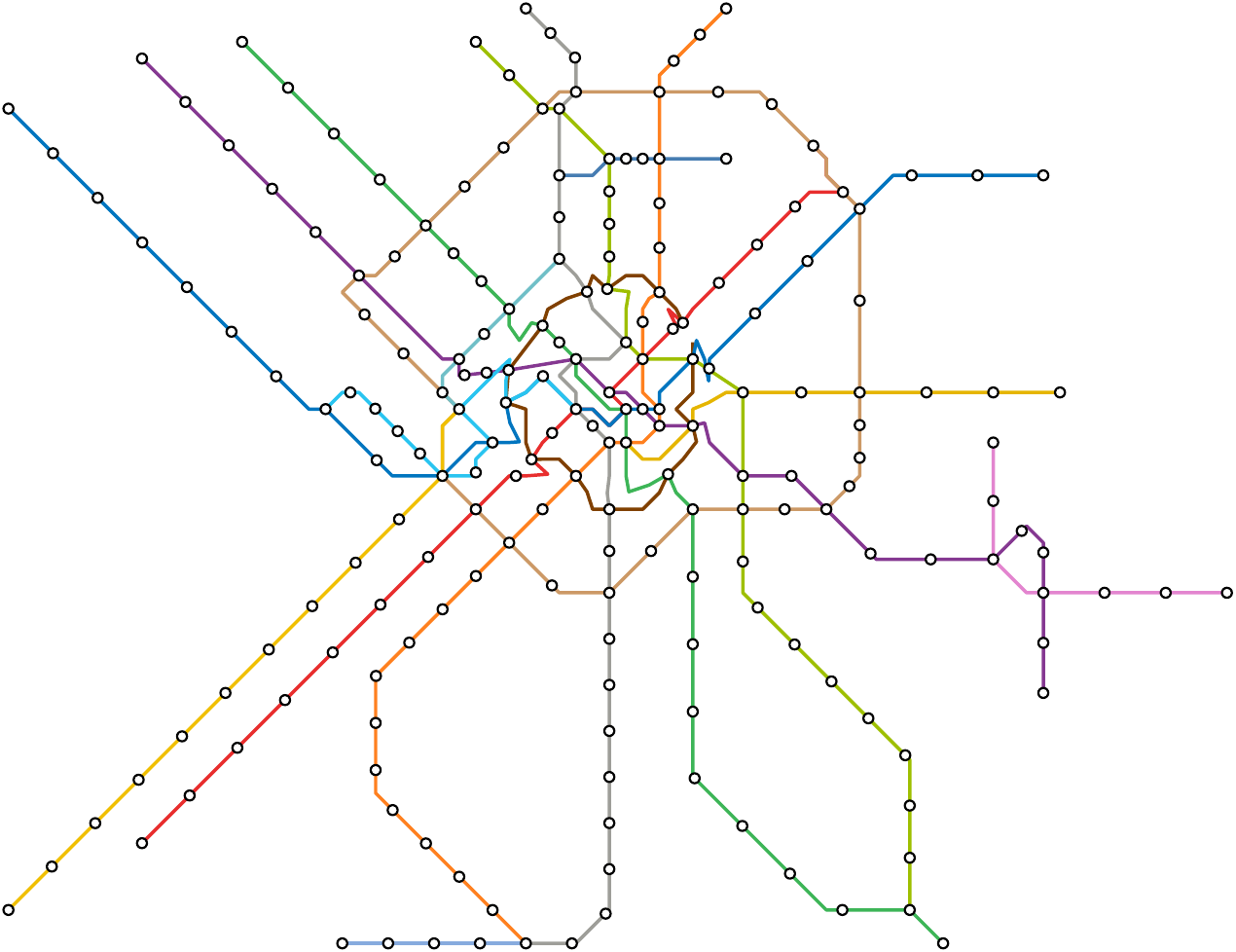} \\
         \multicolumn{3}{c}{\rv{(e) Test-case 4; Parameters for the mixed stage: $w_o = 5$, $w_p = 0.1$, $w_c = 5$.}} \\
        \includegraphics[width=0.33\linewidth]{figures/parameters/moscow/default/smooth/Metro.png} &
        \includegraphics[width=0.33\linewidth]{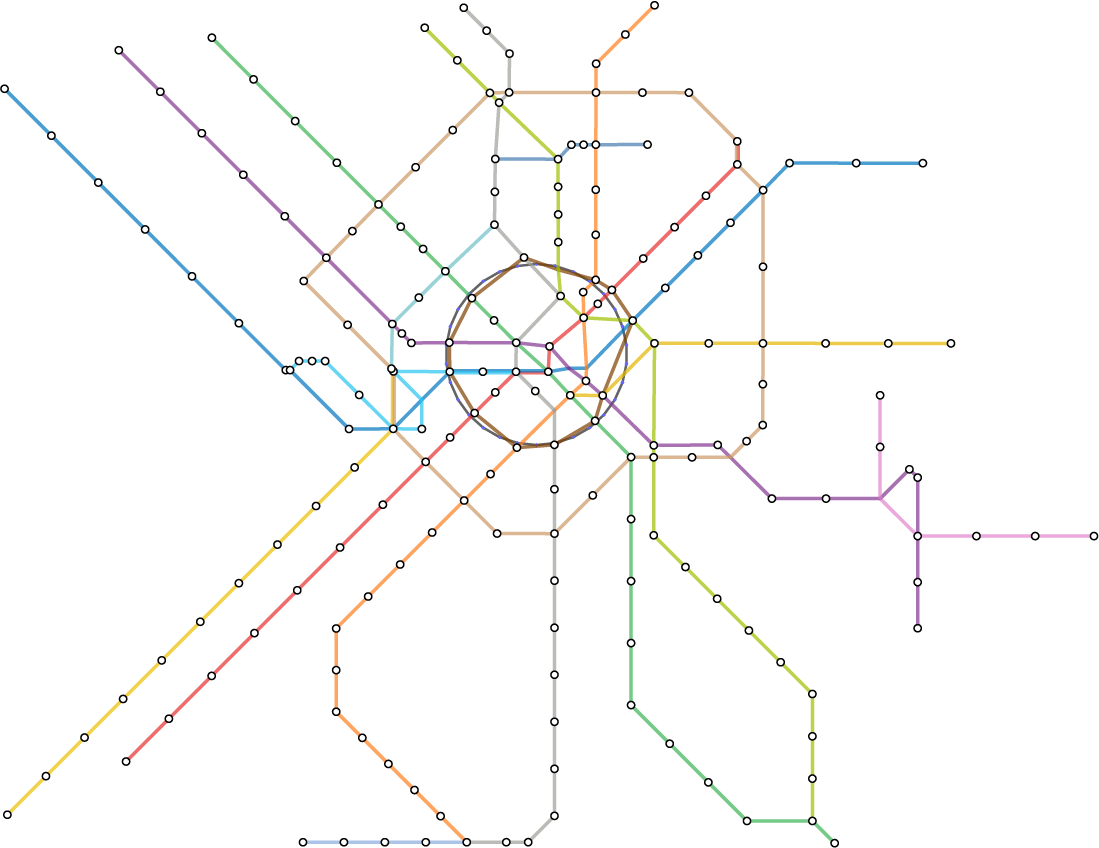} &
        \includegraphics[width=0.33\linewidth]{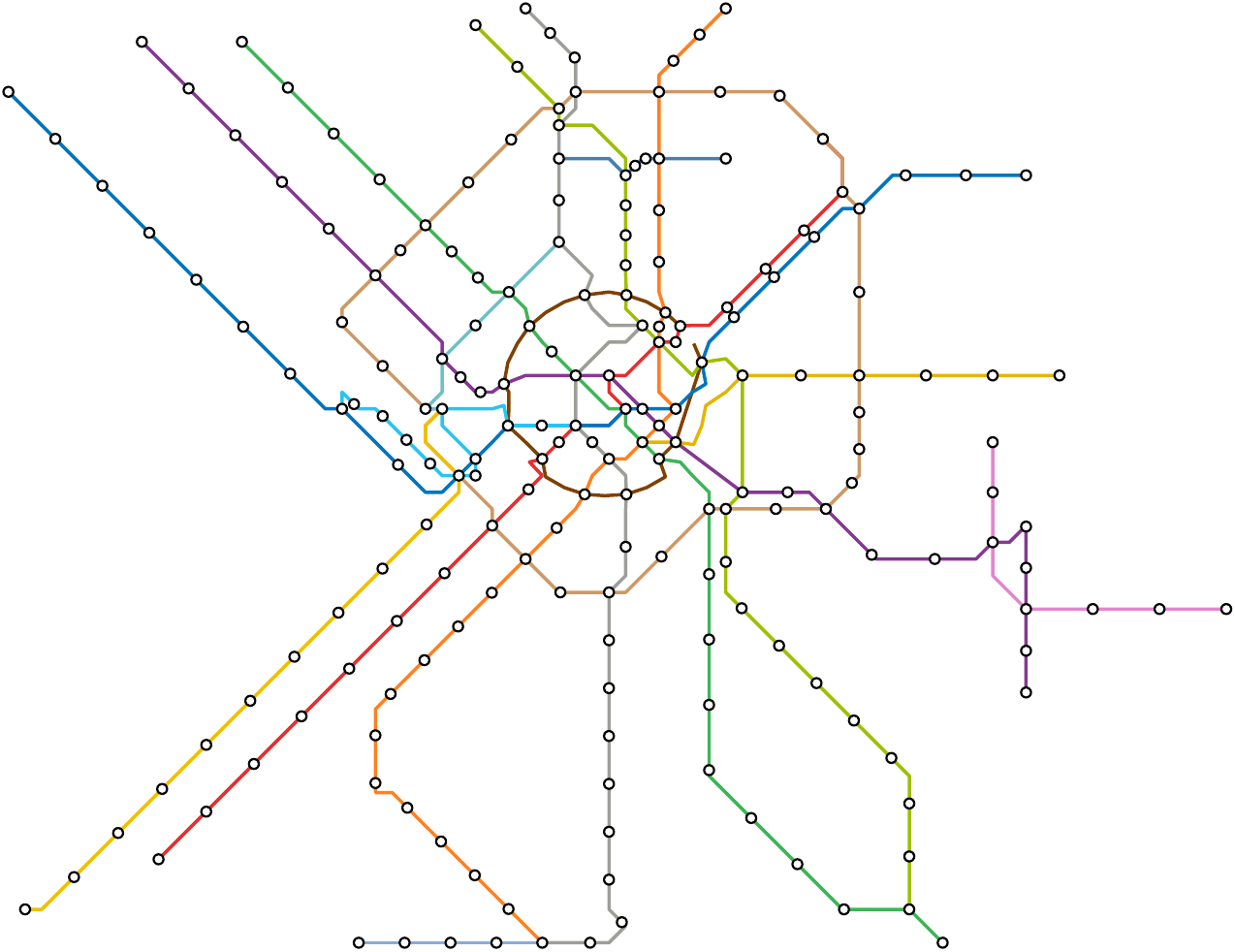} \\
        \multicolumn{3}{c}{\rv{(f) Test-case 5; Parameters for the mixed stage: $w_o = 10$, $w_p = 0.1$, $w_c = 2$.}} \\
        \includegraphics[width=0.33\linewidth]{figures/parameters/moscow/default/smooth/Metro.png} &
        \includegraphics[width=0.33\linewidth]{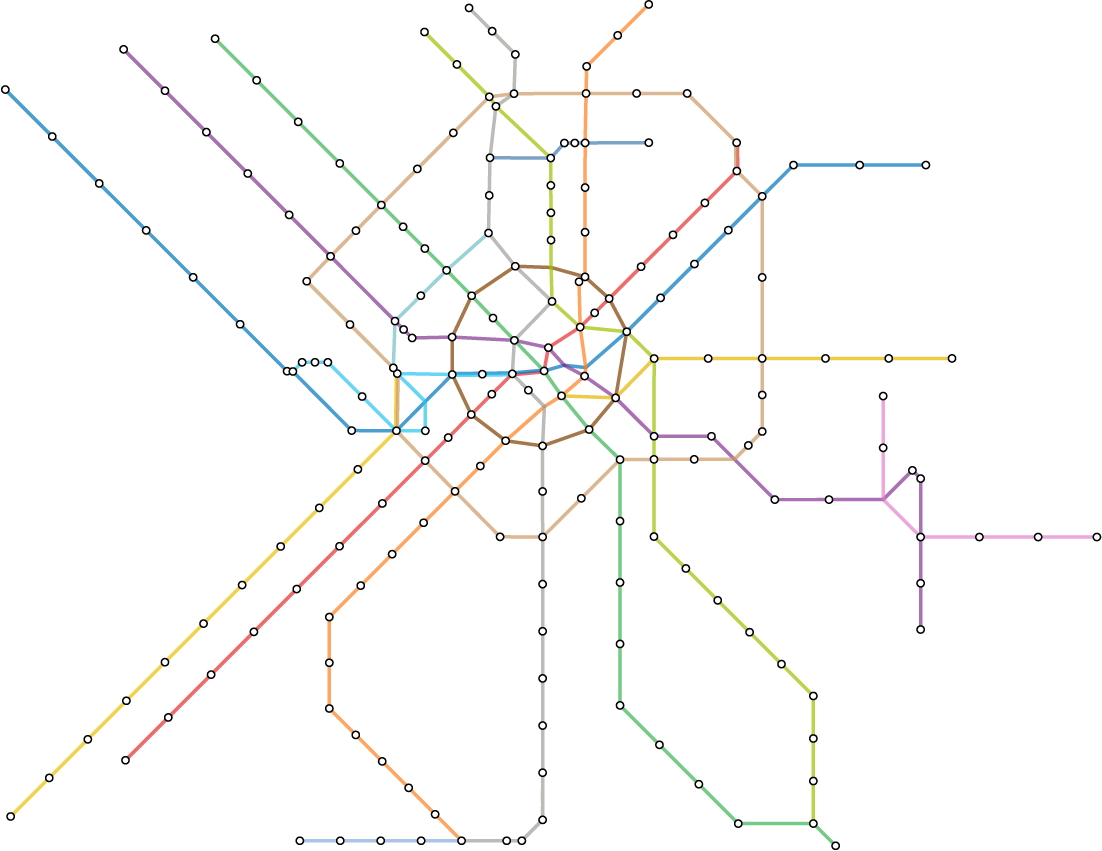} &
        \includegraphics[width=0.33\linewidth]{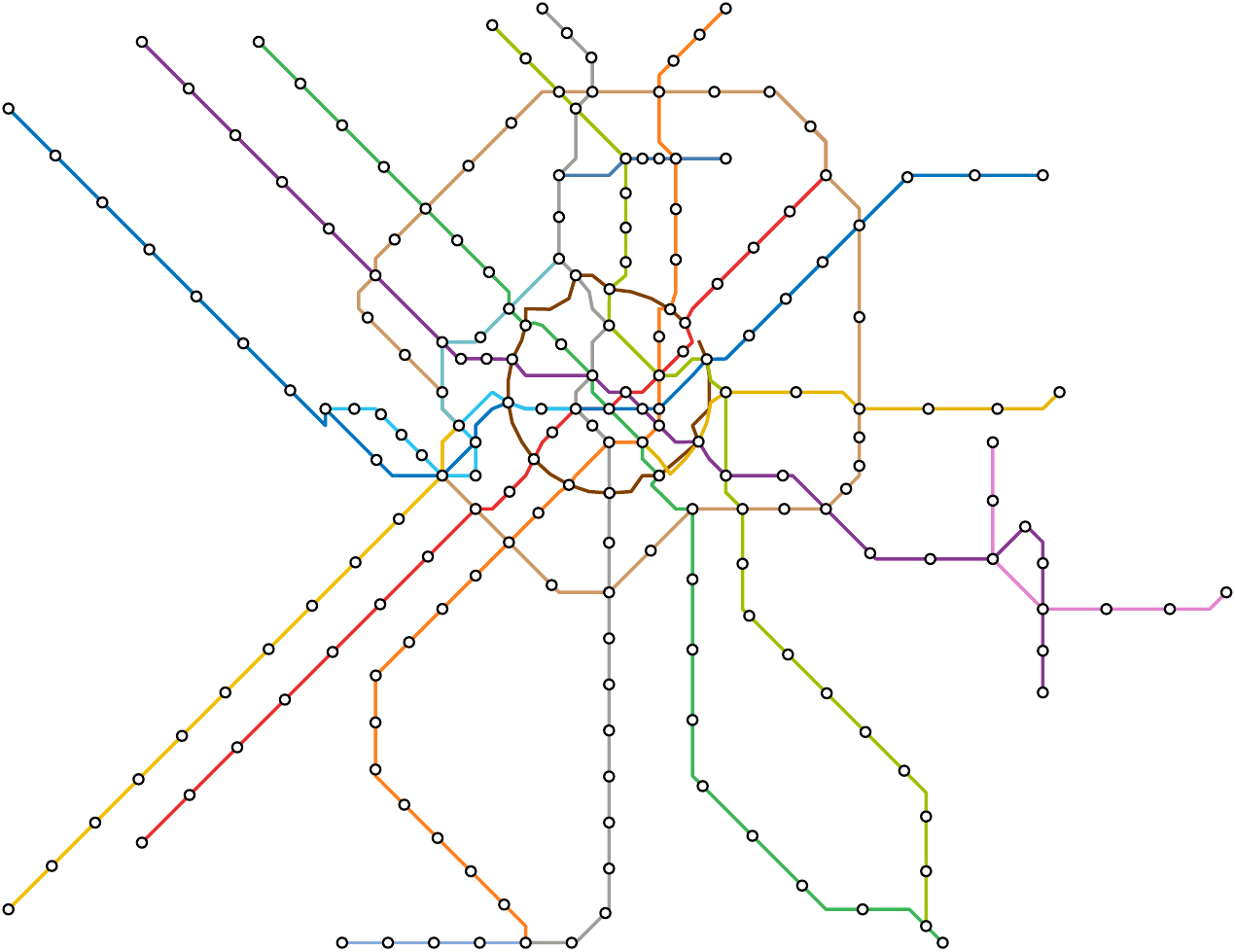} \\
        \multicolumn{3}{c}{\rv{(g) Test-case 6; Parameters for the mixed stage: $w_o = 10$, $w_p = 0.1$, $w_c = 10$.}} \\
        \includegraphics[width=0.33\linewidth]{figures/parameters/moscow/default/smooth/Metro.png} &
        \includegraphics[width=0.33\linewidth]{figures/parameters/moscow/default/mixed/Metro.png} &
        \includegraphics[width=0.33\linewidth]{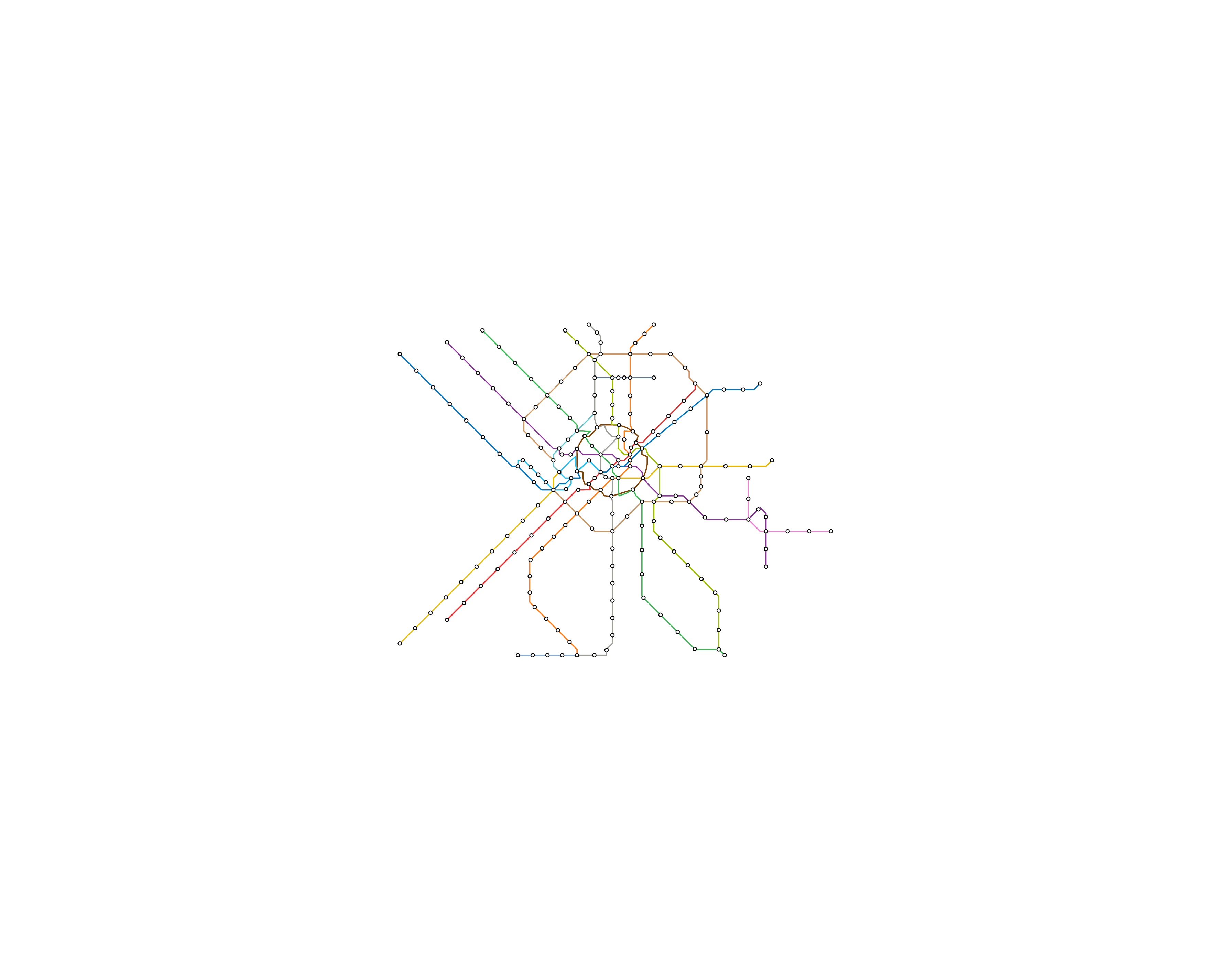} \\
        \multicolumn{3}{c}{\rv{(h) Test-case 7; Parameters for the mixed stage: $w_o = 2$, $w_p = 0.1$, $w_c = 2$.}} \\
    \end{tabular}
    \caption{\rv{Moscow metro system computed with different weights for the mixed optimization process. The smooth stage was computed with the parameters $w_c = 10$, $w_l = 1$, $w_a = 2$ and $w_p = 0.16$. Left: smooth layout, center: mixed layout and right: grid aligned layout. Note that grid alignment layouts can have up to 5 edges, which could not be routed.}}
    \label{fig:moscow-mixed}
    }
\end{figure*}

\rv{To visually show the effects of different weight values in Equations~(\ref{equ:smooth}) and~(\ref{equ:mixed}), we selected different parameter values for demonstration (see Figures~\ref{fig:taipei-smooth}-\ref{fig:moscow-mixed}). 
In Figure~\ref{fig:taipei-smooth}(a) and~\ref{fig:moscow-smooth}(a), we applied the weights $w_c = 4$, $w_l = 1$, $w_a = 2$, and $w_p = 0.16$ for the smooth process. For the test-case 1, we adapted $w_c = 10.0$, for test-case 2, we used $w_a = 0.5$, and for test-case 3, we set $w_a = 5.0$. The corresponding results are shown in Figures~\ref{fig:taipei-smooth}(b)-(d) and ~\ref{fig:moscow-smooth}(b)-(d). For the mixed layout, we applied the default weights $w_o = 2.0$, $w_p = 0.1$, and $w_c = 10.0$. The default test-case and the test cases 1 to 3 are computed with these default mixed weights.}

\rv{The effects of weights for the mixed process are depicted in Figures \ref{fig:taipei-mixed} and \ref{fig:moscow-mixed}. For the test-cases 4 to 7, the smooth layout is computed with the default parameters, and for the test case 4, we adapted $w_o$ and $w_c$ to $5.0$. For the test-case 6, we use $w_o = 10.0$ and for the test-case 7, we applied $w_c = 2$.
The results of the effects are further summarized in the following paragraphs.}

\subsection{Smooth Layout Variables (Test-cases 1-3)}\label{ssec:smoothtests}

\rv{In the smooth results of Taipei with the default parameters (Figure \ref{fig:taipei-smooth}(a)), one can see that the metro lines are straightened and stations are evenly spaced apart. Test-case 2 depicts the effect of $w_a$ clearly (Figure \ref{fig:taipei-smooth}(c) and \ref{fig:moscow-smooth}(c)). By decreasing the $w_a$, the system prioritizes the other objective functions and metro lines are less straight, resulting in a less clear layout after the mixed stage. By increasing $w_a$ (Test-case 3, Figure \ref{fig:taipei-smooth}(d) and \ref{fig:moscow-smooth}(d)) metro lines are even more straightened at the cost of the other objective functions.  This results in a layout that does not preserve the geographic shape of the metro network. For a more complex network like Moscow, the effects of the different parameters are even more significant. This indicates that the potential conflicts between the different objective functions and that the system is not able to find a reasonable solution. For example, if $w_a$ is decreased, the approach does not produce appealing results.} 

\subsection{Mixed Layout Variables (Test-cases 4-7)} \label{ssec:mixedtests}

\rv{For the Taipei test data, the variation of the weights $w_o$ and $w_c$ in the mixed stage have relatively little effect on the results of the final layout, as one can see in Figure~\ref{fig:taipei-mixed}. This implies that the process is relatively stable and the system can find a good solution for the objective function. For the Moscow test cases, the system fails to rotate a large amount of edges to octolinear direction, independently from the parameters, especially in the dense center of the map. Therefore, when the weight $w_c$ is decreased (e.g., Test-case 5 shown in Figure~\ref{fig:moscow-mixed}(e)), the system prioritizes the octolinear edge constraint and pushes stations away from the guide that should be located close to the guide shape.}

\rv{When comparing different results of the smaller and less complex metro systems to the larger and more complex Moscow system, we can see that in the complex systems, the weights have to be balanced carefully, while in simple networks the system produces good potential results for a variety of parameters and finds a reasonable solution that satisfies different objective terms.}

\end{document}